\documentclass[prd]{revtex4-1}

\usepackage{amsmath,amssymb,amsfonts,amsthm}
\usepackage{graphicx}
\usepackage[labelformat=empty]{subfig}
\usepackage{pstricks}

\newcommand{\g}{\gamma}
\newcommand{\T}{\Theta}

\newcommand{\br}{\bar{r}}
\newcommand{\bK}{\bar{K}}
\newcommand{\bC}{\bar{C}}
\newcommand{\bL}{\bar{L}}
\newcommand{\ba}{\bar{a}}
\newcommand{\bQ}{\bar{Q}}
\newcommand{\bP}{\bar{P}}
\newcommand{\bDelta}{\bar{\Delta}}
\newcommand{\mR}{\mathcal{R}}
\newcommand{\mT}{\mathcal{T}}
\newcommand{\mQ}{\mathcal{Q}}
\newcommand{\mP}{\mathcal{P}}
\newcommand{\sgn}{\text{sgn}}

\begin{document}

\title{Charged particle motion in Kerr-Newmann space-times}

\author{Eva Hackmann}
\email{eva.hackmann@zarm.uni-bremen.de}
\affiliation{ZARM, University of Bremen, Am Fallturm, 28359 Bremen, Germany}

\author{Hongxiao Xu}
\email{hxu@uni-bremen.de}
\affiliation{ZARM, University of Bremen, Am Fallturm, 28359 Bremen, Germany}

\begin{abstract}
The motion of charged test-particles in the gravitational field of a rotating and electromagnetically charged black hole as described by the Kerr-Newman metric is considered. We completely classify the colatitudinal and radial motion on the extended manifold $-\infty \leq r \leq \infty$, including orbits crossing the horizons or $r=0$. Analytical solutions of the equations of motion in terms of elliptic functions are presented which are valid for all types of orbits.
\end{abstract}

\maketitle

\section{Introduction}

The Kerr-Newman solution to the Einstein-Maxwell equations describes the gravitational field of a rotating and electromagnetically charged stationary black hole \cite{Newmanetal65}. It generalizes both the static and charged Reissner-Nordstr\"om metric \cite{Reissner16,Nordstrom18} as well as the rotating Kerr metric \cite{Kerr63}. The latter is of very high importance not only for general relativity but also from an astrophysical point of view, as many black hole candidates were found in recent years, which are expected to rotate. Although it is not very likely that they also carry a net charge, some accretion scenarios were studied which may create such black holes \cite{Damouretal1978,Ruffinietal2010}.

On way to explore the gravitational field of a Kerr-Newman black hole is to consider the geodesic motion of (charged) test-particles in this space-time. Already shortly after the discovery of this solution many aspects of the geodesic motion were studied, among others, timelike equatorial and spherical orbits of uncharged particles \cite{Johnstonetal1974} and the last stable orbit of charged particles \cite{Young1976} (see also Sharp \cite{Sharp1979} and references within). Later the motion of charged particles was studied by Bi\v{c}\'{a}k et al \cite{Bicaketal1989,Bicaketal1989_2} including, besides some general discussion for the radial motion, a discussion of a number of special cases like motion along the symmetry axis and circular motion of ultrarelativistic particles. Only recently, Kov\'{a}\v{r} et al \cite{Kovaretal2008} found off-equatorial circular orbits of charged particles which are unstable outside the outer horizon, and Pugliese et al \cite{Pugliese2013} used equatorial circular orbits of neutral test-particles to distinguish between black holes and naked singularities. A comprehensive analysis of photon orbits in Kerr-Newman space-time was presented by Calvani and Turolla \cite{Calvanietal1981} including the extended manifold with negative values of the radial coordinate and naked singularities.

Analogously to the uncharged case, the geodesic equation in Kerr-Newman space-time can be separated by introducing an additional constant of motion (besides the constants associated to the obvious symmetries of the space-time), the Carter constant \cite{Carter68}, which ensures the integrability of the equations of motions. The resulting structure of the equations of motion is essentially the same as in Schwarzschild space-time, where they can be solved analytically in terms of elliptic functions as first demonstrated by Hagihara in 1931 \cite{Hagihara}. However, due to the remaining coupling of radial and colatitudinal equation, the generalization of his method to Kerr(-Newman) space-time was not straightforward. This issue was solved by Mino \cite{Mino03} by introducing a new time parameter, often called the Mino time, which completely decouples the equations of motion and enables a straightforward application of elliptic functions. This was already used to analytically solve the geodesic equation for bound timelike orbits in Kerr space-time by Fujita and Hikida \cite{FujitaHikida09} and for general timelike and lightlike orbits in Kerr space-time in \cite{eva}.

In this paper, we will discuss the geodesic motion of charged test-particles in Kerr-Newman black hole space-times. For the sake of completeness, we will include a magnetic charge of the black hole which was not done in the cited references but has interesting effects on the colatitudinal motion. After introducing the relevant notations and equations of motion in the next section, we proceed with a complete classification of timelike orbits of (charged) particles in Kerr-Newman space-time. This includes off-equatorial orbits, trajectories crossing the horizons, and orbits with negative values of the radial coordinate. In the fourth section, we will present analytical solutions in terms of elliptic functions dependent on the Mino time for all coordinates. The paper is closed by a summary and conclusion.

\section{Geodesics in Kerr-Newman space-time}

\subsection{Kerr-Newman space-time}
The Kerr-Newman spacetime is a stationary, axisymmetrical, and asymptotically flat solution of the Einstein-Maxwell equation
\begin{align*}
G_{\mu\nu}=-2\Big( g^{\alpha\beta} {F_{\mu\alpha}}F_{\nu\beta}-\frac{1}{4} g_{\mu\nu}F_{\alpha\beta}F^{\alpha\beta}\Big)\,,
\end{align*}
where $G_{\mu\nu}$ is the Einstein tensor and $F_{\mu\nu}$ the electromagnetic tensor. Throughout, the units are chosen such that $c=1$ for the speed of light and $G=1$ for the gravitational constant. In Boyer-Lindquist coordinates the metric takes the form
\begin{equation}
ds^2 = \frac{\rho^2}{\Delta} dr^2 + \rho^2 d\theta^2 + \frac{\sin^2\theta}{\rho^2} \left[ (r^2 +a^2) d\phi - a dt \right]^2 - \frac{\Delta}{\rho^2} \left[  a \sin^2(\theta)  d\phi - dt \right]^2 \label{eq:metric}
\end{equation}
with
\begin{align}
\rho^2(r,\theta) & = r^2+a^2 \cos^2 \theta\,, \label{eq:rho}\\
\Delta(r) & = r^2 - 2 M r + a^2+Q^2 +P^2\,,\label{eq:Delta}
\end{align}
and $M>0$ the mass, $a$ the specific angular momentum, $Q$ the electric, and $P$ the magnetic charge of the gravitating source. (The existence of magnetic charges has not been proven yet but it will be considered for the sake of completeness.) We restrict ourselves here to the case that two horizons exist, given by the coordinate singularities $\Delta(r)=0$, $r_\pm = M \pm \sqrt{M^2-a^2-Q^2-P^2}$. The only genuine singularity is for $\rho^2=0$, where $r= 0$ and $\theta = \pi/2$ is fulfilled simultaneously. This means that a test particle approaching $r=0$ from above or below the equatorial plane does not terminate at $r=0$ as it would in Schwarzschild space-time but continues to negative values of $r$. For large negative values of $r$ this can be interpreted as a ``negative universe'', see \cite{HawkingEllis73}.

The Kerr-Newman metric reduces to the Kerr metric for $Q=P=0$ describing the exterior of rotating non charged black holes. It reduces to the Reissner-Nordstr\"om metric for $a=0$ which describes the exterior of a non rotating but charged black hole. In the case $P=Q=a=0$ the Kerr-Newman metric is reduced to the Schwarzschild metric.

The electromagnetic potential is given by
\begin{align}
A = A_\nu  dx^\nu = \frac{Qr}{\rho^2}(dt-a\sin^2\theta d\phi) + \frac{1}{\rho^2}P\cos\theta \left(a dt -(r^2+a^2)d\phi\right)\,, \label{eq:potA}
\end{align}
from which the electromagnetic tensor can be calculated by $F=\frac{1}{2} (\partial_\nu A_\mu -\partial_\mu A_\nu)  dx^\mu \wedge dx^\nu$. By the interchanges $Q\rightarrow P$, $P\rightarrow -Q$ the electromagnetic potential $\check{A}$ of the dual electromagnetic tensor $\check{F}$ can be obtained.

\subsection{Equations of motion}
The equations of motion for a test particle of normalized mass $\epsilon$, electric charge $e$, and magnetic charge $h$ can be obtained by the Hamiltonian
\begin{equation}
H = \frac{1}{2} g^{\mu\nu} (\pi_\mu + e A_\mu+h{\check{A}_\mu})(\pi_\nu + e A_\nu + h {\check{A}_\nu}) \label{eq:H}
\end{equation}
where $\pi_\mu$ describe the generalized momenta. By introducing
\begin{equation}
\hat{Q}=\frac{e Q+h P}{\sqrt{e^2+h^2}}, \qquad \hat{P} = \frac{eP-hQ}{\sqrt{e^2+h^2}}, \qquad \hat{e}=\sqrt{e^2+h^2}
\end{equation}
the Hamiltonian can be reduced to
\begin{equation}
H = \frac{1}{2} \hat{g}^{\mu\nu} (\pi_\mu + \hat{e} \hat{A}_\mu)(\pi_\nu + \hat{e} \hat{A}_\nu )\,, \label{eq:H1}
\end{equation}
where $\hat{g}$ and $\hat{A}$ are defined by \eqref{eq:metric} and \eqref{eq:potA} with $P,Q,e$ replaced by $\hat{P}, \hat{Q}, \hat{e}$. Therefore, the discussion of a test particle without magnetic charge is sufficient. In the following we omit the hats for brevity.

We can obtain three constants of motion directly, since $H$ does not depend on $\tau$, $\phi$, or $t$. The first, $\epsilon^2 = - g_{\nu \mu}\dot{x}^\nu \dot{x}^\mu$ is the normalization condition with $\epsilon=1$ for timelike and $\epsilon = 0$ for lightlike trajectories. (The dot denotes differentiation with respect to an affine parameter $\tau$.) The second and third equation
\begin{align}
E &= -\pi_t = -g_{tt} \dot{t} - g_{t\phi} \dot{\phi} + eA_t\,, \\
L &= \pi_\phi = g_{\phi t} \dot{t} + g_{\phi\phi} \dot{\phi} - e A_\phi\,,
\end{align}
describe the conservation of energy $E$ and angular momentum in $z$ direction, respectively. A fourth constant of motion can be obtained by considering the Hamilton-Jacobi equation
\begin{align}
- \partial_\tau S = \frac{1}{2} g^{\mu\nu} (\partial_\mu S +e A_\nu) (\partial_\nu S+e A_\mu)\,. \label{eq:HJ1}
\end{align}
With the ansatz $S= \frac{1}{2} \epsilon \tau - E t + L\phi + S_1(r) + S_2(\theta)$ it can be shown that the Hamilton-Jacobi equation indeed separates with the Carter constant $K$ as separation constant, see \cite{Carter68}.

With these four constants the equations of motion become
\begin{align}
\left( \frac{d\theta}{d\g} \right)^2 & = \bK-\epsilon \ba^2\cos^2\theta-\frac{\mT^2(\theta)}{\sin^{2}\theta} =: \T(\theta) \,, \label{eq:eomtheta}\\
\left( \frac{d\br}{d\g} \right)^2 & = \mR^2(\br)-(\epsilon \br^2  + \bK) \bDelta(\br) =: R(\br)\,, \label{eq:eomr}\\
\frac{d\phi}{d\g} & = \frac{\ba\mR(\br)}{\bDelta(\br)}  - \frac{\mT(\theta)}{\sin^2\theta}\,, \label{eq:eomphi}\\
\frac{dt}{d\g} & = \frac{(\br^2+\ba^2)\mR(\br)}{\bDelta(\br)} - \ba \mT(\theta) \,, \label{eq:eomt}
\end{align}
where 
\begin{align}
\mR(\br) & = (\br^2+\ba^2)E-\ba\bL-e\bQ\br \,, \label{mathcalR}\\
\mT(\theta) & = \ba E\sin^2\theta -\bL+e\bP\cos\theta\,.
\end{align}
All quantities with a bar are normalized to $M$, i.e.~$x=\bar{x}M$ for $x=r,a,L,Q,P$ as well as $K=\bK M^2$ and $\bDelta(\br)=\br^2-2\br+\ba^2+\bQ^2+\bP^2$. Here $\g$ is the normalized Mino time \cite{Mino03} given by $d\gamma = M \rho^{-2} d\tau$ with the eigentime $\tau$.

\section{Classification of motion}
In this section we will classify the types of orbits in terms of colatitude and radial motion. We will analyze which orbit configuration, i.e.~which set of orbit types, may appear for given parameters $\ba,\bQ,\bP,E,\bL,\bK,e$ and which region in parameter space a given orbit configurations occupies. Here we assume that $\epsilon=1$, that is, we restrict ourselves to test particles with mass, but the discussion for light may be done analogously. The whole analysis will be based on the conditions $\T(\theta)\geq 0$ and $R(\br) \geq 0$ which are necessary for geodesic motion. 

For both colatitudinal and radial motion, we will first give some general properties as symmetries and notation of orbit types. We then proceed with the determination of possible orbit configurations, i.e.~sets of orbit types which are possible for given parameters (for more than one possible orbit type the actual orbit is determined by initial conditions). Each orbit configuration covers a particular region in the parameter space. Finally, we will analyze how these regions look like and determine their boundaries in parameter space.

\subsection{Colatitudinal motion}
The coordinate $\theta$ may only take a specific value $\theta_0 \in [0, \pi]$ if $\T(\theta_0) \geq 0$ is valid. We will analyze in the following whether this is fulfilled for a given parameter set. 

\subsubsection{General properties}
First, we notice that $\T$ does not depend on $\bQ$ and that $e$ and $\bP$ only appear as the product $e\bP$. The function $\T$ has the following symmetries:
\begin{itemize}
\item A change of sign of $e\bP$ has the same effect as reflecting $\theta$ at the equatorial plane: $\T|_{-e\bP}(\theta)=\T|_{e\bP}(\pi-\theta)$. In particular is $\T$ symmetric with respect to the equatorial plane if $e\bP=0$: $\T|_{e\bP=0}(\theta)=\T|_{e\bP=0}(\pi-\theta)$.
\item A simultaneous change of sign of $\bL$ and $E$ result in a reflection at the equatorial plane: $\T|_{-\bL,-E}(\theta)=\T|_{\bL,E}(\pi-\theta)$ and $\T_{\bL=0=E}(\theta) = \T_{\bL=0=E}(\pi-\theta)$.
\end{itemize}
Therefore, we assume w.l.o.g. $e\bP\geq0$ and $E\geq0$ in the following. The condition $\T(\theta)\geq 0$ also shows that $\bK\geq0$ is a necessary condition for geodesic motion as all other (positive) terms are subtracted. 

The Carter constant also encodes some geometrical information if considered in its alternative form $\bC=\bK-(\ba E-\bL)^2$. Because of $\T(\pi/2) = \bK-(\ba E-\bL)^2 = \bC$ a particle may only cross or stay in the equatorial plane if $\bC\geq0$. For equatorial orbits even $\bC=0$ is necessary as $\frac{d\theta}{d\g}\left(\frac{\pi}{2}\right)=0$ needs to be fulfilled. For $\bK=0$ geodesic motion is only possible if $\ba=\bL=0$ or in the equatorial plane with $\ba E=\bL$.

The function $\T(\theta)$ constains a term which diverges for $\theta=0,\pi$ given by $\frac{-(\bL-e\bP\cos\theta)^2}{\sin^2\theta}$. This fact suggests to distinguish between two cases:
\begin{itemize}
\item $\bL \neq \pm e\bP$: In this case $\T(\theta)\to-\infty$ for $\theta \to 0,\pi$, that is, the north or south pole will never be reached.
\item $\bL = \pm e\bP$: In this case $\T(\theta) \to \bK-\ba^2$ for $\theta \to 0$ and $\bL=e\bP$ as well as for $\theta \to \pi$ and $\bL=-e\bP$. Therefore, a particle with $\bL=e\bP$ may reach the north pole and a particle with $\bL=-e\bP$ the south pole if in addition $\bK\geq\ba^2$. For the subcase $\bL=0=e\bP$ both north and south pole may be reached if $\bK\geq\ba^2$.
\end{itemize}
If the parameters are such that the poles $\theta=0,\pi$ can not be reached it is convenient to consider $\nu=\cos\theta$ instead of $\theta$. In terms of $\nu$ the differential equation for colatitudinal motion reads
\begin{align}\label{eq:eomnu}
\left( \frac{d\nu}{d\gamma} \right)^2 & = \sum_{i=0}^4 b_i\nu^i =: \T_\nu(\nu)\,,
\end{align}
where $b_0=\bK-(\bL-\ba E)^2=\bC$, $b_1=2e\bP(\bL-\ba E)$, $b_2=-\bK-\ba^2-e^2\bP^2+2\ba^2E^2-2\ba E\bL$, $b_3=2\ba Ee\bP$, and $b_4=\ba^2(1-E^2)$.

For a given set of parameters of the space-time and the particle different types of orbits may be possible. We call an orbit
\begin{itemize}
\item northern or $N$, if it stays in the northern hemisphere $\theta<\pi/2$,
\item normal or $E$, if it crosses or stays in the equatorial plane $\theta=\pi/2$,
\item southern or $S$, if it stays in the southern hemisphere $\theta>\pi/2$.
\end{itemize}
Equatorial orbits with $\theta \equiv \pi/2$ are a special case of normal orbits. In addition to the above notions we add an index $N$ if the north pole $\theta=0$ and $S$ if the south pole $\theta=\pi$ is reached, for example $N_N$ for a northern orbit reaching the north pole.

\subsubsection{Orbit configurations}
Let us now analyze which orbit configurations, i.e.~which sets of the above introduced orbit types, are possible for given parameters. We use the necessary condition for colatitudinal motion $\T(\theta)\geq0$ for this, which implies to analyze the occurrence of real zeros of $\T$ in $[0,\pi]$ and the behavior of $\T$ at the boundaries $\theta=0,\pi$ giving the sign of $\T$ between its zeros. (The orbit a test particle with the given parameters actually follows in a space-time with the given parameters depends on the initial values.)
\paragraph*{(A) Case $\bL\neq \pm e\bP$} Here $\T(\theta)=-\infty$ at $\theta=0,\pi$ which implies that $\T_\nu$ has an even number of zeros in $(-1,1)$ (counted with multiplicity). If $\T_\nu$ has no real zeros there, no colatitudinal motion is possible, which gives a restriction to the permitted sets of parameters for geodesic motion. In the case of 2 real zeros there is a single orbit of type $N$, $E$, or $S$, which is stable at a constant $\theta$ if the 2 zeros coincide. If all 4 zeros of $\T_\nu$ lie in $(-1,1)$ all combinations of two orbit types except $EE$ are possible. For two or more coinciding zeros this point is stable if it is a maximum of $\T_\nu$ and unstable otherwise.
\paragraph*{(B) Case $\bL=\pm e\bP$} Here we have to consider four subcases: \textit{(B1)} $\bK<\ba^2$: The same orbit types as in (A) are possible. \textit{(B2)} $\bK>\ba^2$, $\bL \neq 0$: In this case $\T$ has different signs at $\theta=0,\pi$ which implies that $\T$ has an odd number of real zeros in $(0,\pi)$. For 1 real zero there is one orbit of type $N_N$ or $E_N$ for $\bL=e\bP$ and one of type $S_S$ or $E_S$ for $\bL=-e\bP$. If $\T$ has 3 real zeros in $(0,\pi)$ there is one additional orbit not reaching a pole. \textit{(B3)} $\bK>\ba^2$, $\bL=0=e\bP$: Here $\T>0$ at $\theta=0,\pi$, i.e.~$\T$ has an even number of zeros in $(0,\pi)$. Also, for $e\bP=0$ the function $\T$ is symmetric with respect to the equatorial plane. For no real zeros there is one orbit of type $E_{NS}$ which reaches both poles, and for 2 real zeros an $N_N$ and an $S_S$ orbit. More zeros in $[0,\pi]$ are not possible. \textit{(B4)} $\bK=\ba^2$: Here an orbit with constant $\theta=0$ ($\theta=\pi$) is possible for $\bL=e\bP$ ($\bL=-e\bP$). For $\bL=0$ no other than the two constant orbits are possible but for $\bL\neq0$ it is $\T \to -\infty$ at the other boundary. In the latter case, if the orbit is stable, there may be one additional orbit of type $E$, $N$, or $S$.\\
For an overview of this different orbit configuration see table \ref{tab:theta}.

\begin{table}
\begin{minipage}{0.48\textwidth}
\subfloat[(A) $\bL\neq \pm e\bP$; (B1) $\bL=\pm e\bP$, $\bK<\ba^2$]{
\begin{tabular}{|c|c|c|}
\hline
zeros & range of $ \theta \in [0,\pi]$ & types of orbits \\ 
\hline\hline
2 &
\begin{pspicture}(-2.3,-0.24)(2.3,0.24)
\psline[linewidth=0.5pt]{|-|}(-2.0,0)(2.0,0)
\psline[linewidth=0.5pt]{-}(0,-0.08)(0,0.08)
\psline[linewidth=1.2pt]{*-*}(-1.6,0)(-0.4,0)
\end{pspicture} & $N$\\
2 &
\begin{pspicture}(-2.3,-0.24)(2.3,0.24)
\psline[linewidth=0.5pt]{|-|}(-2.0,0)(2.0,0)
\psline[linewidth=0.5pt]{-}(0,-0.08)(0,0.08)
\psline[linewidth=1.2pt]{*-*}(-1.1,0)(1.1,0)
\end{pspicture} & $E$\\
2 &
\begin{pspicture}(-2.3,-0.24)(2.3,0.24)
\psline[linewidth=0.5pt]{|-|}(-2.0,0)(2.0,0)
\psline[linewidth=0.5pt]{-}(0,-0.08)(0,0.08)
\psline[linewidth=1.2pt]{*-*}(1.6,0)(0.4,0)
\end{pspicture} & $S$\\
\hline
4 &
\begin{pspicture}(-2.2,-0.2)(2.2,0.2)
\psline[linewidth=0.5pt]{|-|}(-2.0,0)(2.0,0)
\psline[linewidth=0.5pt]{-}(0,-0.08)(0,0.08)
\psline[linewidth=1.2pt]{*-*}(-0.9,0)(-0.2,0)
\psline[linewidth=1.2pt]{*-*}(-1.8,0)(-1.2,0)
\end{pspicture} & $N$, $N$\\
4 &
\begin{pspicture}(-2.2,-0.2)(2.2,0.2)
\psline[linewidth=0.5pt]{|-|}(-2.0,0)(2.0,0)
\psline[linewidth=0.5pt]{-}(0,-0.08)(0,0.08)
\psline[linewidth=1.2pt]{*-*}(1.6,0)(-0.2,0)
\psline[linewidth=1.2pt]{*-*}(-1.5,0)(-0.6,0)
\end{pspicture} & $N$, $E$\\
4 &
\begin{pspicture}(-2.3,-0.24)(2.3,0.24)
\psline[linewidth=0.5pt]{|-|}(-2.0,0)(2.0,0)
\psline[linewidth=0.5pt]{-}(0,-0.08)(0,0.08)
\psline[linewidth=1.2pt]{*-*}(-1.6,0)(-0.4,0)
\psline[linewidth=1.2pt]{*-*}(1.5,0)(0.6,0)
\end{pspicture} & $N$,  $S$\\
4 &
\begin{pspicture}(-2.2,-0.2)(2.2,0.2)
\psline[linewidth=0.5pt]{|-|}(-2.0,0)(2.0,0)
\psline[linewidth=0.5pt]{-}(0,-0.08)(0,0.08)
\psline[linewidth=1.2pt]{*-*}(-1.6,0)(0.2,0)
\psline[linewidth=1.2pt]{*-*}(1.5,0)(0.6,0)
\end{pspicture} & $E$, $S$\\
4 &
\begin{pspicture}(-2.2,-0.2)(2.2,0.2)
\psline[linewidth=0.5pt]{|-|}(-2.0,0)(2.0,0)
\psline[linewidth=0.5pt]{-}(0,-0.08)(0,0.08)
\psline[linewidth=1.2pt]{*-*}(0.9,0)(0.2,0)
\psline[linewidth=1.2pt]{*-*}(1.8,0)(1.2,0)
\end{pspicture} & $S$, $S$\\
\hline
\end{tabular}
}\\
\subfloat[(B2) $\bL=e\bP$, $\bL\neq0$, $\bK>\ba^2$]{
\begin{tabular}{|c|c|c|}
\hline
zeros & range of $ \theta \in [0,\pi]$ & types of orbits \\ 
\hline\hline
1&
\begin{pspicture}(-2.3,-0.24)(2.3,0.24)
\psline[linewidth=0.5pt]{[-|}(-2.0,0)(2.0,0)
\psline[linewidth=0.5pt](2,-0.08)(2,0.08)
\psline[linewidth=0.5pt]{-}(0,-0.1)(0,0.1)
\psline[linewidth=1.2pt]{-*}(-2,0)(-0.6,0)
\end{pspicture} & $N_N$\\
1 &
\begin{pspicture}(-2.3,-0.24)(2.3,0.24)
\psline[linewidth=0.5pt]{[-|}(-2.0,0)(2.0,0)
\psline[linewidth=0.5pt](2,-0.08)(2,0.08)
\psline[linewidth=0.5pt]{-}(0,-0.1)(0,0.1)
\psline[linewidth=1.2pt]{-*}(-2,0)(1.4,0)
\end{pspicture} & $E_N$\\
\hline
3&
\begin{pspicture}(-2.0, -0.2)(2.0, 0.2)
\psline[linewidth=0.5pt]{[-|}(-2.0, 0)(2.0, 0)
\psline[linewidth=0.5pt](2, -0.08)(2, 0.08)
\psline[linewidth=0.5pt]{-}(0, -0.1)(0, 0.1)
\psline[linewidth=1.2pt]{-*}(-2, 0)(-1.2, 0)
\psline[linewidth=1.2pt]{*-*}(-0.2, 0)(-0.9, 0)
\end{pspicture} & $N_N$, $N$\\
3&
\begin{pspicture}(-2.0, -0.2)(2.0, 0.2)
\psline[linewidth=0.5pt]{[-|}(-2.0, 0)(2.0, 0)
\psline[linewidth=0.5pt](2, -0.08)(2, 0.08)
\psline[linewidth=0.5pt]{-}(0, -0.1)(0, 0.1)
\psline[linewidth=1.2pt]{-*}(-2, 0)(-0.6, 0)
\psline[linewidth=1.2pt]{*-*}(+1.4, 0)(-0.2, 0)
\end{pspicture} & $N_N$, $E$\\
3 &
\begin{pspicture}(-2.0, -0.2)(2.0, 0.2)
\psline[linewidth=0.5pt]{[-|}(-2.0, 0)(2.0, 0)
\psline[linewidth=0.5pt](0, -0.08)(0, 0.08)
\psline[linewidth=1.2pt]{*-}(-0.4, 0)(-2, 0)
\psline[linewidth=1.2pt]{*-*}(0.3, 0)(1.5, 0)
\end{pspicture} & $N_N$, $S$\\
3 &
\begin{pspicture}(-2.0, -0.2)(2.0, 0.2)
\psline[linewidth=0.5pt]{[-|}(-2.0, 0)(2.0, 0)
\psline[linewidth=0.5pt](0, -0.08)(0, 0.08)
\psline[linewidth=1.2pt]{-*}(-2.0, 0)(0.4, 0)
\psline[linewidth=1.2pt]{*-*}(1.0, 0)(1.6, 0)
\end{pspicture} & $E_N$, $S$\\
\hline
\end{tabular}
}
\end{minipage}
\begin{minipage}{0.48\textwidth}
\subfloat[(B3) $\bL=0=e\bP$, $\bK>\ba^2$]{
\begin{tabular}{|c|c|c|}
\hline
zeros & range of $ \theta \in [0,\pi]$ & types of orbits \\ 
\hline\hline
0 &
\begin{pspicture}(-2.3,-0.24)(2.3,0.24)
\psline[linewidth=0.5pt]{[-]}(-2.0,0)(2.0,0)
\psline[linewidth=0.5pt]{-}(0,-0.08)(0,0.08)
\psline[linewidth=1.2pt]{-}(-2.0,0)(2.0,0)
\end{pspicture} & $E_{N,S}$\\
\hline
2 &
\begin{pspicture}(-2.3,-0.24)(2.3,0.24)
\psline[linewidth=0.5pt]{[-]}(-2.0,0)(2.0,0)
\psline[linewidth=0.5pt]{-}(0,-0.08)(0,0.08)
\psline[linewidth=1.2pt]{-*}(-2.0,0)(-0.4,0)
\psline[linewidth=1.2pt]{-*}(2.0,0)(0.4,0)
\end{pspicture} & $N_N$, $S_S$\\
\hline
\end{tabular}
}\\
\subfloat[(B4) $\bL=e\bP$, $\bL\neq0$, $\bK=\ba^2$]{
\begin{tabular}{|c|c|c|}
\hline
zeros & range of $ \theta \in [0,\pi]$ & types of orbits \\ 
\hline\hline
2&
\begin{pspicture}(-2.3,-0.24)(2.3,0.24)
\psline[linewidth=0.5pt]{-|}(-2.0,0)(2.0,0)
\psline[linewidth=0.5pt](2,-0.08)(2,0.08)
\psline[linewidth=0.5pt]{-}(0,-0.1)(0,0.1)
\psline[linewidth=1.2pt]{-o}(-2,0)(-2,0)
\end{pspicture} & $N_N$\\
\hline
3 &
\begin{pspicture}(-2.3,-0.24)(2.3,0.24)
\psline[linewidth=0.5pt]{-|}(-2.0,0)(2.0,0)
\psline[linewidth=0.5pt]{-}(0,-0.08)(0,0.08)
\psline[linewidth=1.2pt]{o-*}(-2,0)(-0.4,0)
\end{pspicture} & $N_N$\\
3 &
\begin{pspicture}(-2.3,-0.24)(2.3,0.24)
\psline[linewidth=0.5pt]{-|}(-2.0,0)(2.0,0)
\psline[linewidth=0.5pt]{-}(0,-0.08)(0,0.08)
\psline[linewidth=1.2pt]{o-*}(-2,0)(0.4,0)
\end{pspicture} & $E_N$\\
\hline
4 &
\begin{pspicture}(-2.3,-0.24)(2.3,0.24)
\psline[linewidth=0.5pt]{-|}(-2.0,0)(2.0,0)
\psline[linewidth=0.5pt]{-}(0,-0.08)(0,0.08)
\psline[linewidth=1.2pt]{*-*}(-1.6,0)(-0.4,0)
\psline[linewidth=1.2pt]{-o}(-2,0)(-2,0)
\end{pspicture} & $N_N$, $N$\\
4 &
\begin{pspicture}(-2.3,-0.24)(2.3,0.24)
\psline[linewidth=0.5pt]{|-|}(-2.0,0)(2.0,0)
\psline[linewidth=0.5pt]{-}(0,-0.08)(0,0.08)
\psline[linewidth=1.2pt]{*-*}(-1.1,0)(1.1,0)
\psline[linewidth=1.2pt]{-o}(-2,0)(-2,0)
\end{pspicture} & $N_N$, $E$\\
4 &
\begin{pspicture}(-2.3,-0.24)(2.3,0.24)
\psline[linewidth=0.5pt]{|-|}(-2.0,0)(2.0,0)
\psline[linewidth=0.5pt]{-}(0,-0.08)(0,0.08)
\psline[linewidth=1.2pt]{*-*}(1.6,0)(0.4,0)
\psline[linewidth=1.2pt]{-o}(-2,0)(-2,0)
\end{pspicture} & $N_N$, $S$\\
\hline
\end{tabular}
}\\
\subfloat[(B4) $\bL=0=e\bP$, $\bK=\ba^2$]{
\begin{tabular}{|c|c|c|}
\hline
zeros & range of $ \theta \in [0,\pi]$ & types of orbits \\ 
\hline\hline
4&
\begin{pspicture}(-2.3,-0.24)(2.3,0.24)
\psline[linewidth=0.5pt]{-|}(-2.0,0)(2.0,0)
\psline[linewidth=0.5pt](2,-0.08)(2,0.08)
\psline[linewidth=0.5pt]{-}(0,-0.1)(0,0.1)
\psline[linewidth=1.2pt]{-o}(-2,0)(-2,0)
\psline[linewidth=1.2pt]{-o}(2,0)(2,0)
\end{pspicture} & $N_N$, $S_S$\\
\hline
\end{tabular}
}
\end{minipage}
\caption{Overview of different orbit configurations for colatitudinal motion. The vertical bar of the second column denotes $\theta=\pi/2$ and the thick lines $\T\geq0$, i.e.~regions where a motion is possible. Dots represent single zeros and circles double zeros. If zeros merge, the resulting orbits are stable if a line is reduced to a point and unstable if lines merge. The configurations $(B2)$ and $(B4)$ with $\bL=-e\bP$ are obtained by a reflection at the equatorial plane.}
\label{tab:theta}
\end{table}

\subsubsection{Regions of orbit configurations in parameter space}

It is now of interest for which sets of parameters a given orbit configuration changes. As $\T\geq0$ is necessary for geodesic motion, this happens if the behavior of $\T$ at the boundaries changes, which means a switch from one of the above cases \textit{(A)}, \textit{(B1)}, \ldots, \textit{(B4)} to another, or if the number of real zeros of $\T$ changes. The latter occurs at that parameters for which $\T$ has multiple zeros. With these two conditions the different regions of orbit configurations in parameter space can be completely determined. The first condition was already analyzed above.

For $\theta \in (0,\pi)$ the function $\T$ has the same zeros as the polynomial $\T_\nu$ and we may use $\T_\nu$ instead of $\T$ for all orbits not reaching $\theta=0,\pi$. If $\bL=\pm e\bP$ then $\nu_0=\pm 1$ is a zero of $\T_\nu$ but does not correspond to a turning point of the colatitudinal motion. If in addition $\bK=\ba^2$ then $\nu_0=\pm 1$ is a double zero of $\T_\nu$ and $\theta=0,\pi$ a simple zero of $\T$, which does correspond to a turning point of $\theta$. Keeping this in mind we will also use $\T_\nu$ for these cases but discuss the occurence of multiple zeros at $\theta=0,\pi$ separatly using $\T$.

The condition for a double zero $\nu_0$ is $\frac{d\T_\nu}{d\nu}(\nu_0)=0=\T_\nu(\nu_0)$. This can be read as 2 conditions on 2 of the 5 parameters $E$, $\bL$, $\bK$, $e\bP$, and $a$. Solving these two conditions for $E$ and $\bL$ dependent on the position of the double zero $\nu_0$ and the other parameters yields
\begin{align}
E_{1,2} & = \frac{e\bP}{2 \ba \nu_0} \pm \frac{1}{2} \frac{\sqrt{\nu_0^2(1-\nu_0^2)(\bK-\ba^2\nu_0^2)(\bK+\ba^2-2\ba^2 \nu_0^2)^2}}{(\nu_0^2-1)(\bK-\ba^2\nu_0^2)\ba \nu_0}\,, \label{theta_double_E}\\
\bL_{1,2} & = \frac{e\bP(\nu_0^2+1)}{2\nu_0} \pm \frac{1}{2} \frac{\nu_0(1-\nu_0^2)(\bK-\ba^2)(\bK+\ba^2-2\ba^2 \nu_0^2)}{\sqrt{\nu_0^2(1-\nu_0^2)(\bK-\ba^2\nu_0^2)(\bK+\ba^2-2\ba^2\nu_0^2)^2}}\,. \label{theta_double_L}
\end{align}
The expressions for $E$ and $L$ diverge at $\nu_0=0$ for $e\bP \neq 0$, at $\nu_0=\sqrt{\bK/\ba^2}$, and at $\nu_0=\pm1$ for $E$, what suggests to consider the 2 conditions for double zeros directly for these points:
\paragraph*{Equatorial orbits}
For $\nu_0=0$ the 2 conditions imply that either $\bK=0$ and $\bL=\ba E$ or $\bK=(\ba E-\bL)^2$ and $e\bP=0$ are necessary and sufficient for the existence of equatorial orbits. The asymptotic behavior of $E$ and $\bL$ at $\nu_0=0$ also displays these conditions,
\begin{align}
E_{1,2} & = \frac{e\bP}{2\ba \nu_0} \mp \frac{\bK+\ba^2}{2\ba\sqrt{\bK}} + \mathcal{O}(\nu_0^2)\,, \\
\bL_{1,2} & = \frac{e\bP}{2\nu_0} \pm \frac{\bK-\ba^2}{2\sqrt{\bK}} + \mathcal{O}(\nu_0) = \ba E_{1,2} \pm \sqrt{\bK} + \mathcal{O}(\nu_0)\,.
\end{align}
In the case $\bK=0$, $\bL=\ba E$ the equatorial orbit is the only possible geodesic orbit (for $\T \not\equiv 0$) and, thus, stable, whereas for $\bK=(\ba E-\bL)^2$, $e\bP=0$ the sign of $A:=\ba^2(E^2-1)-\bL^2$ has to be considered: The orbit is stable if $A\leq0$ (for $\T \not\equiv 0$) and unstable if $A>0$. In the case $A=0$ the zero $\theta=\pi/2$ is even fourfold. For $\bK=\ba^2$, $e\bP=0=\bL$, and $E^2=1$ the function $\T$ is identical to zero.
\paragraph*{Orbits with $\theta \equiv 0,\pi$}
Geodesic motion along the axis $\theta=0,\pi$ is possible only if $\theta=0,\pi$ is a double zero of $\T$. The 2 conditions on double zeros show that $\bL=e\bP$ and $\bK=\ba^2$ are necessary and sufficient for $\theta \equiv 0$, and $\bL=-e\bP$, $\bK=\ba^2$ for $\theta\equiv\pi$. This can also be seen by considering the asymptotic behaviour of $\bL_{1,2}$ and $E_{1,2}$ as $\nu_0$ approaches $\pm1$: it is given by $\lim_{\nu_0 \to 1} \bL_{1,2} \to e\bP$ and $\lim_{\nu_0 \to -1} \bL_{1,2} = -e\bP$ whereas $E_{1,2}$ diverges for $\bK \neq \ba^2$. 

Let us discuss the stability of the orbits $\theta \equiv 0,\pi$: The orbits are unstable if $\ba^2-(\ba E-\bL/2)^2>0$ and stable if $\ba^2-(\ba E-\bL/2)^2<0$. For $\ba^2-(\ba E-\bL/2)^2=0$ the poles $\theta=0,\pi$ are fourfold zeros and the orbit $\theta\equiv0$ is stable if $E = \frac{\bL}{2a} - 1$, and the orbit $\theta\equiv\pi$ if $E = \frac{\bL}{2a} + 1$. In the special case of $\bL=0=e\bP$ the two orbits are unstable if $E^2<1$ and stable if $E^2>1$. For $\bL=0=e\bP$, $\bK=\ba^2$, and $E^2=1$ again $\T \equiv 0$. 
\paragraph*{Orbits with $\theta \equiv \pm \sqrt{\bK/\ba^2}$}
The singularity $\nu_0=\pm \sqrt{\bK/\ba^2}$ is located in $(-1,1)$ and is not equal to zero only if $0<\bK<\ba^2$. Assuming this, orbits with constant $\cos \theta \equiv \pm \sqrt{\bK/\ba^2}$ can exist only if $\ba\bL-(\ba^2-\bK)E=\pm e\bP\sqrt{\bK}$ is fulfilled. This can be infered from the asymptotes of $\bL_{1,2}$ in terms of $E_{1,2}$ around $\nu_0=\pm \sqrt{\bK/\ba^2}$,
\begin{align}
\bL_{1,2} & = 
\frac{\ba^2-\bK}{\ba} E_{1,2} \pm \frac{e\bP\sqrt{\bK}}{\ba} + \mathcal{O}\left(\sqrt{\nu\mp\sqrt{\frac{\bK}{\ba^2}}}\right)\,. \label{Asymp_K/a^2}
\end{align}

The expressions \eqref{theta_double_E} and \eqref{theta_double_L} depend linearly on $e\bP$ and a rescaling of parameters ($\bL/\ba$, $\bK/\ba^2$, $e\bP/\ba$) removes the rotation parameter $\ba$ completely from the equations. Only the dependence on $\bK$ is not obvious. If we solve the 2 conditions $\frac{d\T_\nu}{d\nu}(\nu_0)=0=\T_\nu(\nu_0)$ for $E$ and $\bK$ instead of $\bL$ this yields
\begin{align}
E_{1,2} & = \frac{e\bP}{2\ba \nu_0} \pm \frac{1}{2} \frac{\sqrt{(e\bP(\nu_0^2+1)-2\bL\nu_0)^2+4\nu_0^2\ba^2(\nu_0^2-1)^2}}{\ba \nu_0(\nu_0^2-1)} \label{theta_new_double_E}\\
\bK_{1,2} & = \ba^2 - \frac{e\bP(\nu_0^2+1)-2\nu_0\bL}{2\nu_0^2(\nu_0^2-1)} \left[ (e\bP(\nu_0^2+1)-2\nu_0\bL) \pm \sqrt{(e\bP(\nu_0^2+1)-2\bL\nu_0)^2+4\nu_0^2\ba^2(\nu_0^2-1)^2} \right] \label{theta_double_K}
\end{align}
Note that $E_1$ and $\bK_1$ behave regular at $\nu_0=0$ and $\ba E_1 = \bL + \mathcal{O}(\nu_0)$, $\bK_1 =\mathcal{O}(\nu^2)$ whereas $E_2$ and $\bK_2$ are finite at $\nu_0=0$ only if $e\bP=0$. Then $\bK \to (\ba E-\bL)^2$ as expected from the analysis of equatorial orbits above. Also, for $\nu_0 \to \pm 1$ the conditions $\bL=\pm e\bP$, $\bK \to \ba^2$ are recovered.

Triple zeros are also of interest as they correspond to parameters where a stable orbit with constant $\theta$ becomes unstable and vice versa. Therefore, will will study them here. The 3 conditions for such points are $0=\frac{d^2\T_\nu}{d\nu^2}(\nu_0)=\frac{d\T_\nu}{d\nu}(\nu_0)=\T_\nu(\nu_0)$ which we read as 3 conditions on $E$, $\bL$, and $e\bP$ yielding
\begin{align}
E_{1,2} & = \pm \frac{1}{2} \frac{2\nu_0^6\ba^4+6\nu_0^2\bK\ba^2-(3\nu_0^4\ba^2+\bK)(\bK+\ba^2)}{\sqrt{(1-\nu_0^2)(\bK-\ba^2\nu_0^2)}(1-\nu_0^2)(\bK-\ba^2\nu_0^2)\ba}\,, \\
\bL_{1,2} & = \mp \frac{1}{2} \frac{\sqrt{(1-\nu_0^2)(\bK-\ba^2\nu_0^2)}(\bK-\ba^2)(\ba^2\nu_0^6-3\nu_0^4\ba^2+3\nu_0^2\bK-\bK)}{(\nu_0^2-1)^2(\bK-\ba^2\nu_0^2)^2}\,, \\
e\bP_{1,2} & = \pm \frac{\nu_0^3(\bK-\ba^2)^2}{\sqrt{(1-\nu_0^2)(\bK-\ba^2\nu_0^2)}(\nu_0^2-1)(\bK-\ba^2\nu_0^2)}\,.
\end{align}
In particular, triple zeros are also double zeros and the asymptotic behavior of $\bL$ as a function of $E$ and $\bK$ at the singularities $\nu_0=\pm \sqrt{\bK/\ba^2}$ is determined by \eqref{Asymp_K/a^2}. The points $\nu_0=\pm 1$ were studied above but we note here that $\theta=0,\pi$ is a double zero of $\T$ if $\nu=\pm1$ is a triple zeros of $\T_\nu$. 

Besides the equatorial orbits and orbits with constant $\theta=0,\pi$ discussed above, fourfold or even higher order zeros are only possible for parameters corresponding to $\T\equiv0$, which are $\ba=e\bP=\bL=\bK=0$ with arbitrary $E$ or $E=\pm1$, $\bK=\ba^2$, and $\bL=e\bP=0$.

\begin{figure}
\subfloat[$\bK=3$, $e\bP=11.25$]{
\begin{minipage}{0.45\textwidth}
\includegraphics[width=0.45\textwidth]{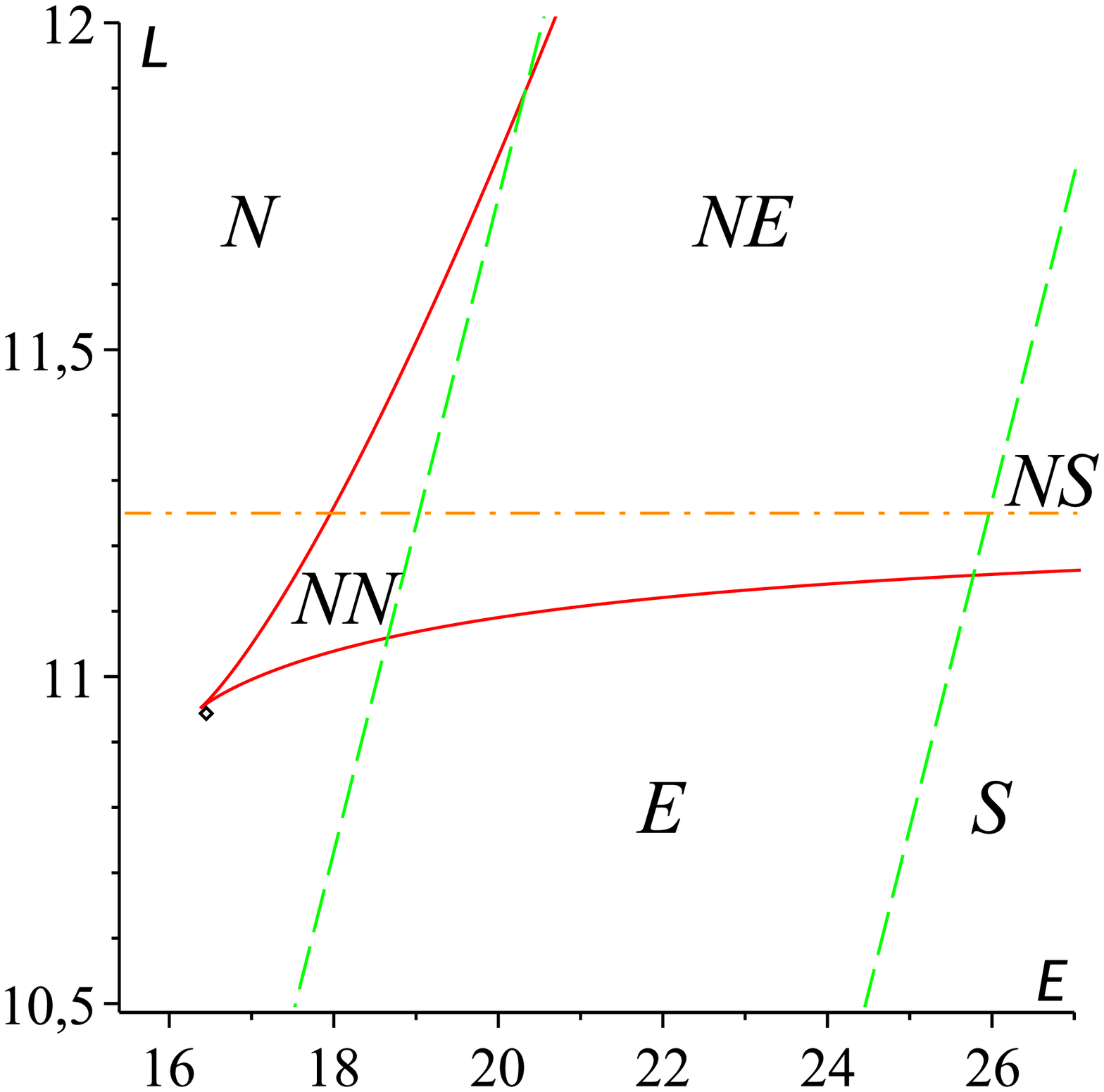}\\
\includegraphics[width=0.99\textwidth]{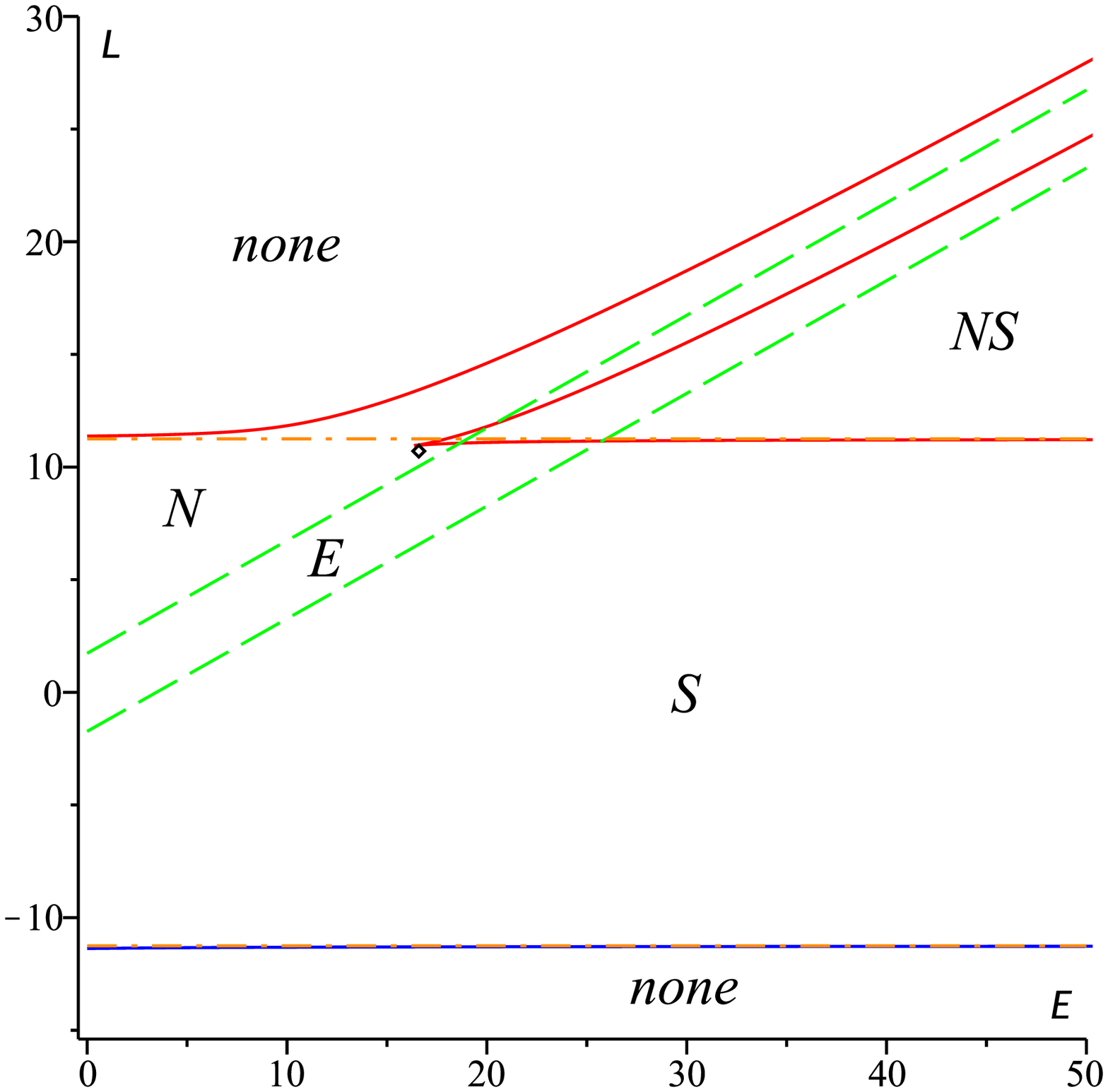}
\end{minipage}
}
\subfloat[$\bK=0.5$, $e\bP=0.3$]{
\begin{minipage}{0.45\textwidth}
\begin{center}
\includegraphics[width=0.45\textwidth]{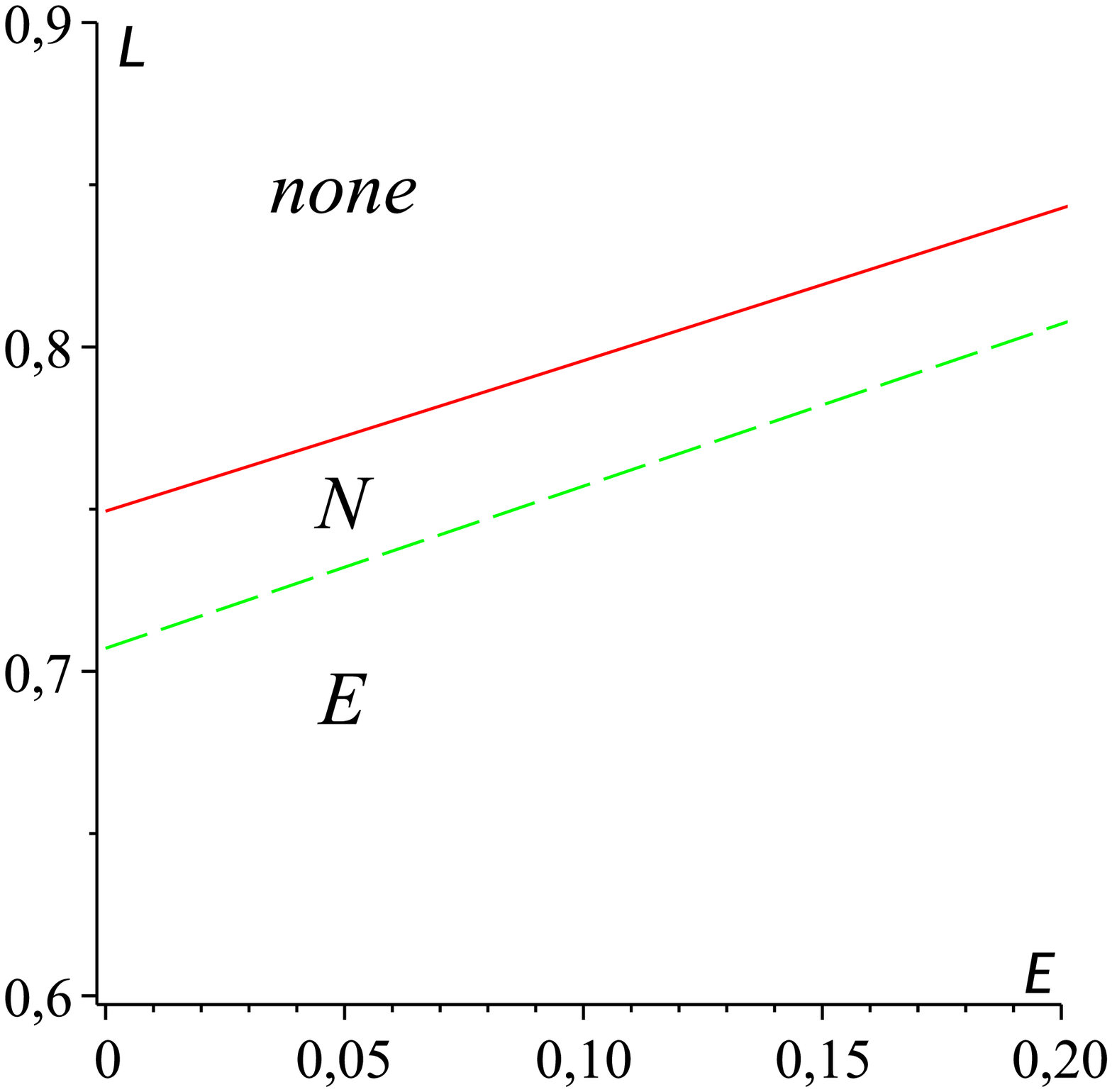}\quad
\includegraphics[width=0.45\textwidth]{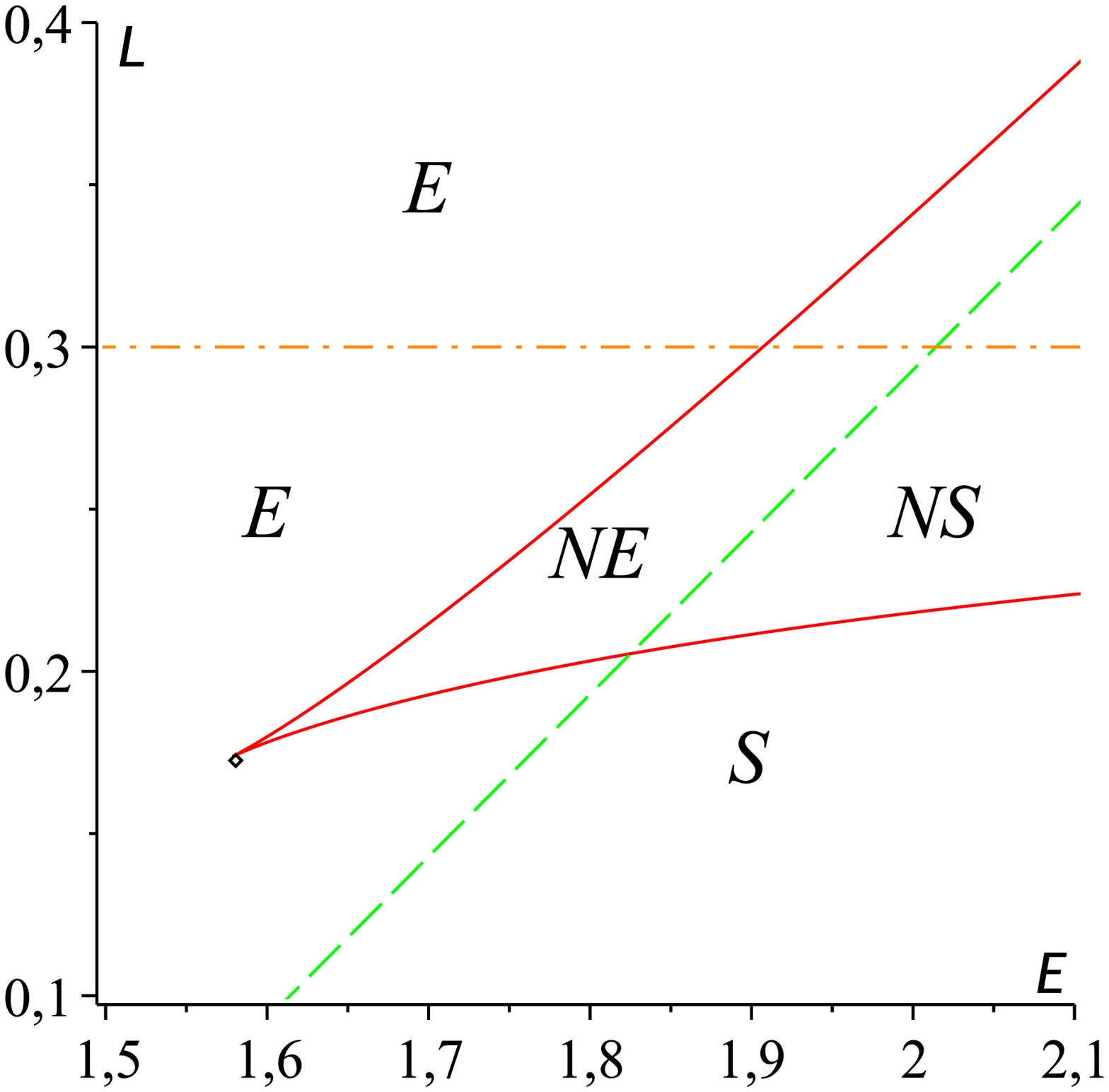}\\
\includegraphics[width=0.99\textwidth]{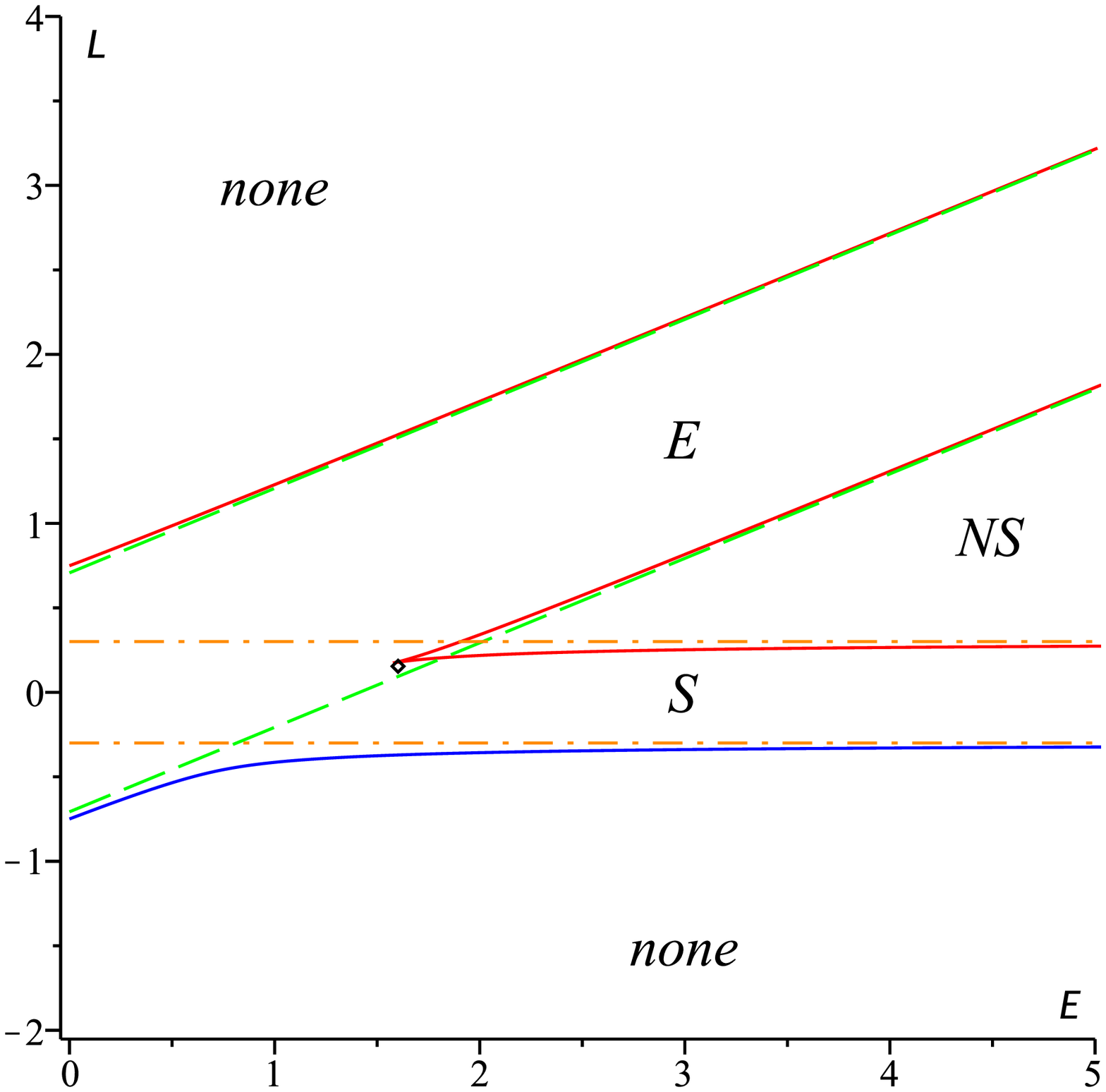}
\end{center}
\end{minipage}
}
\caption{Orbit configurations for the colatitudinal motion with $\ba=0.5$, $\bK>\ba^2$, and $e\bP>0$. For a general description see the text. The orbit configurations on the solid lines contain an orbit with constant $\theta$. They are unstable if marked by the red solid line starting at the dot and approaching the dashed line, and stable otherwise.The dash dotted line at $\bL=e\bP$ ($\bL=-e\bP$) denotes an orbit crossing the north (south) pole and corresponds to the orbit configurations $(B2)$. The other regions correspond to the configurations $(A)$. Small plots on the top are enlarged details of the lower plot.}
\label{fig:theta_LvsE_K>a^2}
\end{figure}

\begin{figure}
\subfloat[(a) $\bK=\ba^2$]{
\begin{minipage}{0.45\textwidth}
\includegraphics[width=0.45\textwidth]{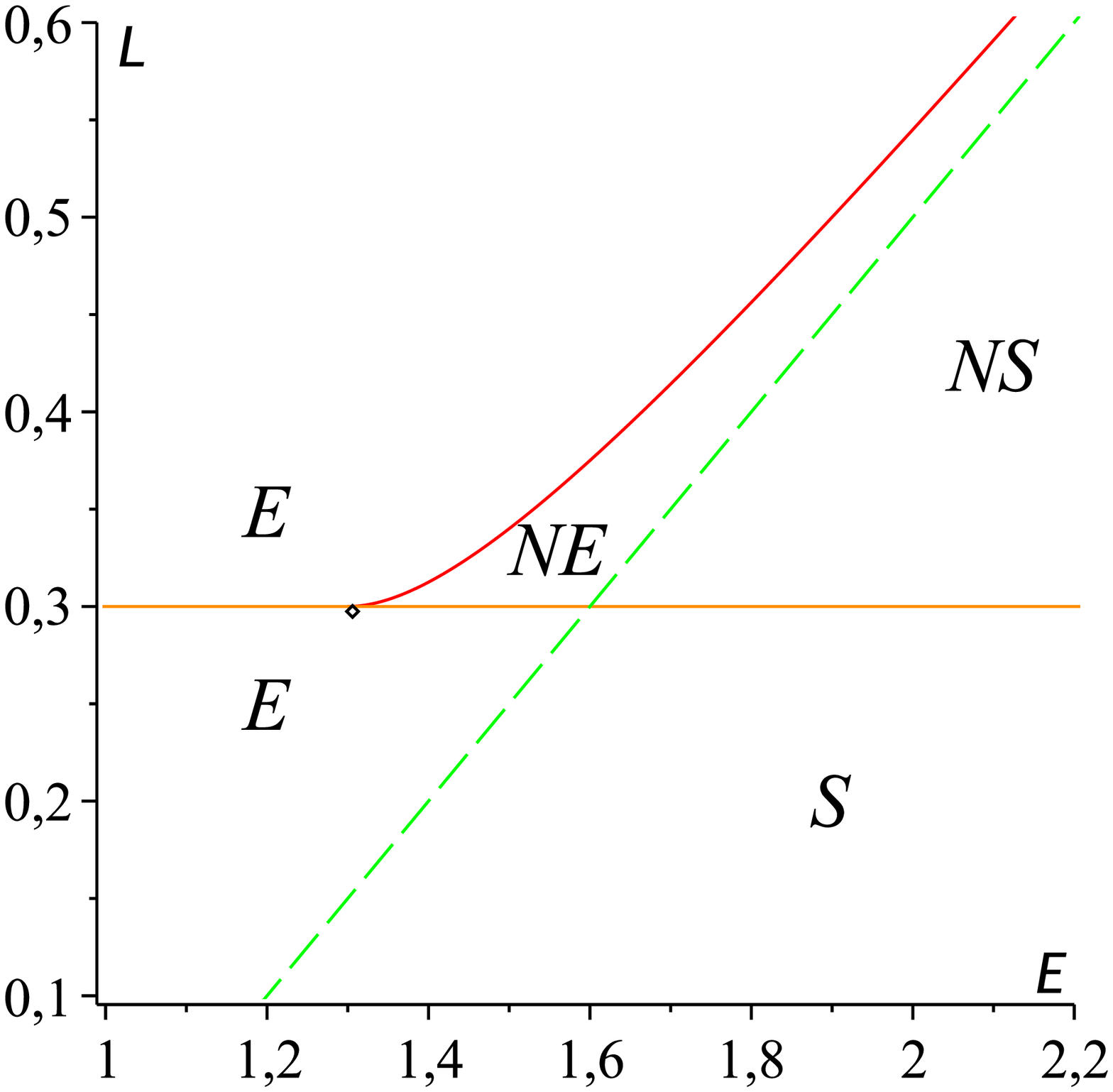}
\includegraphics[width=0.45\textwidth]{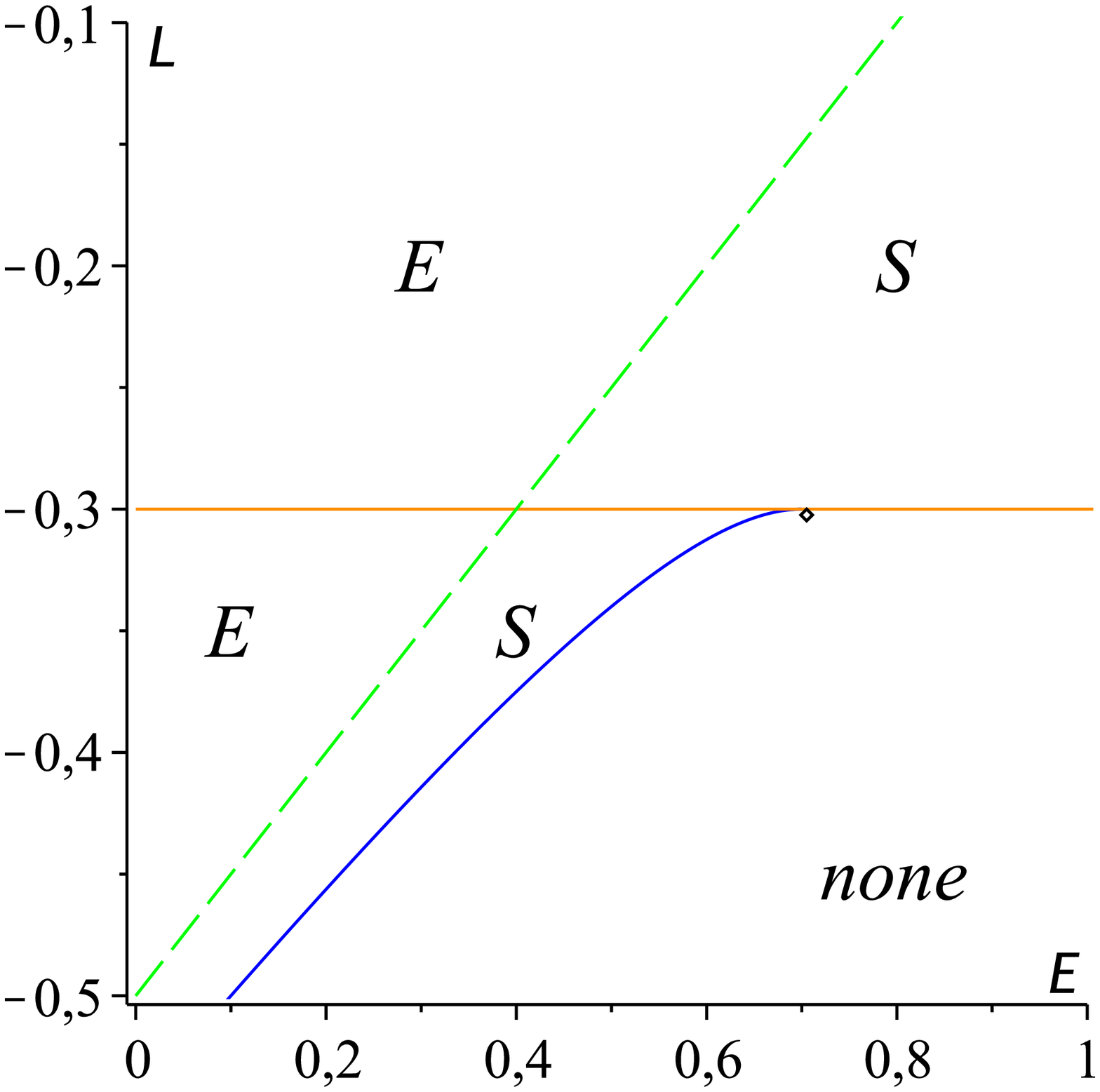}\\
\includegraphics[width=0.99\textwidth]{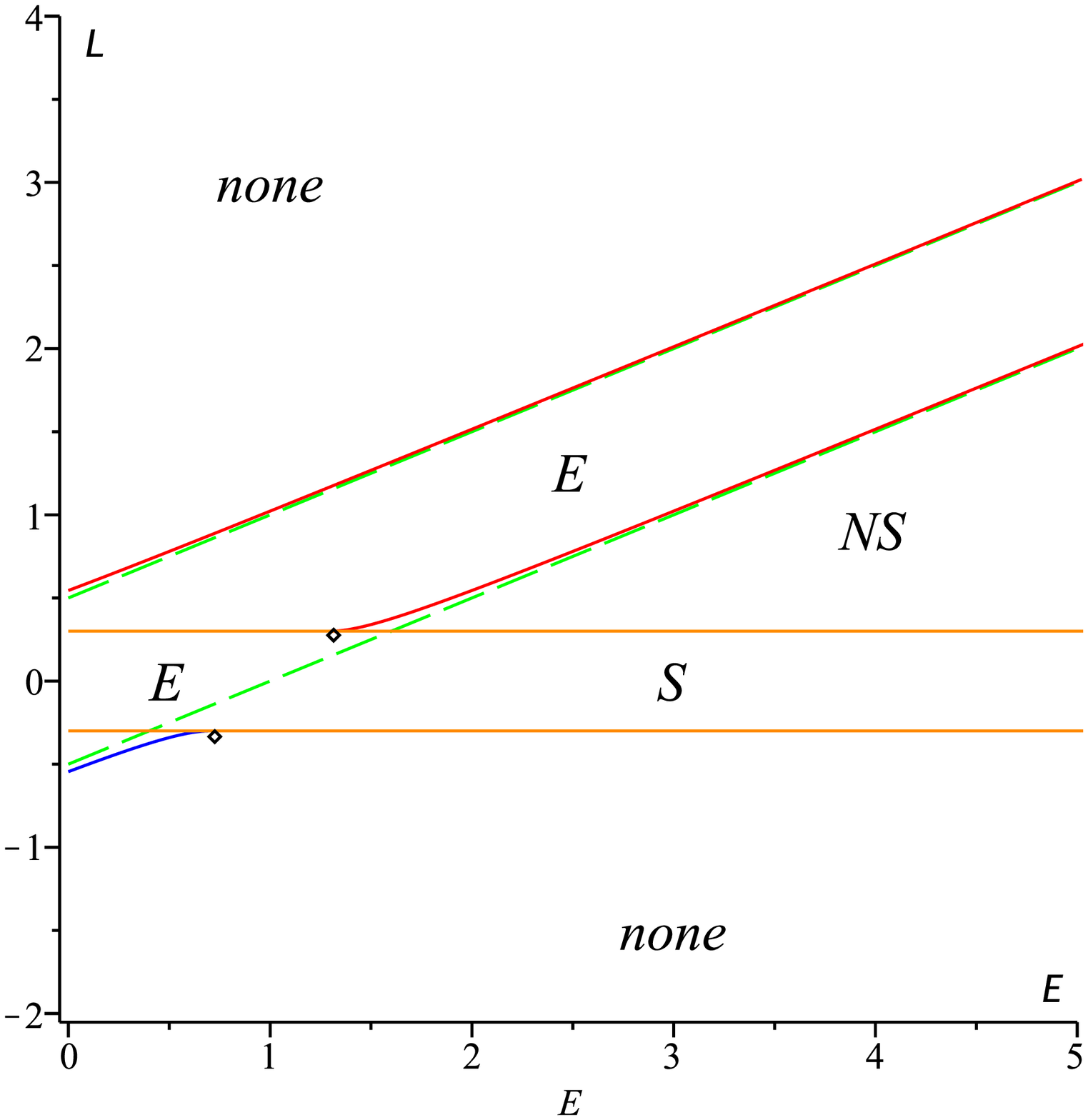}
\end{minipage}
}
\subfloat[(b) $\bK=0.2$]{
\begin{minipage}{0.45\textwidth}
\includegraphics[width=0.45\textwidth]{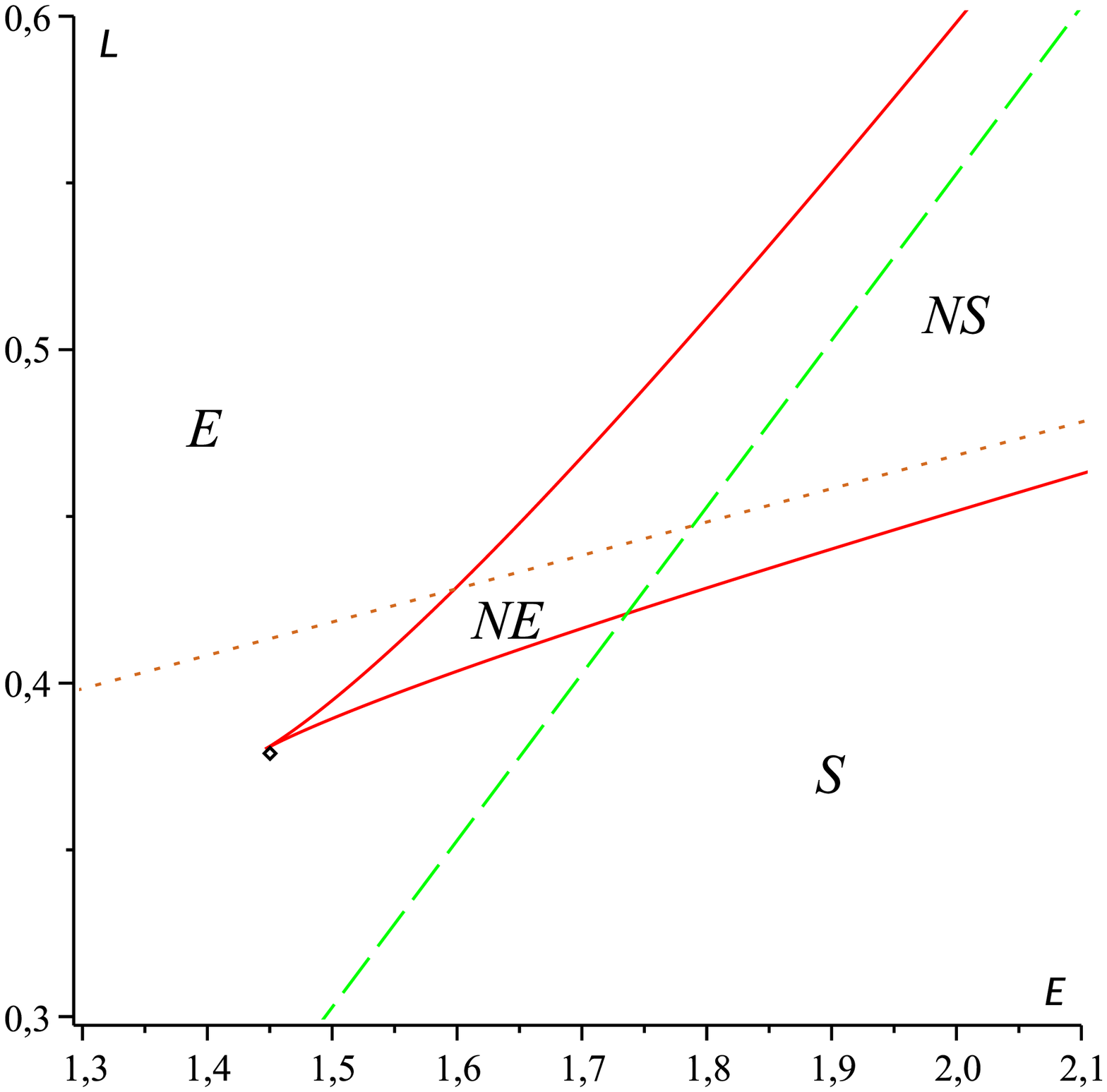}\\
\includegraphics[width=0.99\textwidth]{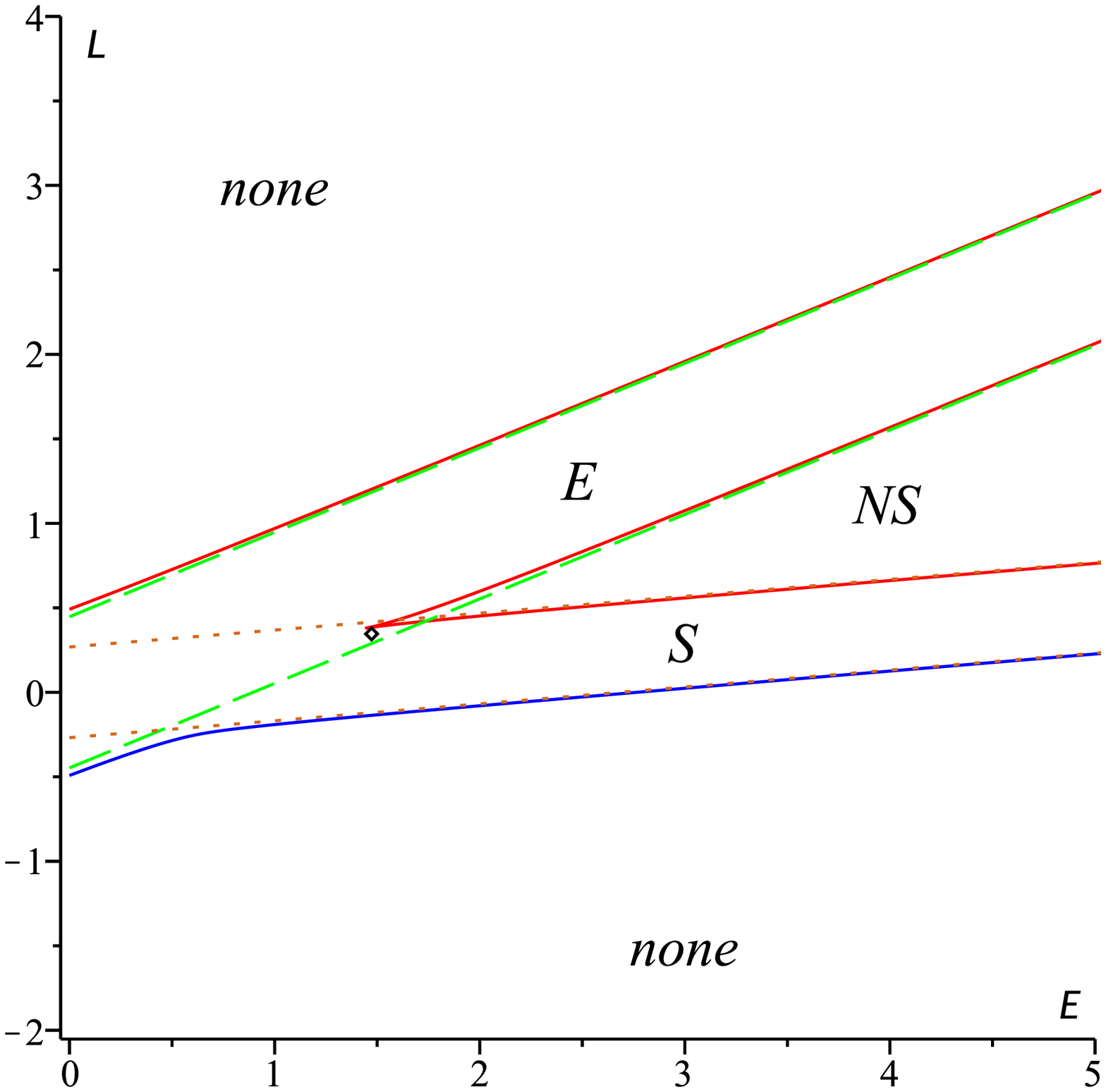}
\end{minipage}
}
\caption{Orbit configurations for the colatitudinal motion with $\ba=0.5$, $\bK\leq\ba^2$, and $e\bP=0.3$. For a general description see the text. The orange solid lines in (a) correspond to orbits of constant $\theta=0,\pi$, which are unstable for energies less than the one marked by the dots. These orbits correspond to the configurations $(B4)$, other regions (in both plots) to the configurations $(A)$ or $(B1)$. In both plots, the orbits labeled by the red solid line starting at the dot and approaching the dashed line are unstable. All other orbits on solid lines are stable. Small plots on the top are enlarged details of the lower plot.}
\label{fig:theta_LvsE_K<=a^2}
\end{figure}

\begin{figure}
\subfloat[(a) $\bK=0.5$]{
\includegraphics[width=0.4\textwidth]{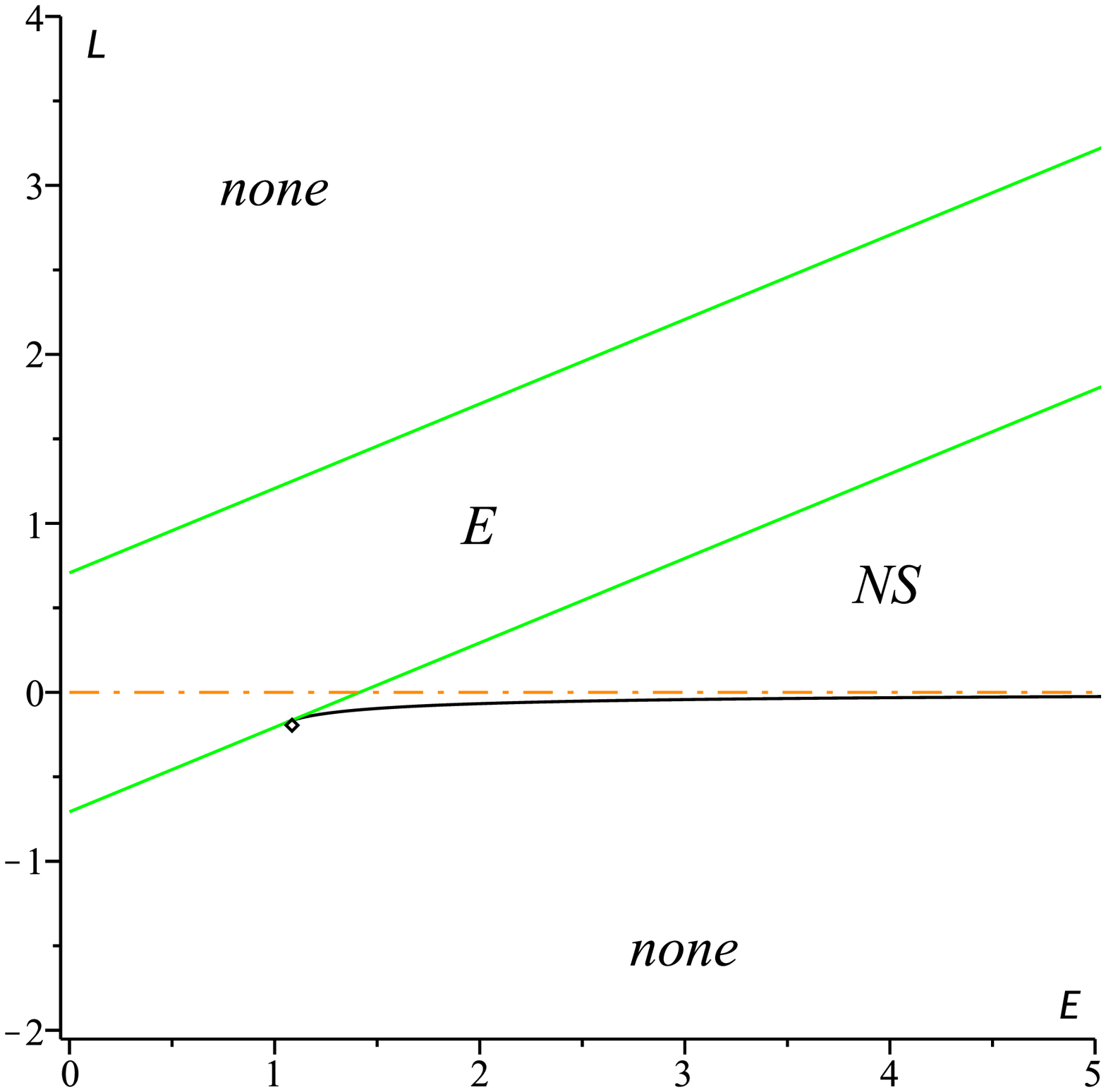}
}\quad
\subfloat[(b) $\bK=0.2$]{
\includegraphics[width=0.4\textwidth]{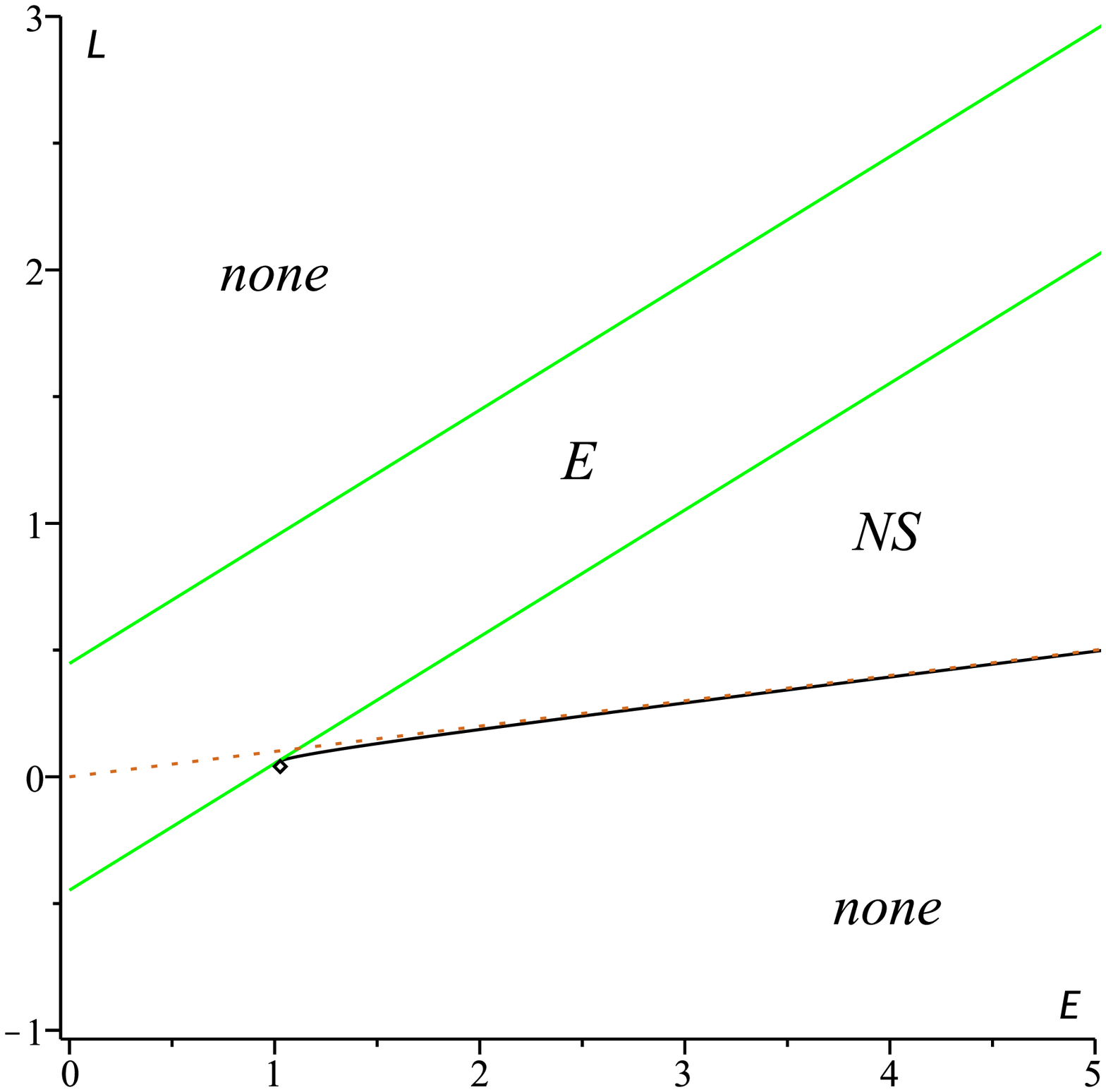}
}
\caption{Orbit configurations for the colatitudinal motion with $\ba=0.5$ and $e\bP=0$. For a general description see the text. The green solid lines correspond to equatorial orbits which are unstable from dot to larger $\bL$ and else stable. The black solid line marks two stable orbits of constant $\theta \neq \frac{\pi}{2}$ which are symmetric with respect to the equatorial plane. The dash dotted line in (a) marks the case $(B3)$ with an $E_{NS}$ orbit between the green solid lines and an $N_NS_S$ configuration else.}
\label{fig:theta_LvsE_eP0}
\end{figure}

\begin{figure}
\subfloat[(a) $\bL=0.5$]{
\begin{minipage}{0.45\textwidth}
\includegraphics[width=0.45\textwidth]{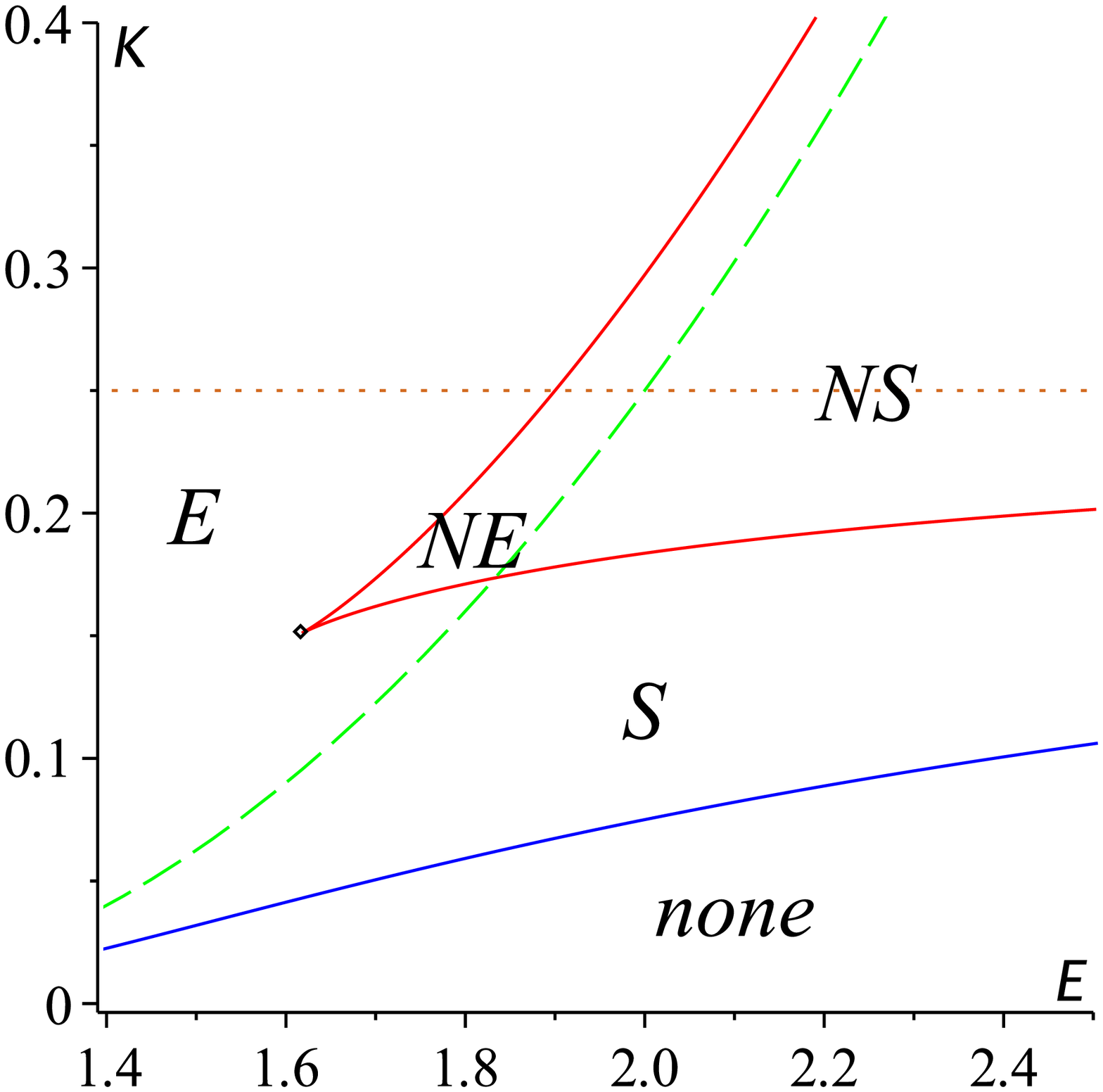}
\includegraphics[width=0.45\textwidth]{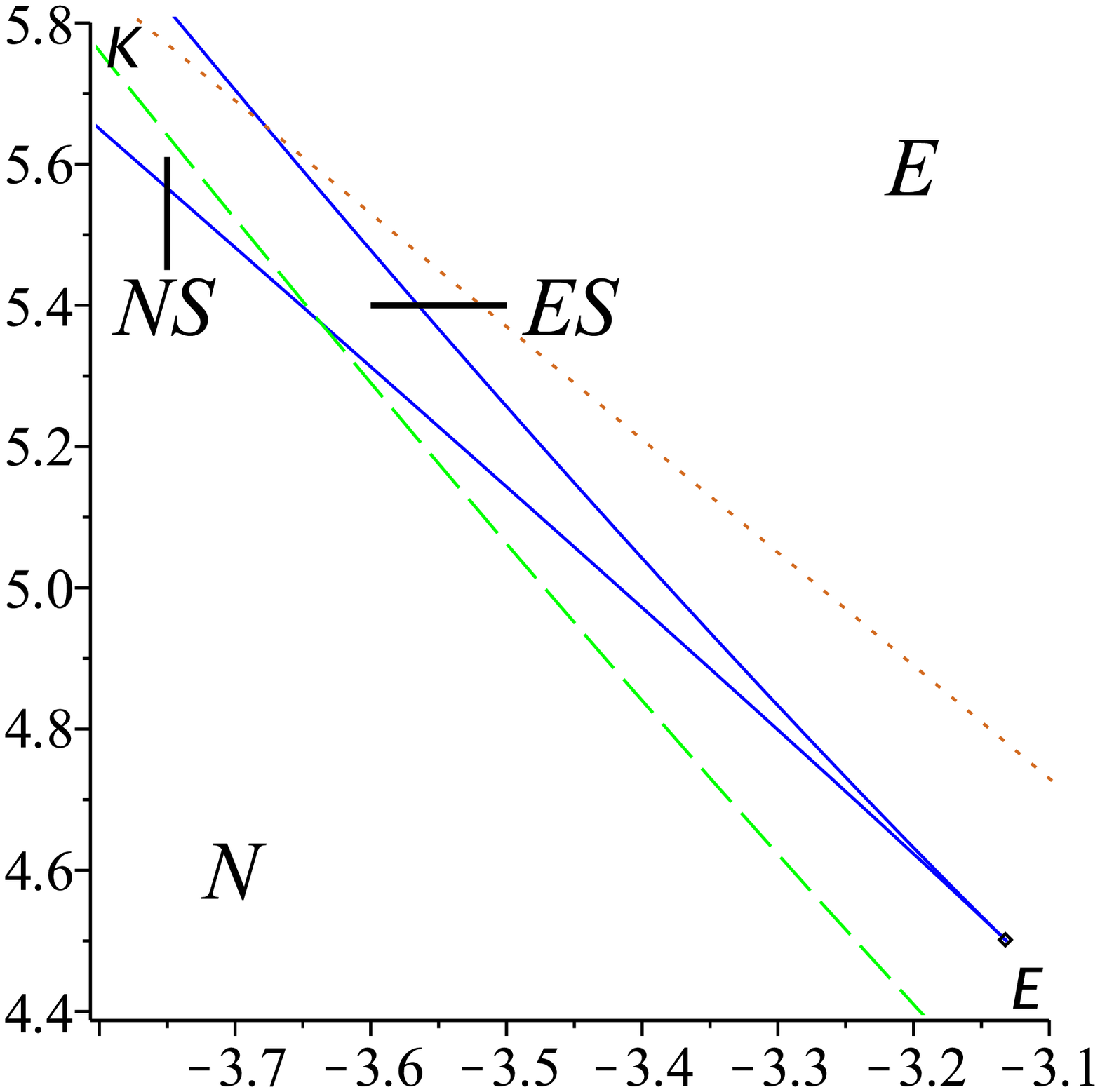}\\
\includegraphics[width=0.95\textwidth]{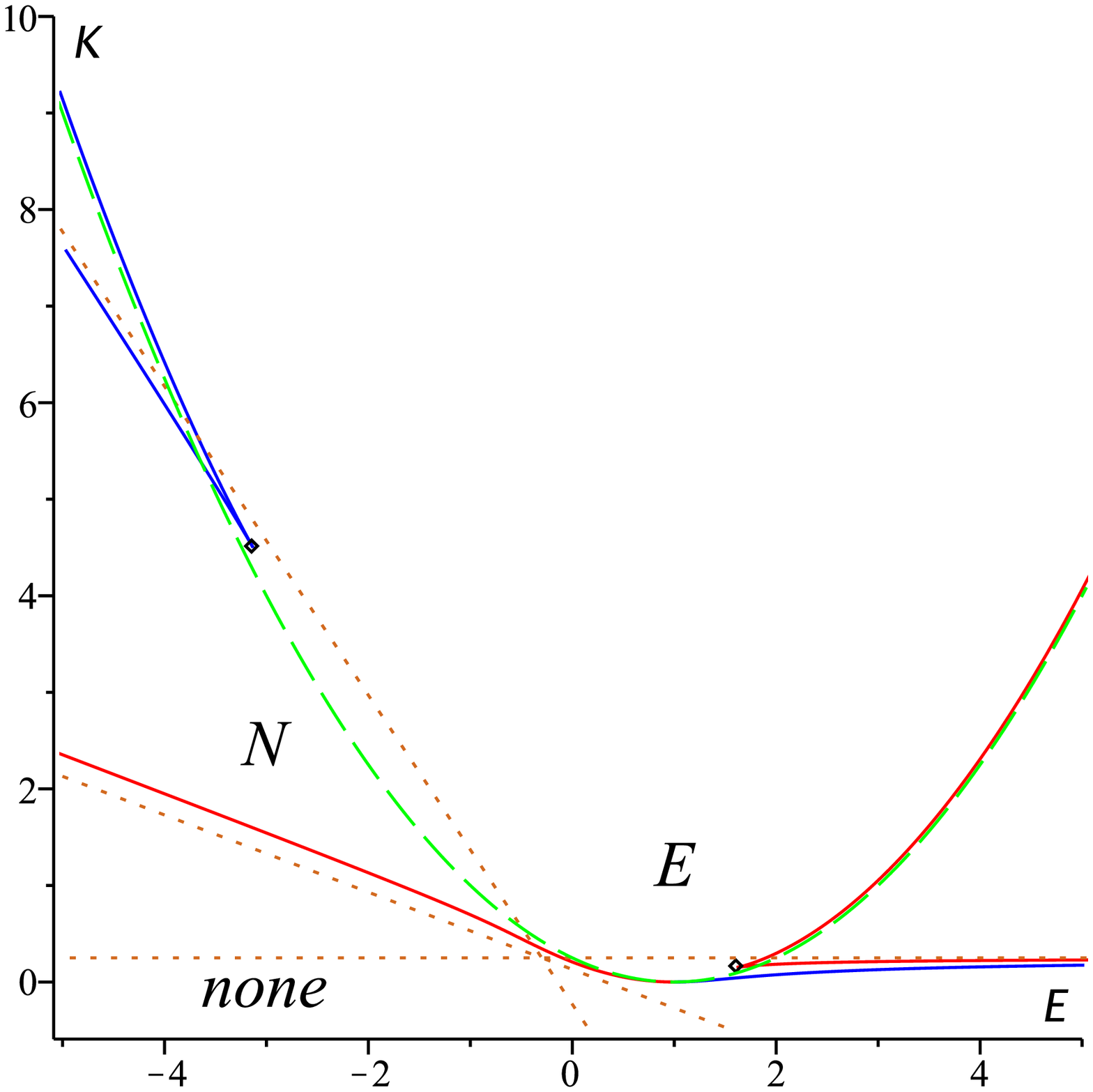}
\end{minipage}
}
\subfloat[(b) $\bL=e\bP$]{
\begin{minipage}{0.45\textwidth}
\includegraphics[width=0.45\textwidth]{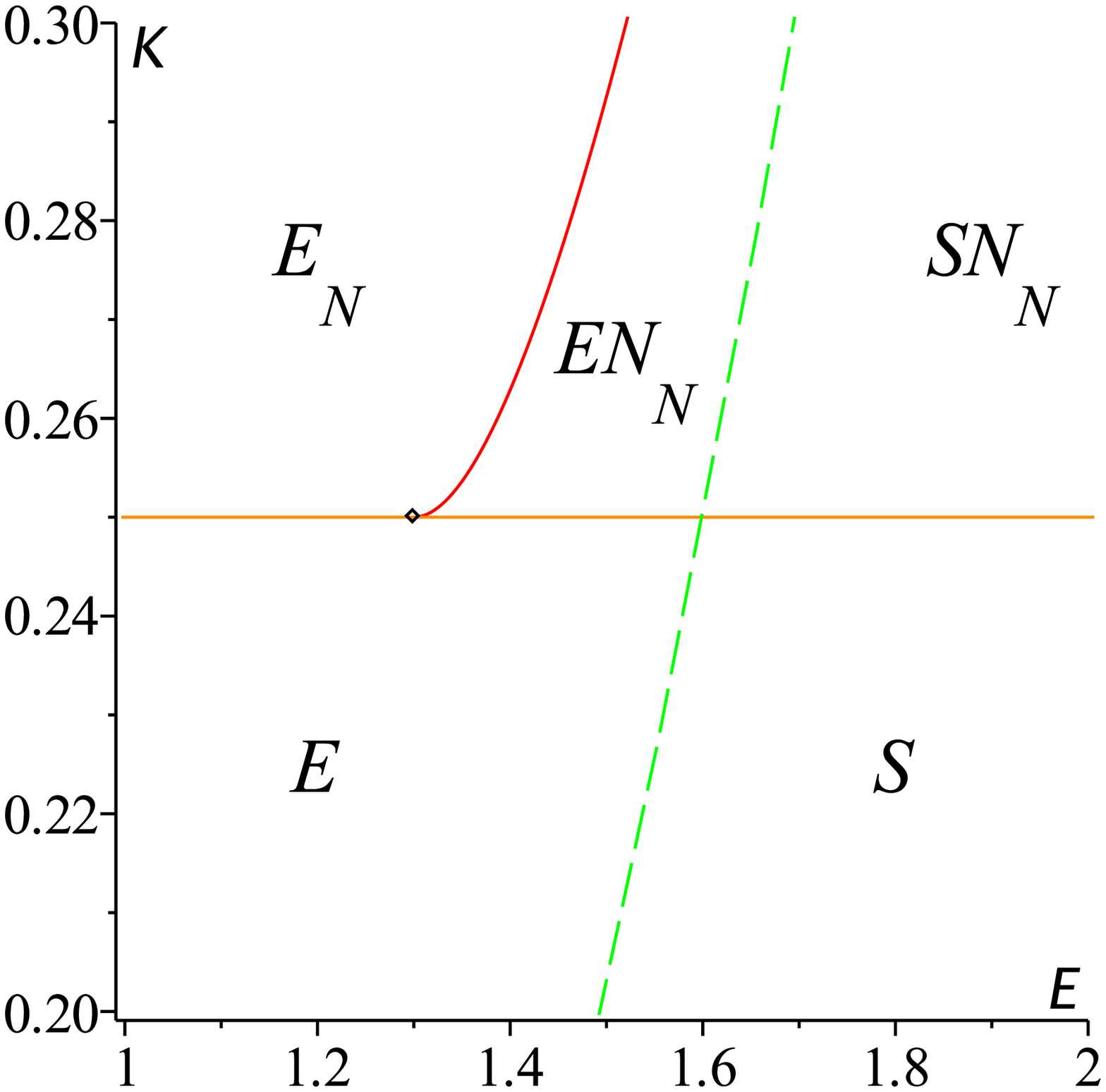}
\includegraphics[width=0.45\textwidth]{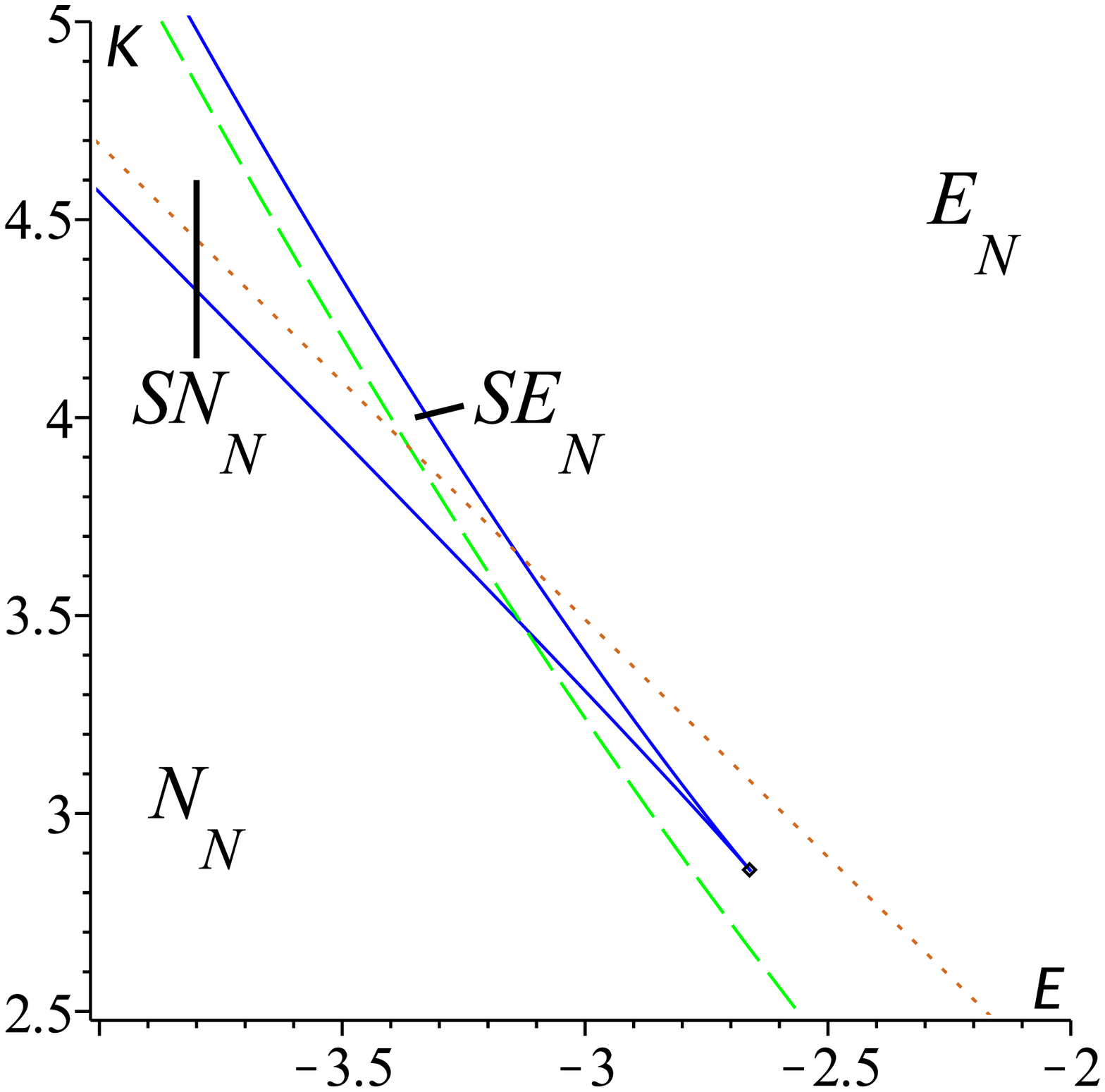}\\
\includegraphics[width=0.95\textwidth]{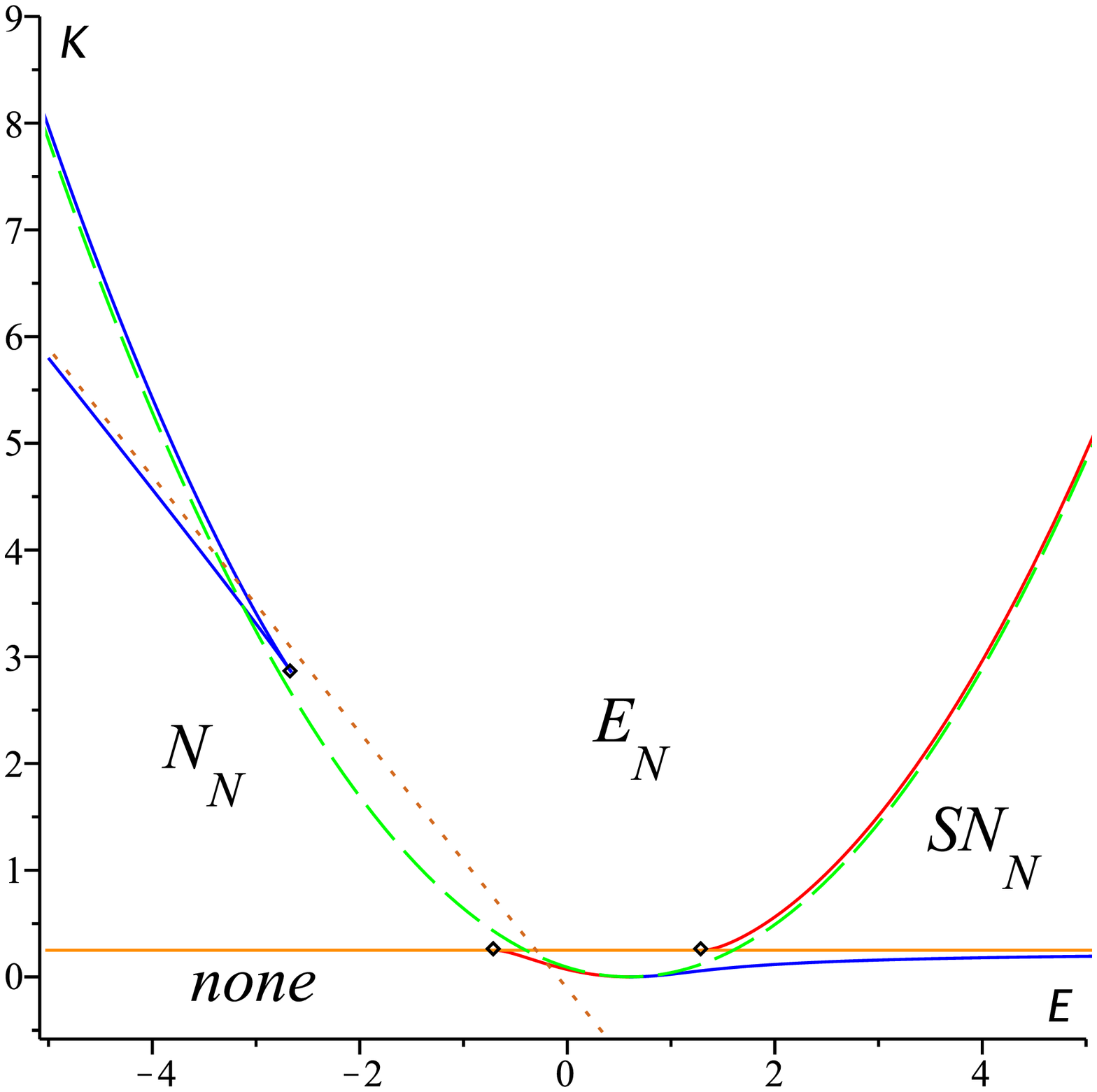}
\end{minipage}
}
\caption{Orbit configurations for the colatitudinal motion with $\ba=0.5$ and $e\bP=0.3$. For a general description see the text. For both plots, solid lines with dashed asymptotes correspond to unstable and all other to stable orbits. In (b) the regions above the solid orange line at $\bK=\ba^2$ correspond to the configurations $(B2)$ and below to $(B1)$. The line itself corresponds to $(B4)$ and covers all possible configurations (leftmost corresponds to topmost in table \ref{tab:theta}) with the exception of the $N_NN$ configuration which is only possible for large values of $e\bP$.}
\label{fig:theta_KvsE}
\end{figure}

The results of this section are visualized in figures \ref{fig:theta_LvsE_K>a^2} to \ref{fig:theta_KvsE}. As the five dimensional parameter space can not be completely pictured we have to fix at least two parameters. For this we choose $\ba$, which can be removed from equations \eqref{theta_double_E}, \eqref{theta_double_L}, \eqref{theta_new_double_E}, and \eqref{theta_double_K} by a rescaling of parameters, and $e\bP$, which enters only linearly in \eqref{theta_double_E} an \eqref{theta_double_L}. As a three dimensional plot of $E$, $\bL$, and $\bK$ is still confusing we present two dimensional plots of $\bL$ over $E$ and $\bK$ over $E$. Special cases to be discussed are then $e\bP=0$, $\bL=\pm e\bP$, $\bK=\ba^2$, and $\bK=0$. For the latter geodesic motion is only possible on a stable orbit of constant $\theta$ (if not $\T\equiv0$), more precise on the equator for $\bL=\ba E$ or on $e\bP \cos \theta = \bL$ for $\ba=0$. In each plot, we use the following conventions:
\begin{itemize}
\item Solid lines indicate double zeros of $\T$ which correspond to stable or unstable orbits of constant $\theta$. We use red lines for orbits with constant $\theta<\frac{\pi}{2}$, blue for constant $\theta>\frac{\pi}{2}$, and green for equatorial orbits.
\item Dashed lines denote orbits with turning points at the equatorial plane and are given by $\bL=\ba E\pm\sqrt{\bK}$. They also mark the transition from $N$ or $S$ to $E$ orbits and are asymptotically approached by solid lines corresponding to orbits with constant $\theta$ near the equatorial plane.
\item Dash dotted lines correspond to orbits which cross a pole. They only appear in $\bL$ over $E$ plots for $\bK\geq\ba^2$ and are located at $\bL=\pm e\bP$. In this case they are asymptotes to the solid lines corresponding to orbits with constant $\theta$ near a pole.
\item Dotted lines are asymptotes to solid lines. They do NOT separate different orbit configurations. For $\bL$ over $E$ plots they only appear for $\bK<\ba^2$ and are approached for $\nu_0 \to \pm \sqrt{\bK}/\ba$. If $\bK$ is plotted versus $E$ they are approached for $\nu_0 \to \pm 1$. 
\item Single dots mark triple zeros which separate stable from unstable orbits.
\item The labels $N$, $E$, and $S$ indicate the orbit configurations summarized in table \ref{tab:theta}. Solid, dashed, and dash dotted lines indicate transitions from one orbit configuration to another. Here dash dotted lines are special in the sense that on both sides there are always the same orbit configurations but only the line itself corresponds to another configuration. Regions where a colatitudinal motion is forbidden are marked with `none`.
\end{itemize}
Note that we plot usually only positive values of $E$ as negative values can be obtained by $E\to-E$, $\bL\to-\bL$ and a reflection at the equatorial plane (e.g.~an $N$ orbit becomes an $S$ orbit).

\subsection{Radial motion}
The discussion of the radial motion will be analogous to that of the $\theta$-motion. An orbit can only take a specific value in $[-\infty,\infty]$ if the radicand of the right hand side of eq. \eqref{eq:eomr} is larger than or equal to zero at that point, i.e.~if
\begin{equation*}
R(\br) =  ((\br^2+\ba^2) E -\ba\bL -e\bQ\br)^2 - (\epsilon \br^2  + \bK)(\br^2-2\br+\ba^2+\bQ^2+\bP^2) \geq 0 \,.
\end{equation*}
We will analyze for which values of the parameters $E$, $\bL$, $\bK$, $e$, $\bP$ and $\bQ$ this inequality is satisfied. 

\subsubsection{General properties}
The three parameters $e$, $\bQ$, and $\bP$ appear only in the two combinations $e\bQ=:\mQ$ and $\bQ^2+\bP^2=:\mP^2$ which may be considered instead. The function $R$ has the following symmetries:
\begin{itemize}
\item It depends quadratically on $\bP$ (and $\mP$): $R|_{-\bP}=R|_{\bP}$ ($R|_{-\mP}=R|_{\mP}$).
\item A simultaneous change of sign of $\bQ$ and $e$ results in the same motion: $R|_{-\bQ,-e} = R|_{\bQ,e}$.
\item Also, a simultaneous change of sign of $\bL$, $E$ and $e$(or $\mQ$) results in the same motion: $R|_{-E,-\bL,-e}=R|_{E,\bL,e}$ ($R|_{-E,-\bL,-\mQ}=R|_{E,\bL,\mQ}$).
\end{itemize}
Thus it suffices to consider $\bP\geq0$, $\bQ \geq 0$, and $e \geq 0$ (or, equivalently, $\mP\geq0$ and $\mQ\geq0$). Note that $\bK \geq 0$ was a necessary condition for the colatitudinal motion to be possible at all and, therefore, this condition remains valid.

The sign of $\bC=\bK-(\ba E-\bL)^2$ again encodes some geometrical information. At $\br=0$ the polynomial $R(0)=-(\bQ^2+\bP^2)(\bL-\ba E)^2-(\ba^2+\bQ^2+\bP^2)\bC$ can only be positive if $\bC\leq0$. Since $\bC\geq0$ needs to be satisfied for an orbit to reach the equatorial plane this implies that $\bC=0$ is necessary for an orbit to hit the ring singularity. For $\bQ^2+\bP^2\neq 0$ also $\bL-\ba E=0$ is neccessary to hit the singularity and, thus, also $\bK=0$.

The zeros of the parabola $\bDelta(\br)$ are the horizons $\br_{\pm}=1\pm \sqrt{1-\ba^2+\bQ^2+\bP^2}$ with $\bDelta(\br)<0$ in between. Therefore, the polynomial $R(\br)$ is always postive for $\br \in [\br_-,\br_+]$. This implies that there can not be any turning points or spherical orbits between the horizons. There is a turning point at a horizon if $\mR(\br_{\pm})=0$ (see eq.~\eqref{mathcalR}).

For a given set of parameters of the space-time and the particle different types of orbits may be possible, for which we use the terminologies
\begin{itemize}
\item transit or $T$, if $\br$ starts at $\pm \infty$ and ends at $\mp \infty$,
\item flyby or $F$, if $\br$ starts and ends at $+\infty$ or $-\infty$,
\item bound or $B$, if $\br$ remains in a finite interval $[\br_{\rm min},\br_{\rm max}]$.
\end{itemize}
We add an index $+$, $0$, or $-$ to a flyby or bound orbit if it stays at $\br>0$, crosses $\br=0$, or stays at $\br<0$. Also, a superscript $^*$ will be added if the orbit crosses the horizons, i.e.~contains the interval $[\br_-,\br_+]$. For example, the orbit $F_{+}^*$ comes from infinity, crosses the two horizons, turns at some $0<\br<\br_-$, and goes back to infinity. If more than one orbit is possible for a given set of parameters, the actual orbit of the test particle is determined by the initial conditions.

\subsubsection{Orbit configurations}

\begin{table}
\begin{minipage}{0.48\textwidth}
\subfloat[(I) $E^2>1$]{
\begin{tabular}{|c|c|c|}
\hline
real zeros & range of $\br$ & types of orbits \\ 
\hline\hline
0 & 
\begin{pspicture}(-1.5,-0.24)(3.0,0.24)
\psline[linewidth=1.5pt](0,-0.12)(0,0.12)
\psline[linewidth=0.5pt](1,-0.08)(1,0.08)
\psline[linewidth=0.5pt](1.5,-0.08)(1.5,0.08)
\psline[linewidth=0.5pt](-1.5,0)(3.0,0)
\psline[linewidth=1.2pt]{-}(-1.5,0)(3.0,0)
\end{pspicture} & T\\
\hline
2 &
\begin{pspicture}(-1.5,-0.24)(3.0,0.24)
\psline[linewidth=1.5pt](0,-0.12)(0,0.12)
\psline[linewidth=0.5pt](1,-0.08)(1,0.08)
\psline[linewidth=0.5pt](1.5,-0.08)(1.5,0.08)
\psline[linewidth=0.5pt](-1.5,0)(3.0,0)
\psline[linewidth=1.2pt]{-*}(-1.5,0)(1.8,0)
\psline[linewidth=1.2pt]{*-}(2.2,0)(3.0,0)
\end{pspicture} & $F_0^*$, $F_+$\\
2 &
\begin{pspicture}(-1.5,-0.24)(3.0,0.24)
\psline[linewidth=1.5pt](0,-0.12)(0,0.12)
\psline[linewidth=0.5pt](1,-0.08)(1,0.08)
\psline[linewidth=0.5pt](1.5,-0.08)(1.5,0.08)
\psline[linewidth=0.5pt](-1.5,0)(3.0,0)
\psline[linewidth=1.2pt]{-*}(-1.5,0)(0.5,0)
\psline[linewidth=1.2pt]{*-}(0.8,0)(3.0,0)
\end{pspicture} & $F_0$, $F_+^*$\\
2 &
\begin{pspicture}(-1.5,-0.24)(3.0,0.24)
\psline[linewidth=1.5pt](0,-0.12)(0,0.12)
\psline[linewidth=0.5pt](1,-0.08)(1,0.08)
\psline[linewidth=0.5pt](1.5,-0.08)(1.5,0.08)
\psline[linewidth=0.5pt](-1.5,0)(3.0,0)
\psline[linewidth=1.2pt]{-*}(-1.5,0)(-0.3,0)
\psline[linewidth=1.2pt]{*-}(0.3,0)(3.0,0)
\end{pspicture} & $F_-$, $F_+^*$\\
2 &
\begin{pspicture}(-1.5,-0.24)(3.0,0.24)
\psline[linewidth=1.5pt](0,-0.12)(0,0.12)
\psline[linewidth=0.5pt](1,-0.08)(1,0.08)
\psline[linewidth=0.5pt](1.5,-0.08)(1.5,0.08)
\psline[linewidth=0.5pt](-1.5,0)(3.0,0)
\psline[linewidth=1.2pt]{-*}(-1.5,0)(-0.7,0)
\psline[linewidth=1.2pt]{*-}(-0.3,0)(3.0,0)
\end{pspicture} & $F_-$, $F_0^*$\\
\hline
4 &
\begin{pspicture}(-1.5,-0.24)(3.0,0.24)
\psline[linewidth=1.5pt](0,-0.12)(0,0.12)
\psline[linewidth=0.5pt](1,-0.08)(1,0.08)
\psline[linewidth=0.5pt](1.5,-0.08)(1.5,0.08)
\psline[linewidth=0.5pt](-1.5,0)(3.0,0)
\psline[linewidth=1.2pt]{-*}(-1.5,0)(1.75,0)
\psline[linewidth=1.2pt]{*-*}(2.0,0)(2.3,0)
\psline[linewidth=1.2pt]{*-}(2.55,0)(3,0)
\end{pspicture} & $F_0^*$, $B_+$, $F_+$\\
4&
\begin{pspicture}(-1.5,-0.24)(3.0,0.24)
\psline[linewidth=1.5pt](0,-0.12)(0,0.12)
\psline[linewidth=0.5pt](1,-0.08)(1,0.08)
\psline[linewidth=0.5pt](1.5,-0.08)(1.5,0.08)
\psline[linewidth=0.5pt](-1.5,0)(3.0,0)
\psline[linewidth=1.2pt]{-*}(-1.5,0)(0.5,0)
\psline[linewidth=1.2pt]{*-*}(0.8,0)(1.8,0)
\psline[linewidth=1.2pt]{*-}(2.2,0)(3,0)
\end{pspicture} & $F_0$, $B_+^*$, $F_+$\\
4&
\begin{pspicture}(-1.5,-0.24)(3.0,0.24)
\psline[linewidth=1.5pt](0,-0.12)(0,0.12)
\psline[linewidth=0.5pt](1,-0.08)(1,0.08)
\psline[linewidth=0.5pt](1.5,-0.08)(1.5,0.08)
\psline[linewidth=0.5pt](-1.5,0)(3.0,0)
\psline[linewidth=1.2pt]{-*}(-1.5,0)(-0.5,0)
\psline[linewidth=1.2pt]{*-*}(0.5,0)(1.8,0)
\psline[linewidth=1.2pt]{*-}(2.2,0)(3,0)
\end{pspicture} & $F_-$, $B_+^*$, $F_+$\\
4&
\begin{pspicture}(-1.5,-0.24)(3.0,0.24)
\psline[linewidth=1.5pt](0,-0.12)(0,0.12)
\psline[linewidth=0.5pt](1,-0.08)(1,0.08)
\psline[linewidth=0.5pt](1.5,-0.08)(1.5,0.08)
\psline[linewidth=0.5pt](-1.5,0)(3.0,0)
\psline[linewidth=1.2pt]{-*}(-1.5,0)(-0.8,0)
\psline[linewidth=1.2pt]{*-*}(-0.5,0)(1.8,0)
\psline[linewidth=1.2pt]{*-}(2.2,0)(3,0)
\end{pspicture} & $F_-$, $B_0^*$, $F_+$\\
4&
\begin{pspicture}(-1.5,-0.24)(3.0,0.24)
\psline[linewidth=1.5pt](0,-0.12)(0,0.12)
\psline[linewidth=0.5pt](1,-0.08)(1,0.08)
\psline[linewidth=0.5pt](1.5,-0.08)(1.5,0.08)
\psline[linewidth=0.5pt](-1.5,0)(3.0,0)
\psline[linewidth=1.2pt]{-*}(-1.5,0)(0.15,0)
\psline[linewidth=1.2pt]{*-*}(0.37,0)(0.63,0)
\psline[linewidth=1.2pt]{*-}(0.85,0)(3,0)
\end{pspicture} & $F_0$, $B_+$, $F_+^*$\\
4&
\begin{pspicture}(-1.5,-0.24)(3.0,0.24)
\psline[linewidth=1.5pt](0,-0.12)(0,0.12)
\psline[linewidth=0.5pt](1,-0.08)(1,0.08)
\psline[linewidth=0.5pt](1.5,-0.08)(1.5,0.08)
\psline[linewidth=0.5pt](-1.5,0)(3.0,0)
\psline[linewidth=1.2pt]{-*}(-1.5,0)(-0.5,0)
\psline[linewidth=1.2pt]{*-*}(0.2,0)(0.5,0)
\psline[linewidth=1.2pt]{*-}(0.8,0)(3,0)
\end{pspicture} & $F_-$, $B_+$, $F_+^*$\\
4&
\begin{pspicture}(-1.5,-0.24)(3.0,0.24)
\psline[linewidth=1.5pt](0,-0.12)(0,0.12)
\psline[linewidth=0.5pt](1,-0.08)(1,0.08)
\psline[linewidth=0.5pt](1.5,-0.08)(1.5,0.08)
\psline[linewidth=0.5pt](-1.5,0)(3.0,0)
\psline[linewidth=1.2pt]{-*}(-1.5,0)(-0.5,0)
\psline[linewidth=1.2pt]{*-*}(-0.2,0)(0.5,0)
\psline[linewidth=1.2pt]{*-}(0.8,0)(3,0)
\end{pspicture} & $F_-$, $B_0$, $F_+^*$\\
4&
\begin{pspicture}(-1.5,-0.24)(3.0,0.24)
\psline[linewidth=1.5pt](0,-0.12)(0,0.12)
\psline[linewidth=0.5pt](1,-0.08)(1,0.08)
\psline[linewidth=0.5pt](1.5,-0.08)(1.5,0.08)
\psline[linewidth=0.5pt](-1.5,0)(3.0,0)
\psline[linewidth=1.2pt]{-*}(-1.5,0)(-1.0,0)
\psline[linewidth=1.2pt]{*-*}(-0.6,0)(-0.2,0)
\psline[linewidth=1.2pt]{*-}(0.8,0)(3,0)
\end{pspicture} & $F_-$, $B_-$, $F_+^*$\\
4&
\begin{pspicture}(-1.5,-0.24)(3.0,0.24)
\psline[linewidth=1.5pt](0,-0.12)(0,0.12)
\psline[linewidth=0.5pt](1,-0.08)(1,0.08)
\psline[linewidth=0.5pt](1.5,-0.08)(1.5,0.08)
\psline[linewidth=0.5pt](-1.5,0)(3.0,0)
\psline[linewidth=1.2pt]{-*}(-1.5,0)(-1.0,0)
\psline[linewidth=1.2pt]{*-*}(-0.8,0)(-0.45,0)
\psline[linewidth=1.2pt]{*-}(-0.2,0)(3,0)
\end{pspicture} & $F_-$, $B_-$, $F_0^*$\\
\hline
\end{tabular}
}\\
\subfloat[(III) $E^2=1$, $e\bQ>1$]{
\begin{tabular}{|c|c|c|}
\hline
real zeros & range of $\br$ & types of orbits \\ 
\hline\hline
1 &
\begin{pspicture}(-1.5,-0.24)(3.0,0.24)
\psline[linewidth=1.5pt](0,-0.12)(0,0.12)
\psline[linewidth=0.5pt](1,-0.08)(1,0.08)
\psline[linewidth=0.5pt](1.5,-0.08)(1.5,0.08)
\psline[linewidth=0.5pt](-1.5,0)(3.0,0)
\psline[linewidth=1.2pt]{-*}(-1.5,0)(2.0,0)
\end{pspicture} & $F_0^*$\\
\hline
3 &
\begin{pspicture}(-1.5,-0.24)(3.0,0.24)
\psline[linewidth=1.5pt](0,-0.12)(0,0.12)
\psline[linewidth=0.5pt](1,-0.08)(1,0.08)
\psline[linewidth=0.5pt](1.5,-0.08)(1.5,0.08)
\psline[linewidth=0.5pt](-1.5,0)(3.0,0)
\psline[linewidth=1.2pt]{-*}(-1.5,0)(-1.0,0)
\psline[linewidth=1.2pt]{*-*}(-0.5,0)(2.0,0)
\end{pspicture} & $F_-$, $B_0^*$\\
3 &
\begin{pspicture}(-1.5,-0.24)(3.0,0.24)
\psline[linewidth=1.5pt](0,-0.12)(0,0.12)
\psline[linewidth=0.5pt](1,-0.08)(1,0.08)
\psline[linewidth=0.5pt](1.5,-0.08)(1.5,0.08)
\psline[linewidth=0.5pt](-1.5,0)(3.0,0)
\psline[linewidth=1.2pt]{-*}(-1.5,0)(-1.0,0)
\psline[linewidth=1.2pt]{*-*}(0.5,0)(2.0,0)
\end{pspicture} & $F_-$, $B_+^*$\\
3 &
\begin{pspicture}(-1.5,-0.24)(3.0,0.24)
\psline[linewidth=1.5pt](0,-0.12)(0,0.12)
\psline[linewidth=0.5pt](1,-0.08)(1,0.08)
\psline[linewidth=0.5pt](1.5,-0.08)(1.5,0.08)
\psline[linewidth=0.5pt](-1.5,0)(3.0,0)
\psline[linewidth=1.2pt]{-*}(-1.5,0)(0.3,0)
\psline[linewidth=1.2pt]{*-*}(0.6,0)(2.0,0)
\end{pspicture} & $F_0$, $B_+^*$\\
3 &
\begin{pspicture}(-1.5,-0.24)(3.0,0.24)
\psline[linewidth=1.5pt](0,-0.12)(0,0.12)
\psline[linewidth=0.5pt](1,-0.08)(1,0.08)
\psline[linewidth=0.5pt](1.5,-0.08)(1.5,0.08)
\psline[linewidth=0.5pt](-1.5,0)(3.0,0)
\psline[linewidth=1.2pt]{-*}(-1.5,0)(1.8,0)
\psline[linewidth=1.2pt]{*-*}(2.2,0)(2.6,0)
\end{pspicture} & $F_0^*$, $B_+$\\
\hline
\end{tabular}
}
\end{minipage}\quad
\begin{minipage}{0.48\textwidth}
\subfloat[(II) $E^2<1$]{
\begin{tabular}{|c|c|c|}
\hline
real zeros & range of $\br$ & types of orbits \\ 
\hline\hline
2&
\begin{pspicture}(-1.5,-0.24)(3.0,0.24)
\psline[linewidth=1.5pt](0,-0.12)(0,0.12)
\psline[linewidth=0.5pt](1,-0.08)(1,0.08)
\psline[linewidth=0.5pt](1.5,-0.08)(1.5,0.08)
\psline[linewidth=0.5pt](-1.5,0)(3.0,0)
\psline[linewidth=1.2pt]{*-*}(0.2,0)(1.8,0)
\end{pspicture} & $B_+^*$\\
2&
\begin{pspicture}(-1.5,-0.24)(3.0,0.24)
\psline[linewidth=1.5pt](0,-0.12)(0,0.12)
\psline[linewidth=0.5pt](1,-0.08)(1,0.08)
\psline[linewidth=0.5pt](1.5,-0.08)(1.5,0.08)
\psline[linewidth=0.5pt](-1.5,0)(3.0,0)
\psline[linewidth=1.2pt]{*-*}(-0.5,0)(1.8,0)
\end{pspicture} & $B_0^*$\\
\hline\hline
4&
\begin{pspicture}(-1.5,-0.24)(3.0,0.24)
\psline[linewidth=1.5pt](0,-0.12)(0,0.12)
\psline[linewidth=0.5pt](1,-0.08)(1,0.08)
\psline[linewidth=0.5pt](1.5,-0.08)(1.5,0.08)
\psline[linewidth=0.5pt](-1.5,0)(3.0,0)
\psline[linewidth=1.2pt]{*-*}(0.5,0)(1.8,0)
\psline[linewidth=1.2pt]{*-*}(2.2,0)(2.8,0)
\end{pspicture} & $B_+^*$, $B_+$\\
4&
\begin{pspicture}(-1.5,-0.24)(3.0,0.24)
\psline[linewidth=1.5pt](0,-0.12)(0,0.12)
\psline[linewidth=0.5pt](1,-0.08)(1,0.08)
\psline[linewidth=0.5pt](1.5,-0.08)(1.5,0.08)
\psline[linewidth=0.5pt](-1.5,0)(3.0,0)
\psline[linewidth=1.2pt]{*-*}(-0.5,0)(1.8,0)
\psline[linewidth=1.2pt]{*-*}(2.2,0)(2.8,0)
\end{pspicture} & $B_0^*$, $B_+$\\
4&
\begin{pspicture}(-1.5,-0.24)(3.0,0.24)
\psline[linewidth=1.5pt](0,-0.12)(0,0.12)
\psline[linewidth=0.5pt](1,-0.08)(1,0.08)
\psline[linewidth=0.5pt](1.5,-0.08)(1.5,0.08)
\psline[linewidth=0.5pt](-1.5,0)(3.0,0)
\psline[linewidth=1.2pt]{*-*}(0.85,0)(1.8,0)
\psline[linewidth=1.2pt]{*-*}(0.2,0)(0.55,0)
\end{pspicture} & $B_+$, $B_+^*$\\
4&
\begin{pspicture}(-1.5,-0.24)(3.0,0.24)
\psline[linewidth=1.5pt](0,-0.12)(0,0.12)
\psline[linewidth=0.5pt](1,-0.08)(1,0.08)
\psline[linewidth=0.5pt](1.5,-0.08)(1.5,0.08)
\psline[linewidth=0.5pt](-1.5,0)(3.0,0)
\psline[linewidth=1.2pt]{*-*}(0.5,0)(1.8,0)
\psline[linewidth=1.2pt]{*-*}(-1.2,0)(0.2,0)
\end{pspicture} & $B_0$, $B_+^*$\\
4&
\begin{pspicture}(-1.5,-0.24)(3.0,0.24)
\psline[linewidth=1.5pt](0,-0.12)(0,0.12)
\psline[linewidth=0.5pt](1,-0.08)(1,0.08)
\psline[linewidth=0.5pt](1.5,-0.08)(1.5,0.08)
\psline[linewidth=0.5pt](-1.5,0)(3.0,0)
\psline[linewidth=1.2pt]{*-*}(-1.2,0)(-0.4,0)
\psline[linewidth=1.2pt]{*-*}(0.4,0)(1.8,0)
\end{pspicture} & $B_-$, $B_+^*$\\
4&
\begin{pspicture}(-1.5,-0.24)(3.0,0.24)
\psline[linewidth=1.5pt](0,-0.12)(0,0.12)
\psline[linewidth=0.5pt](1,-0.08)(1,0.08)
\psline[linewidth=0.5pt](1.5,-0.08)(1.5,0.08)
\psline[linewidth=0.5pt](-1.5,0)(3.0,0)
\psline[linewidth=1.2pt]{*-*}(-1.2,0)(-0.8,0)
\psline[linewidth=1.2pt]{*-*}(-0.2,0)(1.8,0)
\end{pspicture} & $B_-$, $B_0^*$\\
\hline
\end{tabular}
}\\
\subfloat[(III) $E^2=1$, $e\bQ<1$]{
\begin{tabular}{|c|c|c|}
\hline
real zeros & range of $\br$ & types of orbits \\ 
\hline\hline
1 &
\begin{pspicture}(-1.5,-0.24)(3.0,0.24)
\psline[linewidth=1.5pt](0,-0.12)(0,0.12)
\psline[linewidth=0.5pt](1,-0.08)(1,0.08)
\psline[linewidth=0.5pt](1.5,-0.08)(1.5,0.08)
\psline[linewidth=0.5pt](-1.5,0)(3.0,0)
\psline[linewidth=1.2pt]{*-}(0.6,0)(3.0,0)
\end{pspicture} & $F_+^*$\\
1 &
\begin{pspicture}(-1.5,-0.24)(3.0,0.24)
\psline[linewidth=1.5pt](0,-0.12)(0,0.12)
\psline[linewidth=0.5pt](1,-0.08)(1,0.08)
\psline[linewidth=0.5pt](1.5,-0.08)(1.5,0.08)
\psline[linewidth=0.5pt](-1.5,0)(3.0,0)
\psline[linewidth=1.2pt]{*-}(-0.4,0)(3.0,0)
\end{pspicture} & $F_0^*$\\
\hline
3 &
\begin{pspicture}(-1.5,-0.24)(3.0,0.24)
\psline[linewidth=1.5pt](0,-0.12)(0,0.12)
\psline[linewidth=0.5pt](1,-0.08)(1,0.08)
\psline[linewidth=0.5pt](1.5,-0.08)(1.5,0.08)
\psline[linewidth=0.5pt](-1.5,0)(3.0,0)
\psline[linewidth=1.2pt]{*-*}(0.6,0)(1.8,0)
\psline[linewidth=1.2pt]{*-}(2.2,0)(3.0,0)
\end{pspicture} & $B_+^*$, $F_+$\\
3 &
\begin{pspicture}(-1.5,-0.24)(3.0,0.24)
\psline[linewidth=1.5pt](0,-0.12)(0,0.12)
\psline[linewidth=0.5pt](1,-0.08)(1,0.08)
\psline[linewidth=0.5pt](1.5,-0.08)(1.5,0.08)
\psline[linewidth=0.5pt](-1.5,0)(3.0,0)
\psline[linewidth=1.2pt]{*-*}(-0.4,0)(1.8,0)
\psline[linewidth=1.2pt]{*-}(2.2,0)(3.0,0)
\end{pspicture} & $B_0^*$, $F_+$\\
3 &
\begin{pspicture}(-1.5,-0.24)(3.0,0.24)
\psline[linewidth=1.5pt](0,-0.12)(0,0.12)
\psline[linewidth=0.5pt](1,-0.08)(1,0.08)
\psline[linewidth=0.5pt](1.5,-0.08)(1.5,0.08)
\psline[linewidth=0.5pt](-1.5,0)(3.0,0)
\psline[linewidth=1.2pt]{*-*}(0.2,0)(0.55,0)
\psline[linewidth=1.2pt]{*-}(0.8,0)(3.0,0)
\end{pspicture} & $B_+$, $F_+^*$\\
3 &
\begin{pspicture}(-1.5,-0.24)(3.0,0.24)
\psline[linewidth=1.5pt](0,-0.12)(0,0.12)
\psline[linewidth=0.5pt](1,-0.08)(1,0.08)
\psline[linewidth=0.5pt](1.5,-0.08)(1.5,0.08)
\psline[linewidth=0.5pt](-1.5,0)(3.0,0)
\psline[linewidth=1.2pt]{*-*}(-0.2,0)(0.55,0)
\psline[linewidth=1.2pt]{*-}(0.8,0)(3.0,0)
\end{pspicture} & $B_0$, $F_+^*$\\
3 &
\begin{pspicture}(-1.5,-0.24)(3.0,0.24)
\psline[linewidth=1.5pt](0,-0.12)(0,0.12)
\psline[linewidth=0.5pt](1,-0.08)(1,0.08)
\psline[linewidth=0.5pt](1.5,-0.08)(1.5,0.08)
\psline[linewidth=0.5pt](-1.5,0)(3.0,0)
\psline[linewidth=1.2pt]{*-*}(-1.2,0)(-0.6,0)
\psline[linewidth=1.2pt]{*-}(0.6,0)(3.0,0)
\end{pspicture} & $B_-$, $F_+^*$\\
3 &
\begin{pspicture}(-1.5,-0.24)(3.0,0.24)
\psline[linewidth=1.5pt](0,-0.12)(0,0.12)
\psline[linewidth=0.5pt](1,-0.08)(1,0.08)
\psline[linewidth=0.5pt](1.5,-0.08)(1.5,0.08)
\psline[linewidth=0.5pt](-1.5,0)(3.0,0)
\psline[linewidth=1.2pt]{*-*}(-1.2,0)(-0.6,0)
\psline[linewidth=1.2pt]{*-}(-0.2,0)(3.0,0)
\end{pspicture} & $B_-$, $F_0^*$\\
\hline
\end{tabular}
}
\end{minipage}
\caption{Overview of different orbit configurations for radial motion. The vertical bars of the second column mark $\br=0$, $\br=\br_-$, and $\br=\br_+$ (from left to right). The dots represent the real zeros of $R$ (turning points) and the thick lines $R\geq0$, i.e.~regions where a motion is possible. If zeros merge, the resulting orbits are stable if a line is reduced to a point and unstable if lines merge.}
\label{tab:radial}
\end{table}

We will now analyze which sets of the above introduced orbit types are possible for given parameters. Geodesic motion is only possible in regions of $R(\br) \geq 0$ and, therefore, the possible orbit configurations are fully determined by the number of real zeros of the polynomial $R$ and its sign at $\pm \infty$. The latter is determined by the sign of $E^2-1$ (or the lower order coefficients if $E^2=1$, what we will consider separately). This suggest to introduce the following classes of orbit configurations.
\paragraph*{(I) Case $E^2>1$} Here $R(\br) \to \infty$ for $\br \to \pm \infty$ and $R$ may have none, two, or four real zeros. For no zeros there is a transit orbit, for two zeros there are two flyby orbits, and for four zeros there are two flyby and a bound orbit.
\paragraph*{(II) Case $E^2<1$} For such energies a test particle can not reach $\pm \infty$, $R(\br)\to - \infty$ for $\br \to \pm \infty$. As $R(\br)>0$ between the horizons it has at least 2 real zeros and there is always one bound orbit crossing the horizons. If $R$ has four real zeros there is an additional bound orbit.
\paragraph*{(III) Case $E^2=1$} Here the behaviour of $R$ at infinity depends on the sign of $1-e\bQ$. For $1-e\bQ>0$ it is $R(\br) \to \pm \infty$ for $\br \to \pm \infty$ and the other way around for $1-e\bQ<0$. In both cases $R$ has one or three real zeros. For one real zero there is a flyby orbit which crosses the horizons and for three real zeros there is a flyby and a bound orbit. The flyby orbit reaches $+\infty$ for $1-e\bQ>0$ and $-\infty$ for $1-e\bQ<0$. If also $1-e\bQ=0$ one has to consider the sign of the second order coefficient and so on.\\
For an overview of this different orbit configuration for the radial motion see table \ref{tab:radial}.

\subsubsection{Regions of orbit configurations in parameter space}
After considering the possible orbit configurations for the radial motion we will now find the sets of parameters for which a given orbit configuration changes to another. The orbit configurations are fully determined by the signs of $R(\pm\infty)$ as categorized above and the number of real zeros of $R$, which changes if two zeros of $R$ merge. The latter occurs if the two conditions on double zeros $R(\br_0)=0$ and $\frac{dR}{d\br}(\br_0)=0$ are fulfilled. Read as two conditions on $E$ and $\bL$ this implies
\begin{align}
E_{1,2} & = \frac{e\bQ}{2\br_0} \pm \frac{1}{2} \frac{\sqrt{\bDelta(\br_0)(\br_0^2+\bK)(\br_0\bDelta(\br_0)+(\br_0-1)(\bK+\br_0^2))^2}}{\bDelta(\br_0)(\br_0^2+\bK)\br_0}\,, \label{R_double_E} \\
\bL_{1,2} & = \frac{e\bQ(\ba^2-\br_0^2)}{2\ba \br_0}  \pm \frac{1}{2} \frac{\sqrt{\bDelta(\br_0)(\br_0^2+\bK)(\br_0\bDelta(\br_0)+(\br_0-1)(\bK+\br_0^2))^2}}{\bDelta(\br_0)(\br_0^2+\bK)(\br_0\bDelta(\br_0)+(\br_0-1)(\bK+\br_0^2))\br_0 \ba} \nonumber \\
& \qquad \times 
(\br_0\bDelta(\br_0)(\ba^2-\bK)+(\bK+\br_0^2)(\br_0^2-\br_0(\bP^2+\bQ^2)-\ba^2))\,. \label{R_double_L}
\end{align}
These expressions diverge at $\br_0=0$ and at the horizons $\br_0=\br_\pm$. We consider these points separately, see below. In addition, $\bL$ diverges also at $\br_0=\pm \infty$, $\bL_{1,2} \to \frac{\pm{\rm sign}(\br_0)-e\bQ}{2\ba} \br_0 + \mathcal{O}(1)$ for $\br_0 \to \pm\infty$. This implies that for $e\bQ>1$ other orbit configurations than for $e\bQ<1$ may appear. At the limits $\br_0 \to \pm \infty$ the energy $E$ remains finite, $E_{1,2} \to \pm{\rm sign}(\br_0)$ there.
\paragraph*{Orbits at $\br=0$}
As the ring singularity is located at $\br=0$, $\theta=\frac{\pi}{2}$ only those orbits which are not equatorial do not terminate at $\br=0$. From the discussion of the colatitudinal motion equatorial orbits occur for $\bK=0$, $\ba E=\bL$ or $\bK=(\ba E-\bL)^2$, $e\bP=0$. If we exclude these parameters $\br=0$ is a multiple zero if and only if $e\bQ\neq 0$, $\bK=e^2\bQ^2\bDelta(0)$, and $\ba E-\bL=\frac{e\bQ\bDelta(0)}{\ba}$. This can also be infered from the asymptotic behaviour of $E_{1,2}$ and $\bL_{1,2}$ at $\br_0=0$, which is given by
\begin{align}
E_{1,2} & = \frac{e\bQ\bDelta(0)\pm\sqrt{\bK\bDelta(0)}}{2 \bDelta(0) \br_0} \mp \frac{\bDelta(0)(1+e^2\bQ^2)-e^2\bQ^2}{2e\bQ\bDelta(0)} + \mathcal{O}(\br_0)\,,\\
\bL_{1,2} & = \ba E_{1,2}\pm\frac{\sqrt{K\bDelta(0)}}{\ba} + \mathcal{O}(\br_0)\,.
\end{align}
Here the equation for $E_2$ implies two facts: First, at $\bK=e^2\bQ^2\bDelta(0)$ the regions of orbit configurations essentially change because the sign of the $r_0^{-1}$ term changes and $E_2$ switches between $\pm \infty$. Second, for $\bK=e^2\bQ^2\bDelta(0)$ the expressions for $E_2$, $L_2$, and $e\bQ$ are the same as \eqref{r_trip_E}-\eqref{TripleQ} with $\br_0=0$, which correspond to triple zeros.

\paragraph*{Orbits at the horizons}
It was noted above that a horizon is a turning point of the radial motion if $\mathcal{R}(\br_\pm)=0$. This implies that the horizons can not be multiple zeros because of $\frac{dR}{d\br}(\br_\pm) = -2(\br_\pm-1)(\br_\pm^2+\bK) \neq 0$ for $\mathcal{R}(r_\pm)=0$. (This is valid for massive test particles only. For light rays a horizon is a multiple zero if $\bK=0$ and $\ba\bL=E(\br_\pm^2+\ba^2)-e\bQ\br_\pm$ but the corresponding orbit is always unstable.) The asymptotic behaviour of $E_{1,2}$ and $\bL_{1,2}$ near the horizons is given by
\begin{align}
\lim_{\br_0\to\br_+}E_{1,2} & \to \pm \frac{\sqrt{2}}{4} \frac{\sqrt{\br_+-1}\sqrt{\br_+^2+K}}{\br_+ \sqrt{\br_0-\br_+}} + \mathcal{O}(1)\,, \\
\lim_{\br_0\to\br_-}E_{1,2} & \to \mp \frac{\sqrt{2}}{4} \frac{\sqrt{\br_--1}\sqrt{\br_-^2+K}}{\br_- \sqrt{\br_0-\br_-}} + \mathcal{O}(1)\,, \\
\lim_{\br_0\to\br_{\pm}}\bL_{1,2} & \to \frac{E_{1,2}}{\ba} (2\br_{\pm}-\bP^2-\bQ^2) + \mathcal{O}(1)\,.
\end{align}

\begin{figure}
\subfloat[(a) $e\bQ=0.3$, $e\bQ<1<\sqrt{\frac{\bK}{\bDelta(0)}}$]{
\begin{minipage}{0.45\textwidth}
\begin{center}
\includegraphics[width=0.47\textwidth]{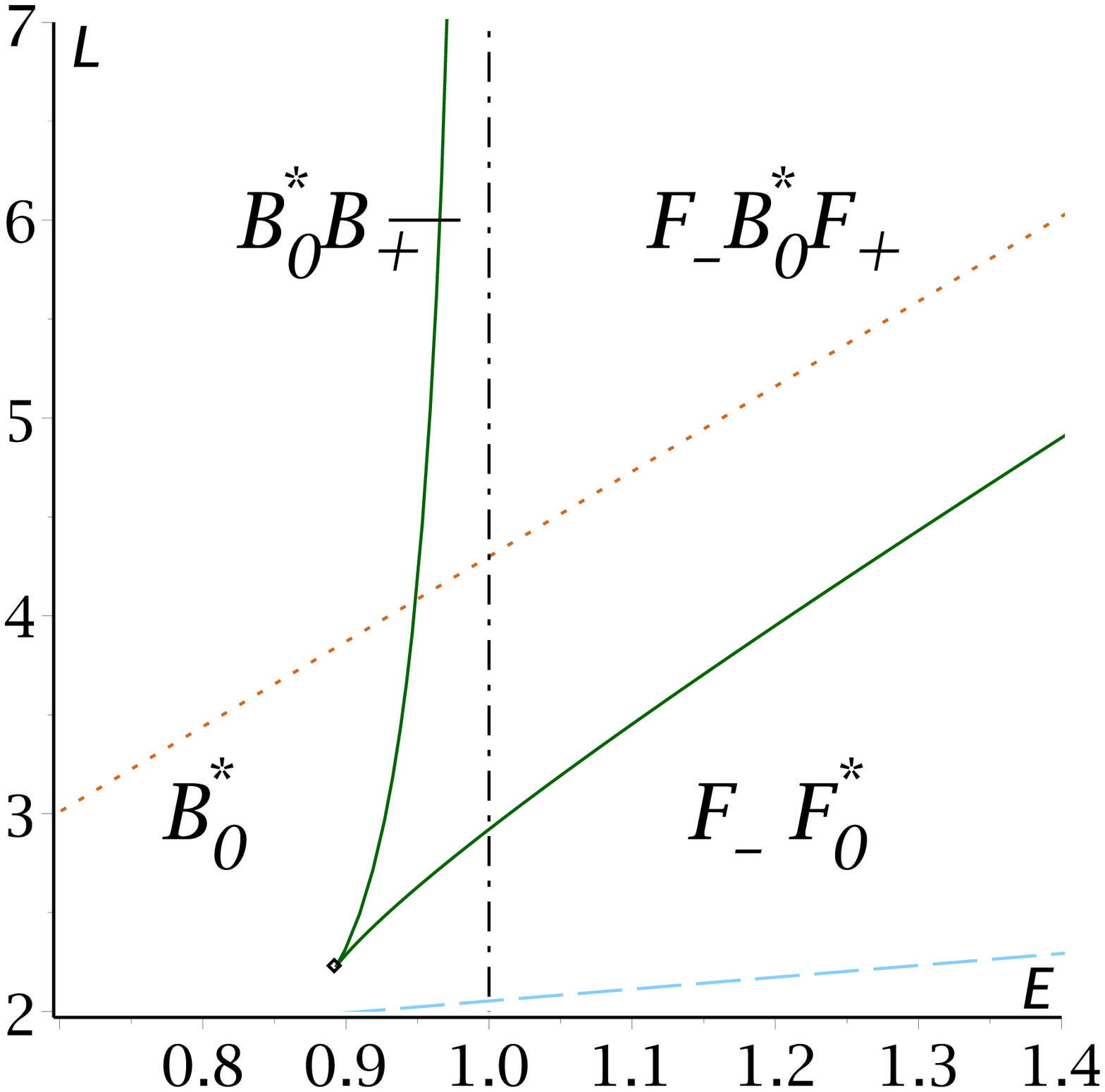}
\includegraphics[width=0.47\textwidth]{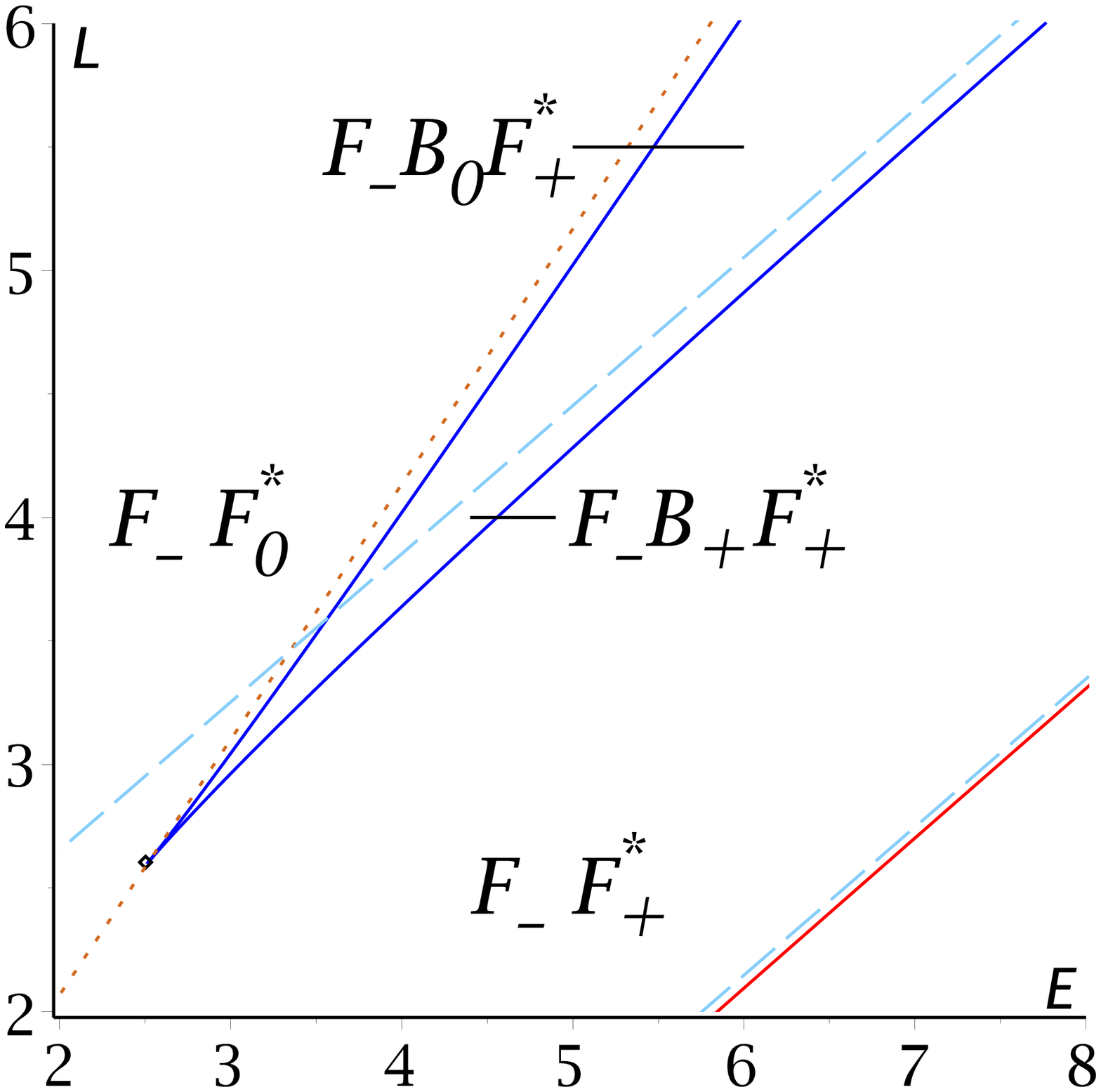}\\
\includegraphics[width=0.99\textwidth]{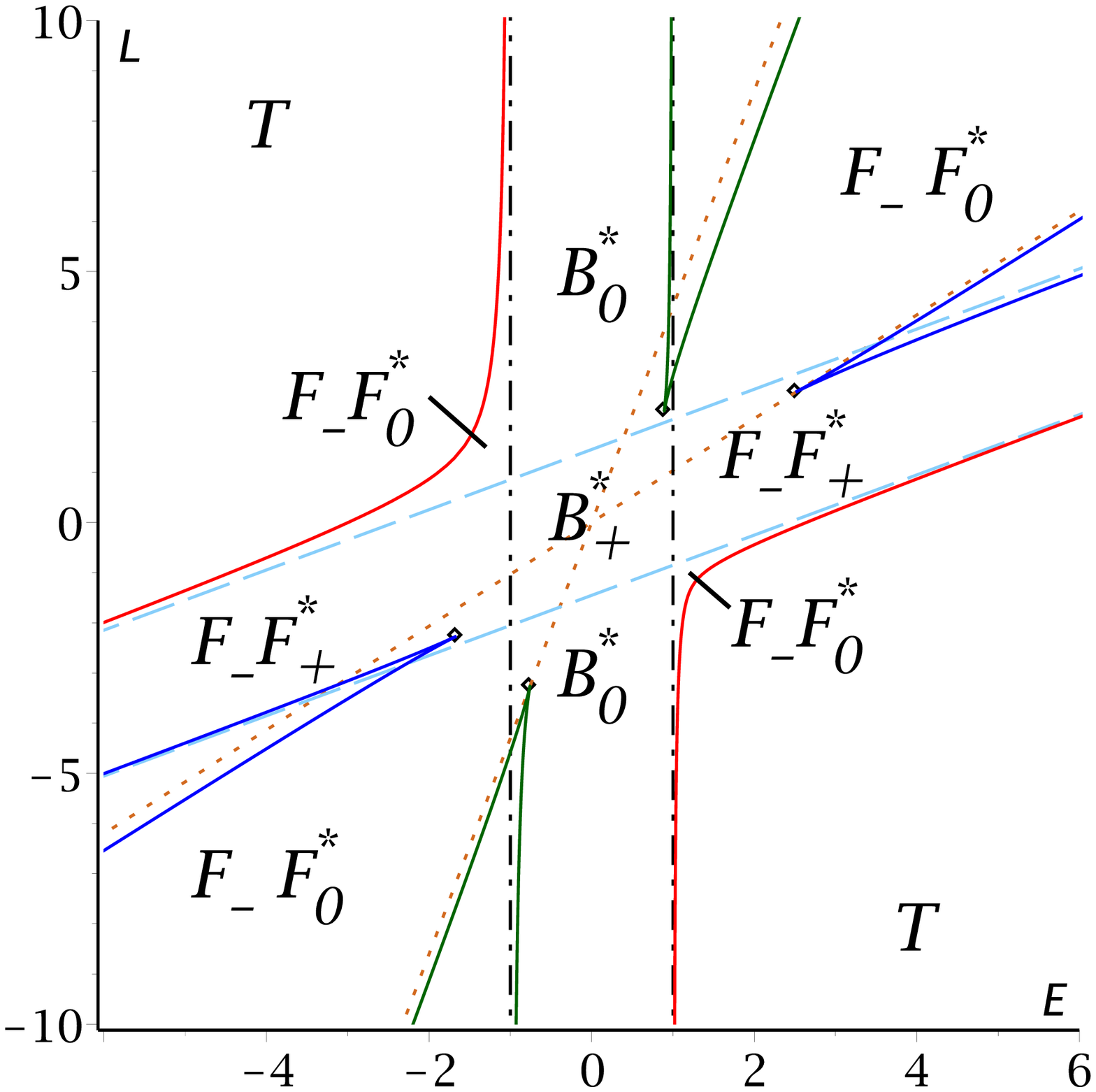}
\end{center}
\end{minipage}
}
\subfloat[(b) $e\bQ=0.9$, $e\bQ<1<\sqrt{\frac{\bK}{\bDelta(0)}}$]{
\begin{minipage}{0.45\textwidth}
\includegraphics[width=0.47\textwidth]{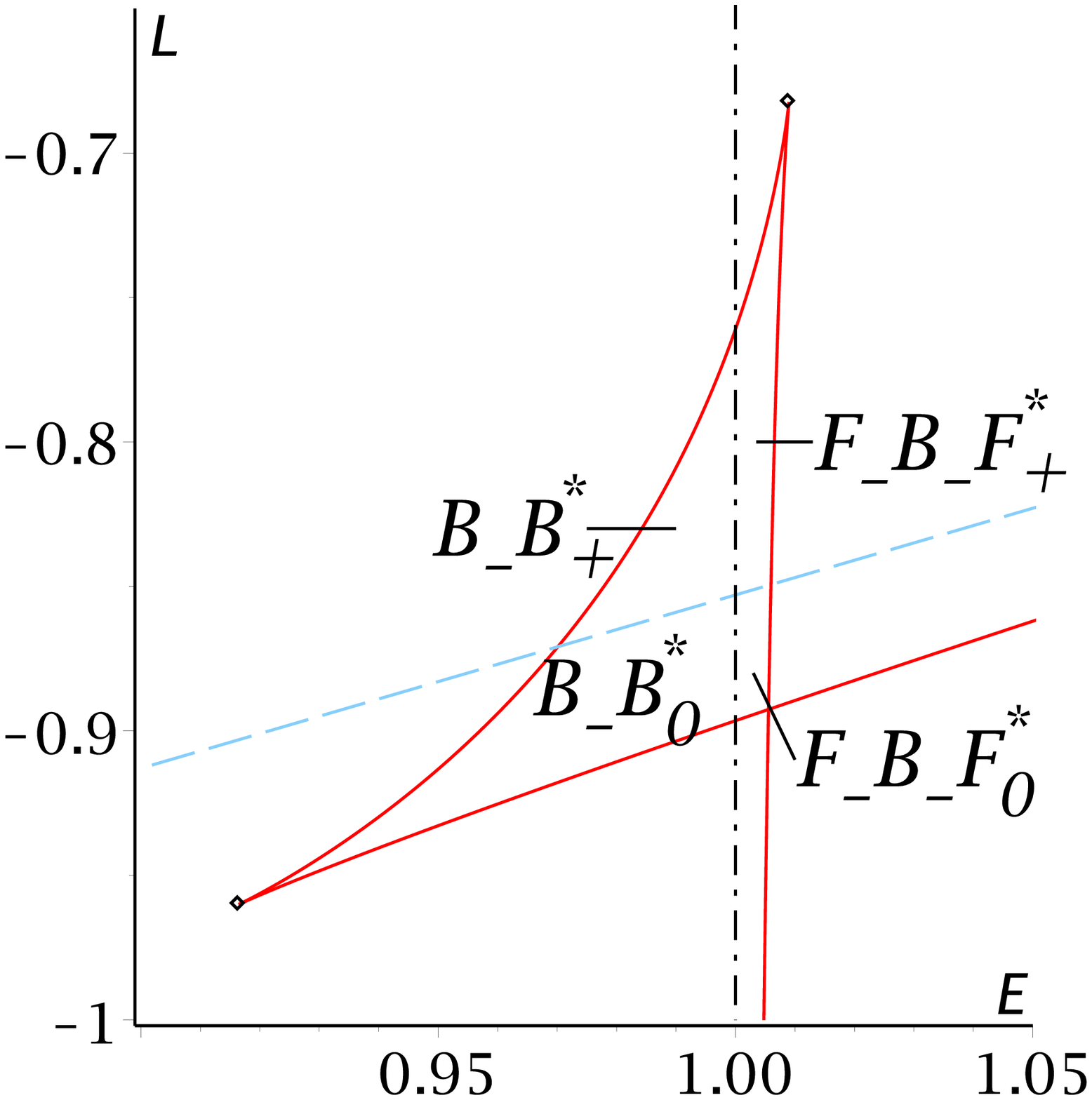}
\includegraphics[width=0.47\textwidth]{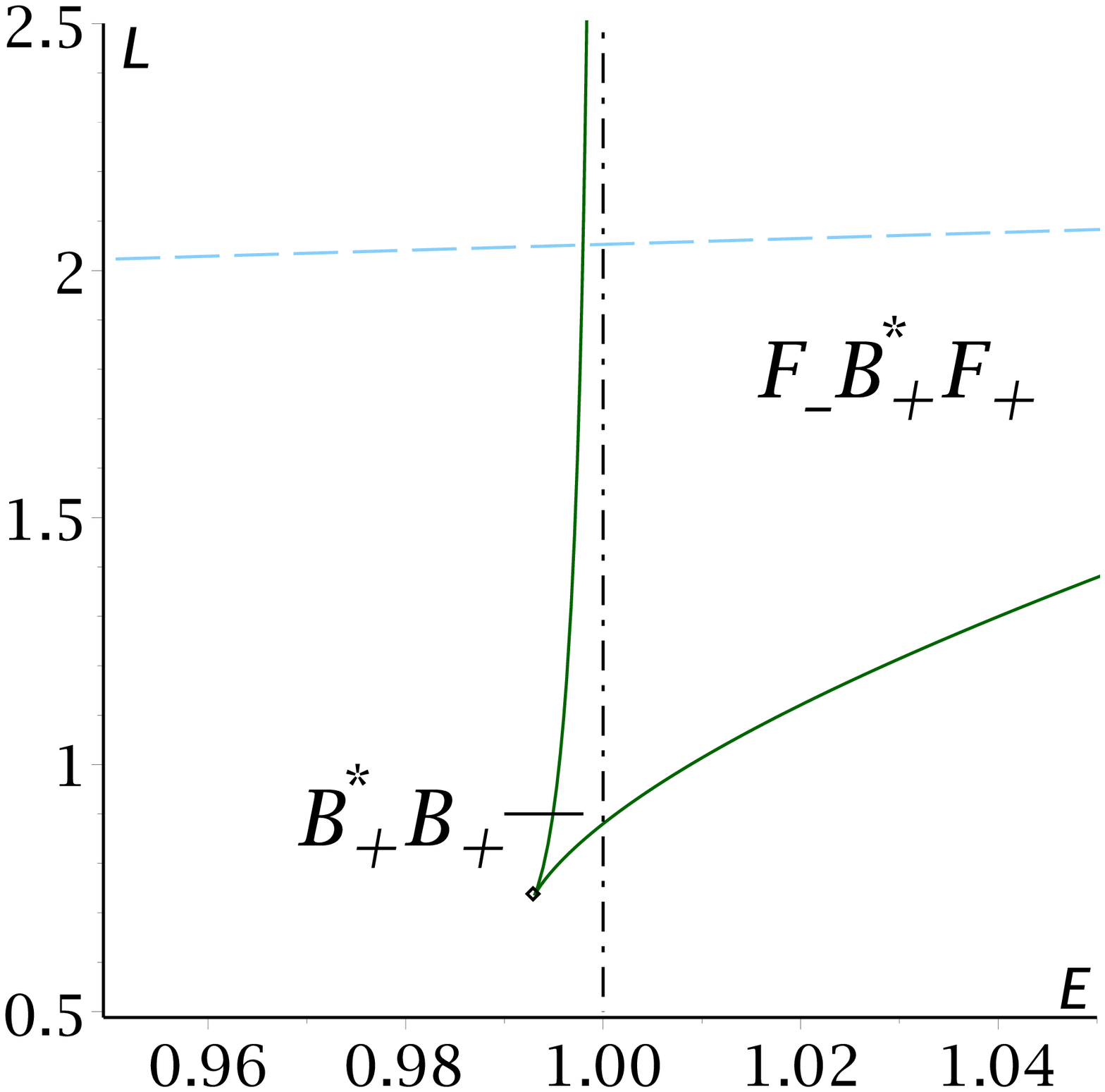}\\
\includegraphics[width=0.47\textwidth]{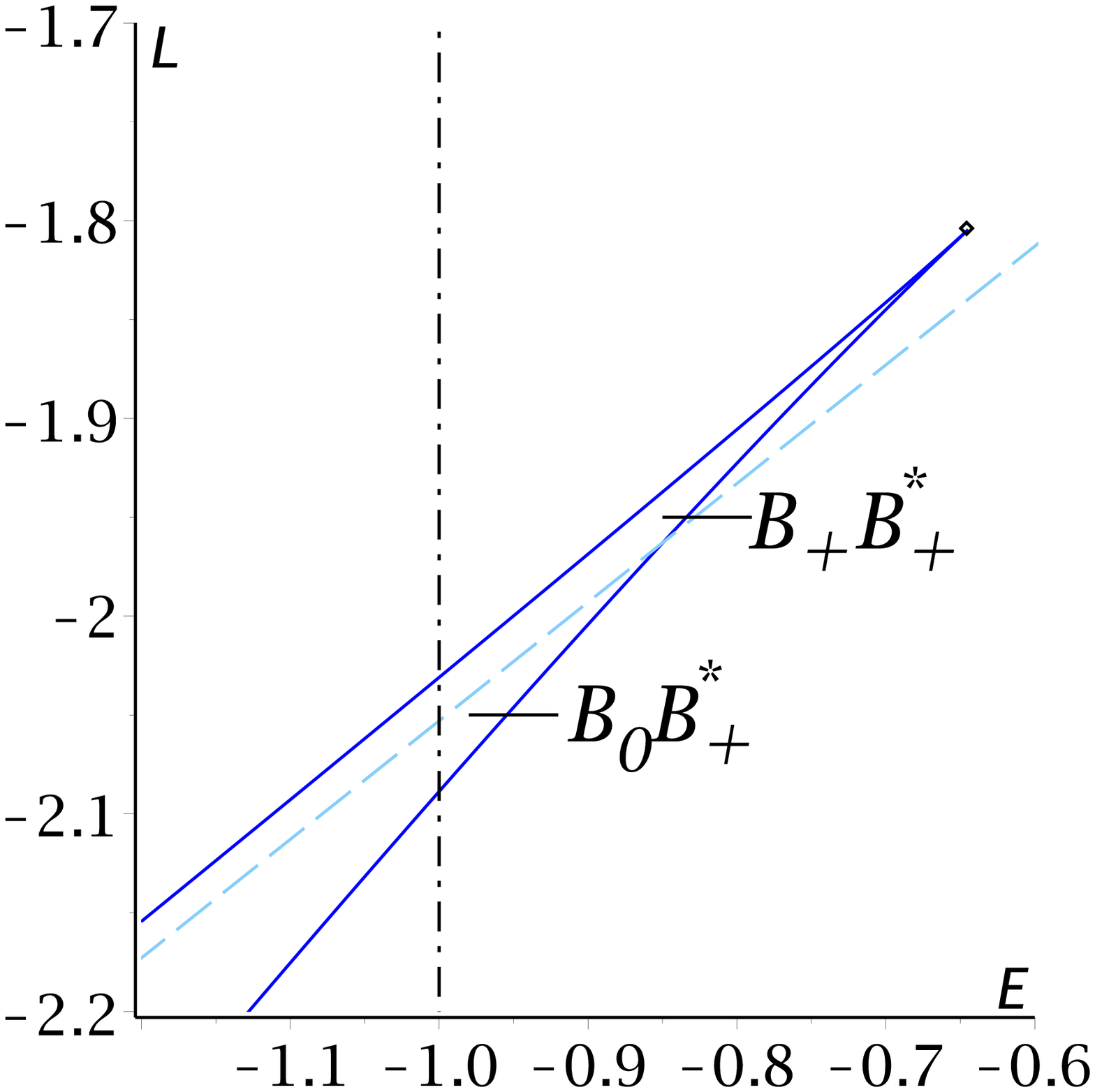}\\
\includegraphics[width=0.99\textwidth]{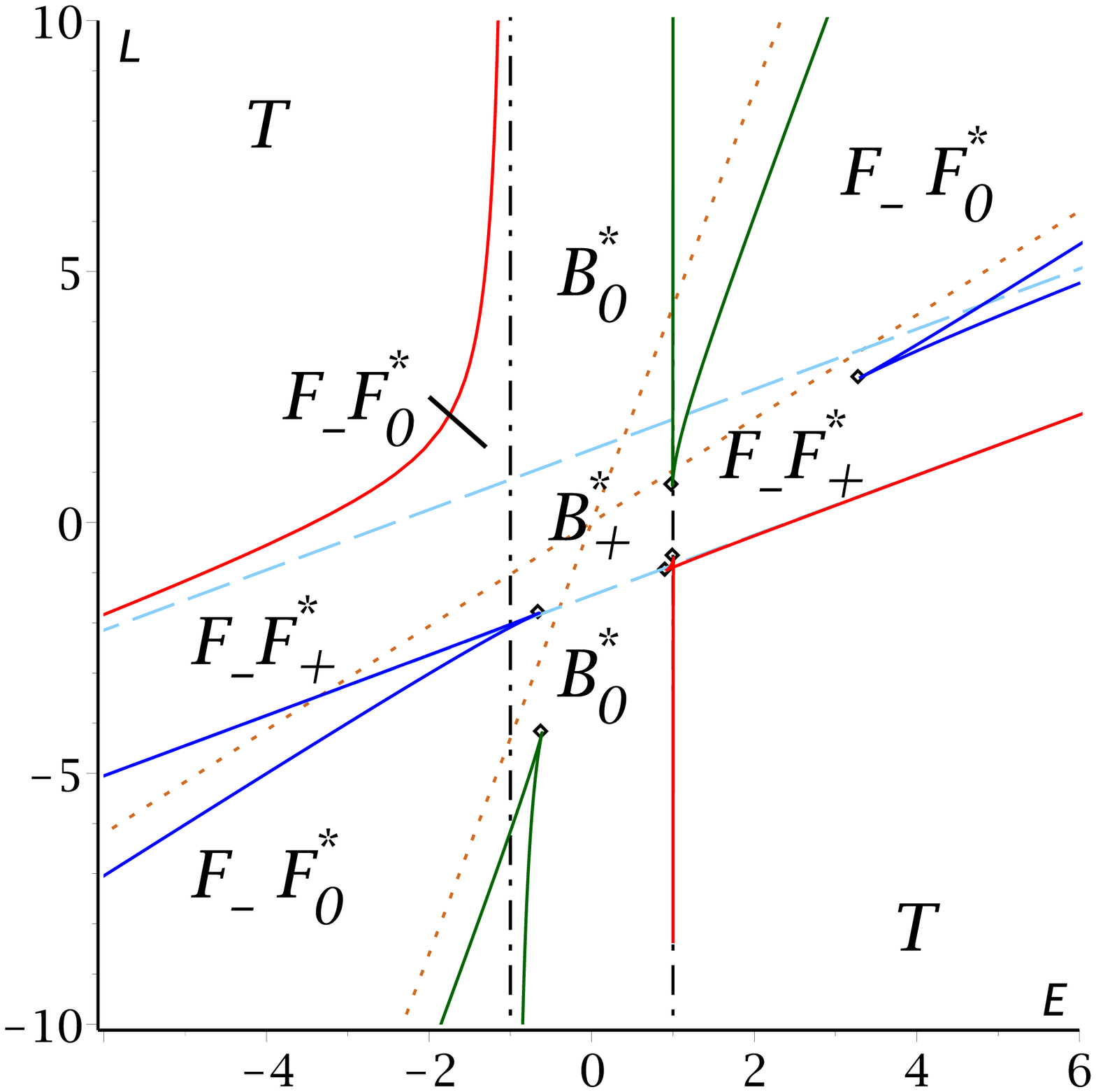}
\end{minipage}
\label{fig:r_LvsE_i_large}}
\caption{Orbit configurations for the radial motion with $\ba=0.6$, $\bK=1$, and $\mP^2=0.4$ for case \textit{(i)}. For a general description of colours and linestyles see the text. The spherical orbits marked by the dark blue (green) solid lines starting at the dots and approaching the light blue dashed (the black dash dotted) lines are stable. In (b) also the orbits on the red solid line between the two dots are stable. All other spherical orbits are unstable. Note that for the small $e\bQ$ of (a) regions of orbit configurations are only slightly deformed by the transformation $E\to-E$, $\bL\to-\bL$ but remain unchanged otherwise. Small plots on the top are enlarged details of the lower plot.}
\label{fig:r_LvsE_i}
\end{figure}

\begin{figure}
\begin{minipage}{0.5\textwidth}
\begin{center}
\includegraphics[width=0.99\textwidth]{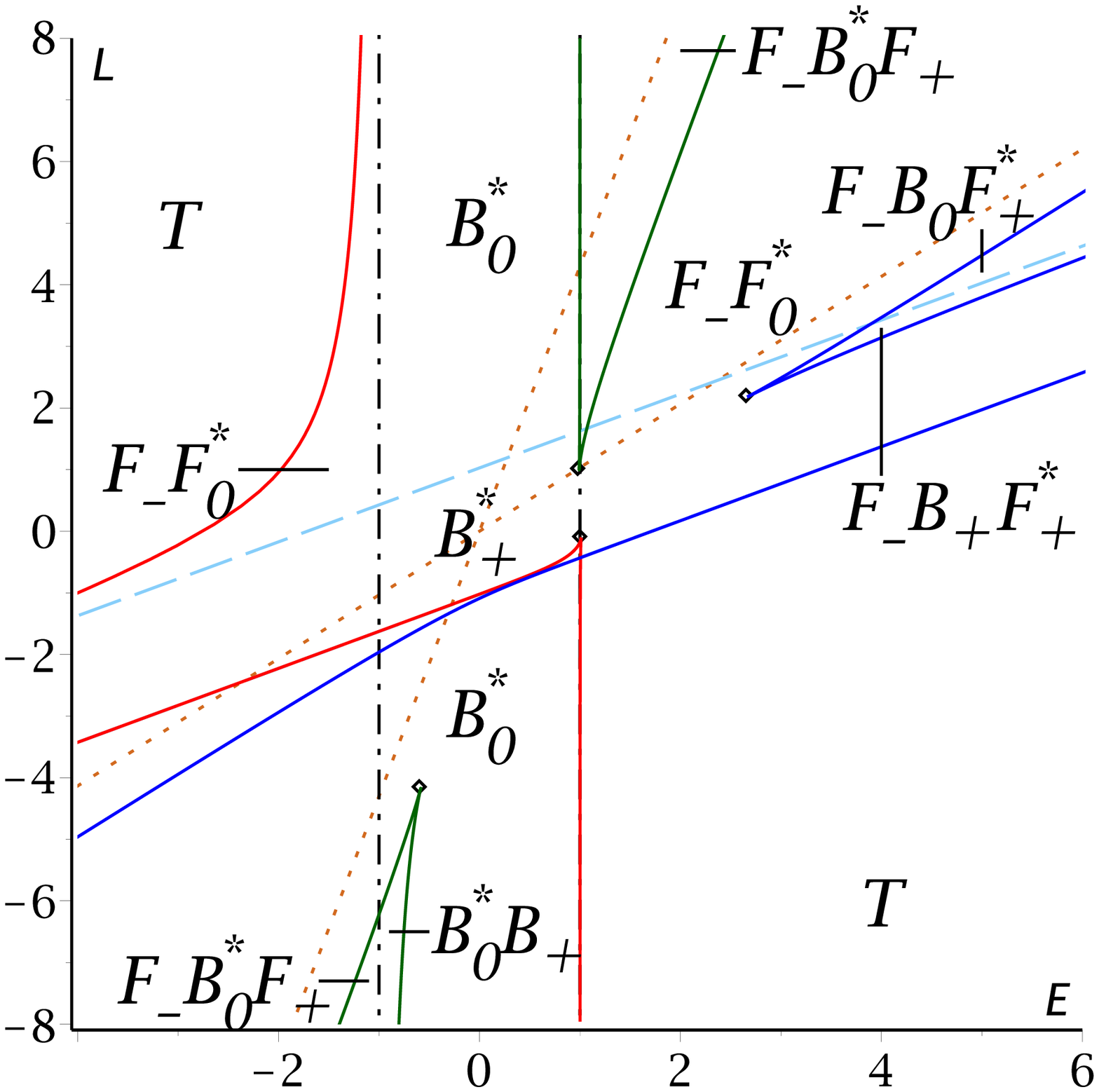}
\end{center}
\end{minipage}
\begin{minipage}{0.45\textwidth}
\includegraphics[width=0.7\textwidth]{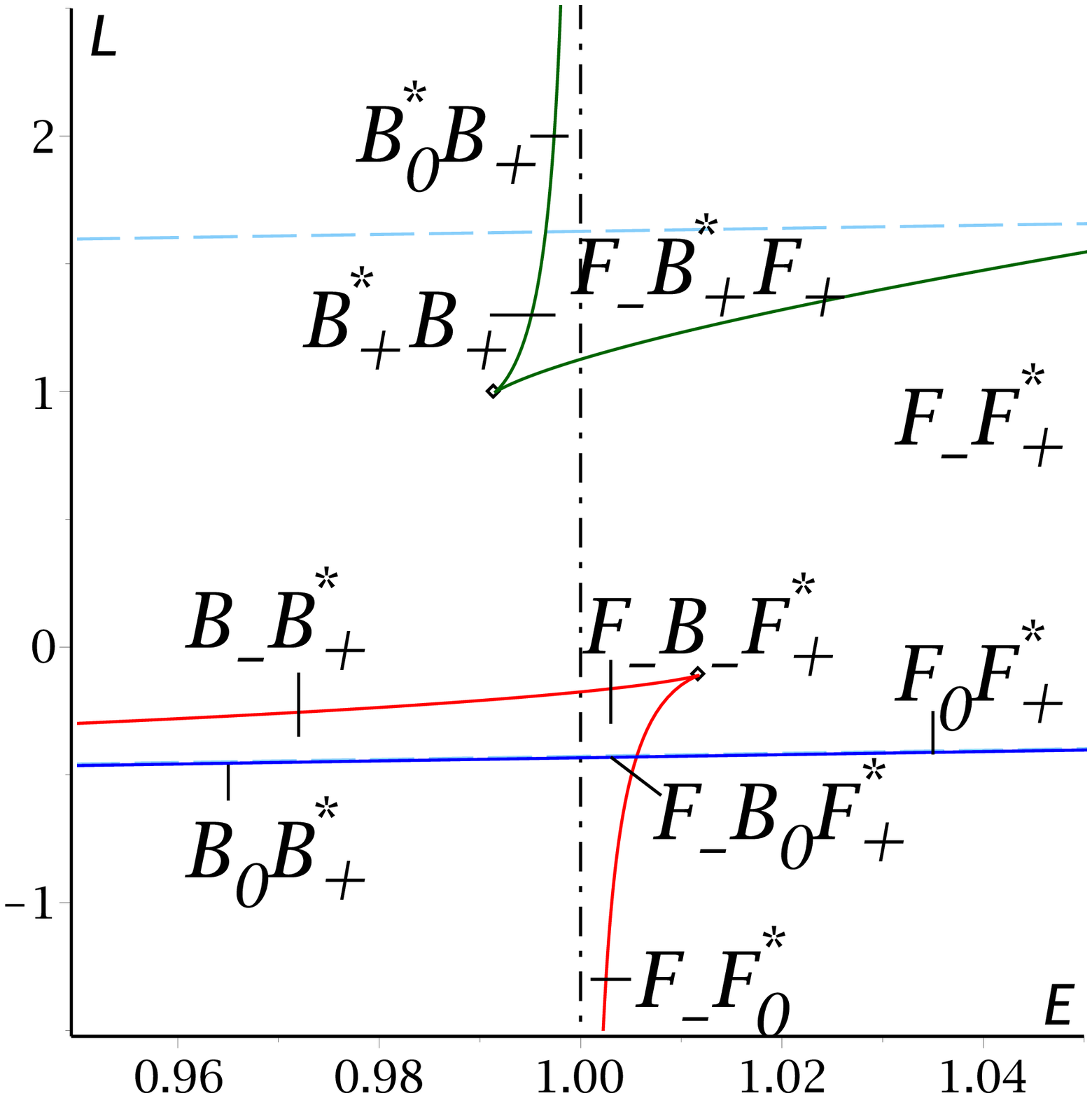}\\
\includegraphics[width=0.7\textwidth]{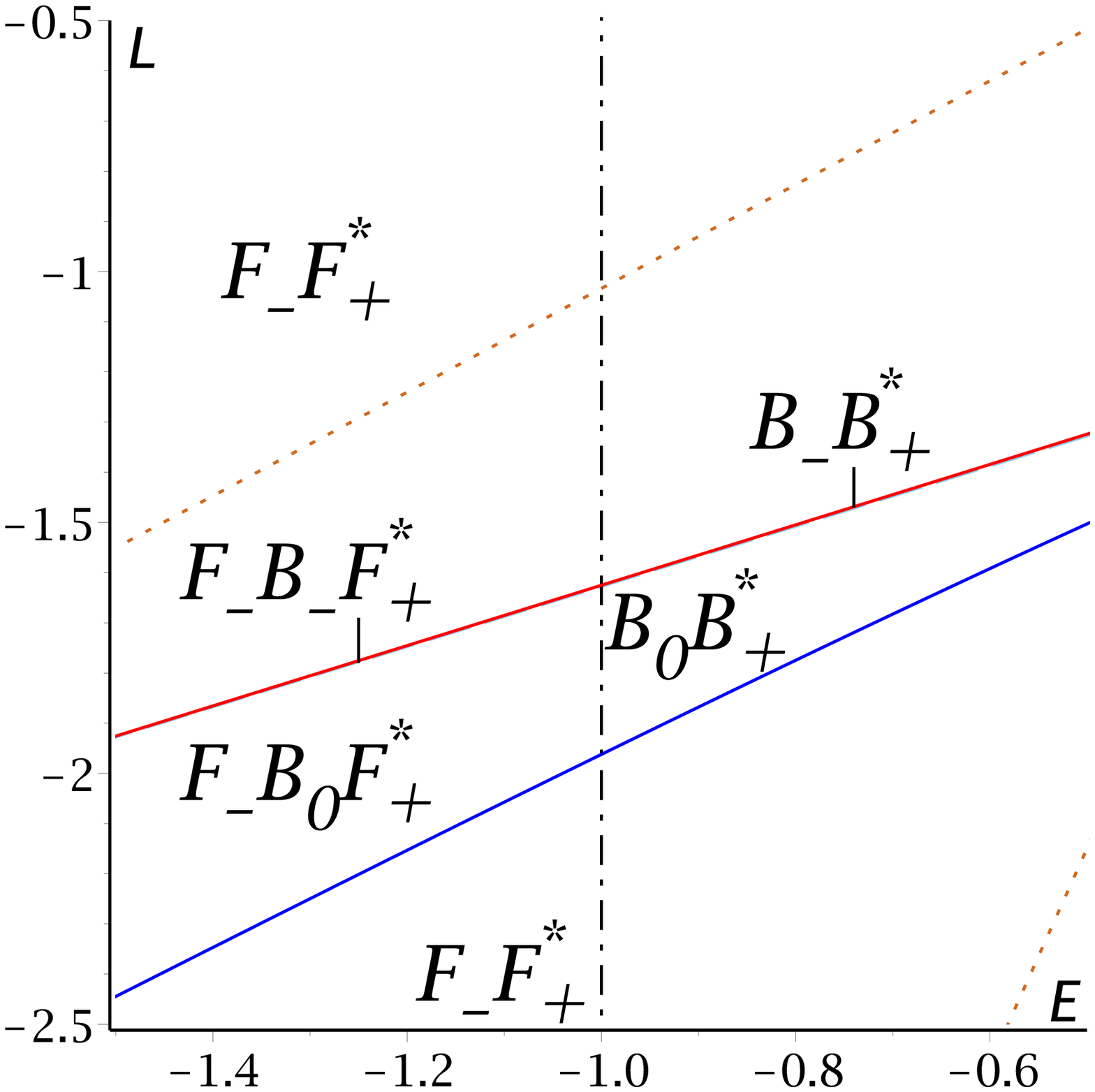}
\end{minipage}
\caption{Orbit configurations for the radial motion with $\ba=0.6$, $\bK=0.5$, $\mP^2=0.4$, and $e\bQ=0.9$, $\sqrt{\frac{\bK}{\bDelta(0)}}<e\bQ<1$. For a general description of colours and linestyles see the text. The spherical orbits on solid lines starting at a dot and approaching the black dash dotted lines or the light blue dashed lines are stable (two green lines, a dark blue line, and a red line). All other spherical orbits are unstable. In the detailed plot on the upper right the solid blue line approaches the light blue dashed line so close from below that they are hard to distinguish; the regions indicated there are meant to be between them. The same holds for the lower right plot with the red solid line approaching from above.}
\label{fig:r_LvsE_ii}
\end{figure}

\begin{figure}
\begin{minipage}{0.5\textwidth}
\begin{center}
\includegraphics[width=0.99\textwidth]{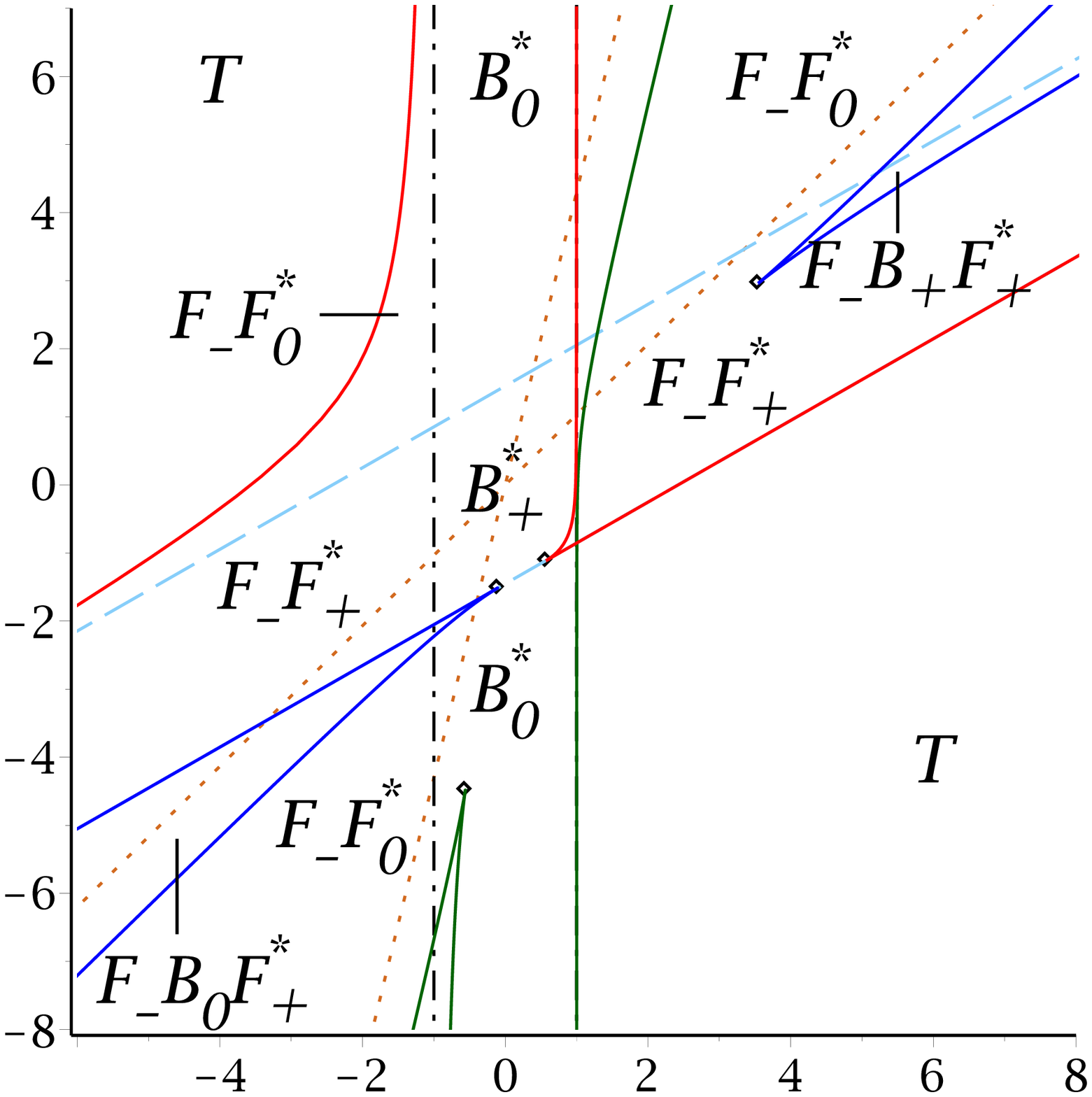}
\end{center}
\end{minipage}
\begin{minipage}{0.45\textwidth}
\includegraphics[width=0.7\textwidth]{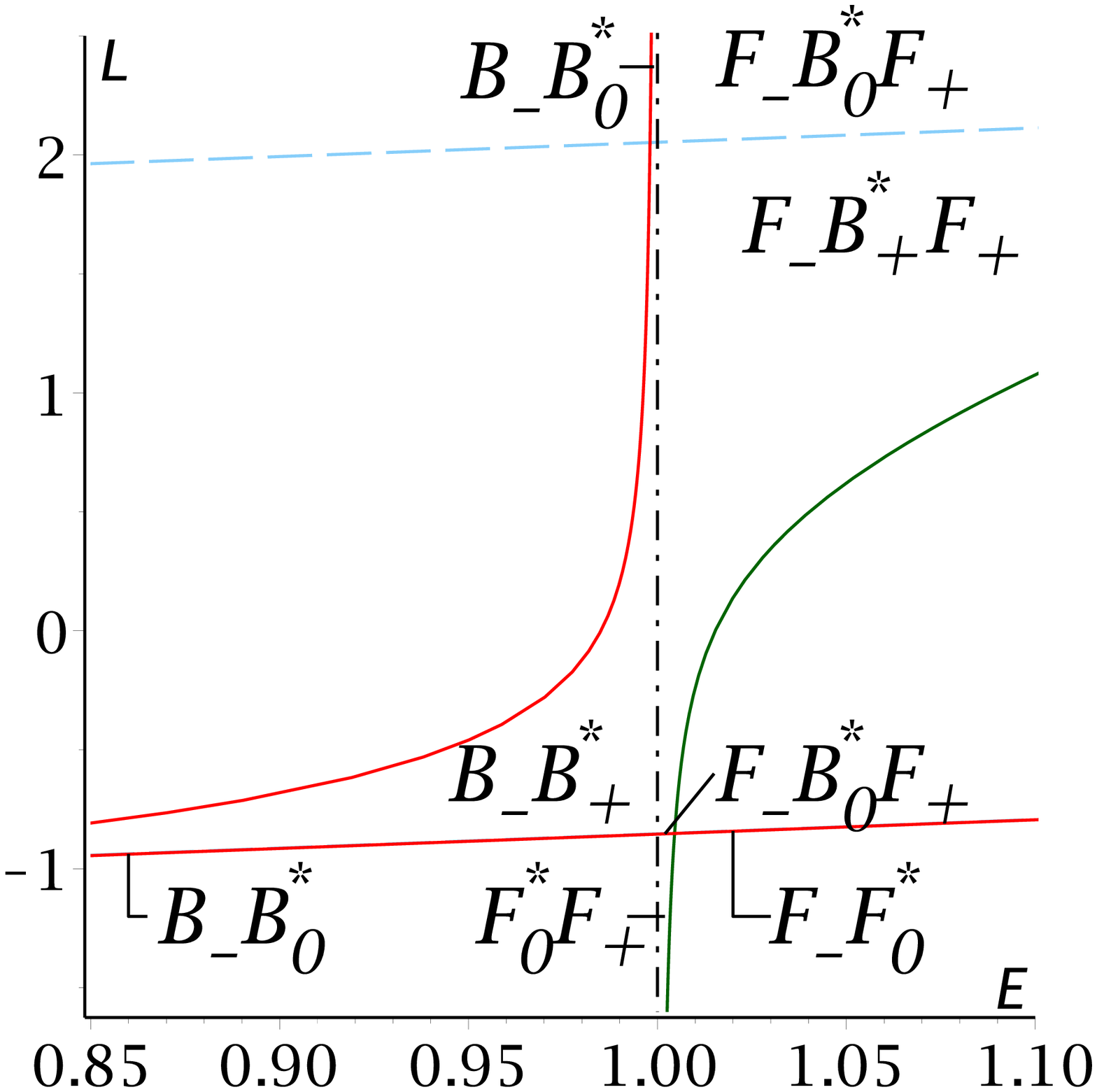}\\
\includegraphics[width=0.7\textwidth]{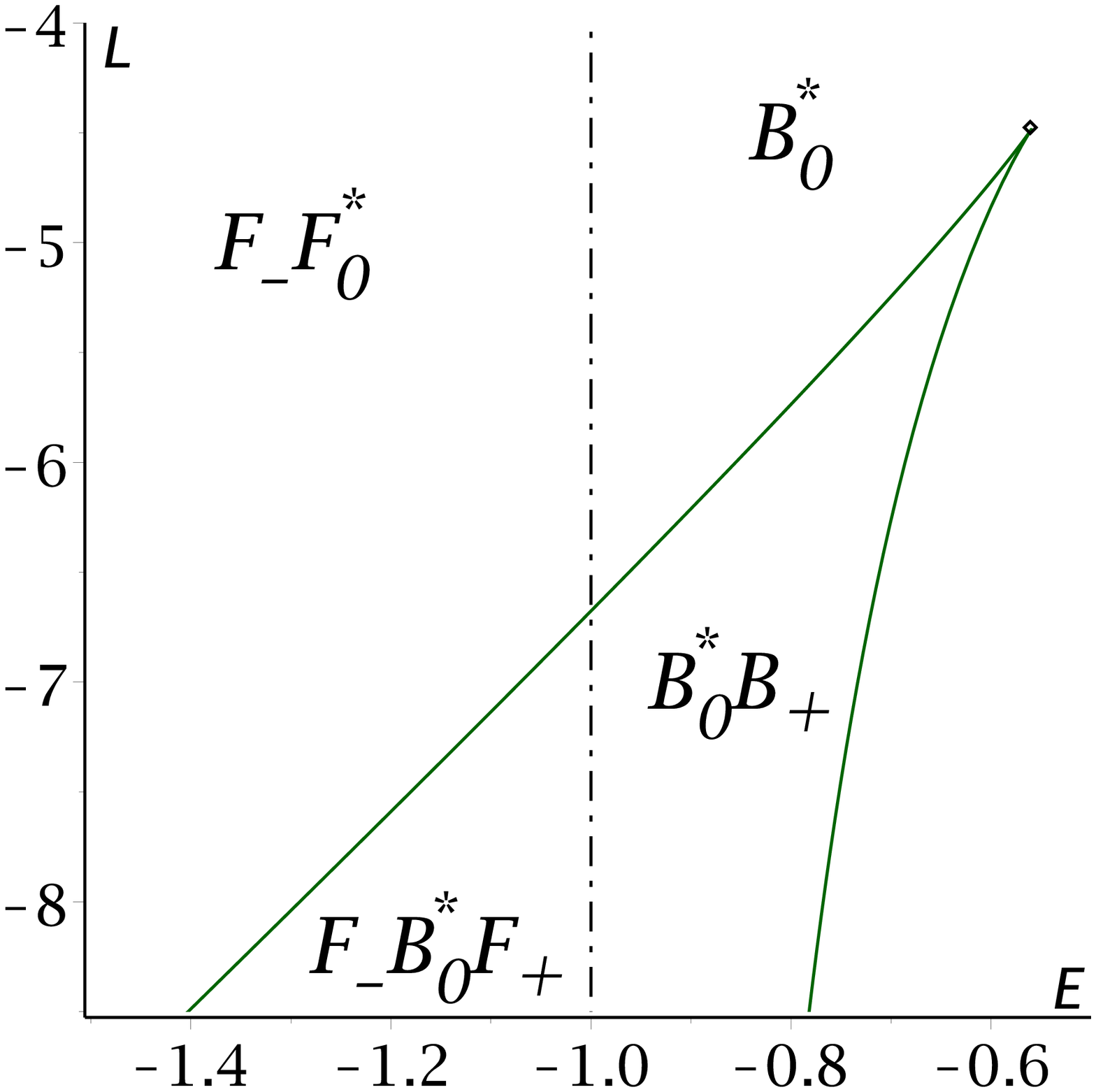}
\end{minipage}
\caption{Orbit configurations for the radial motion with $\ba=0.6$, $\mP^2=0.4$, $\bK=1$, and $e\bQ=1.1$, $1<e\bQ<\sqrt{\frac{\bK}{\bDelta(0)}}$. For a general description of colours and linestyles see the text. The spherical orbits on solid lines starting at a dot and approaching the black dash dotted lines or the light blue dashed lines are stable (two dark blue lines, a green line, and a red line). All other spherical orbits are unstable. In the detailed plot on the upper right one of the solid red lines approaches the light blue dashed line so close from below that they are hard to distinguish; the regions indicated there are meant to be between them.}
\label{fig:r_LvsE_iii}
\end{figure}

\begin{figure}
\subfloat[(a) $e\bQ=2$]{
\begin{minipage}{0.45\textwidth}
\begin{center}
\includegraphics[width=0.47\textwidth]{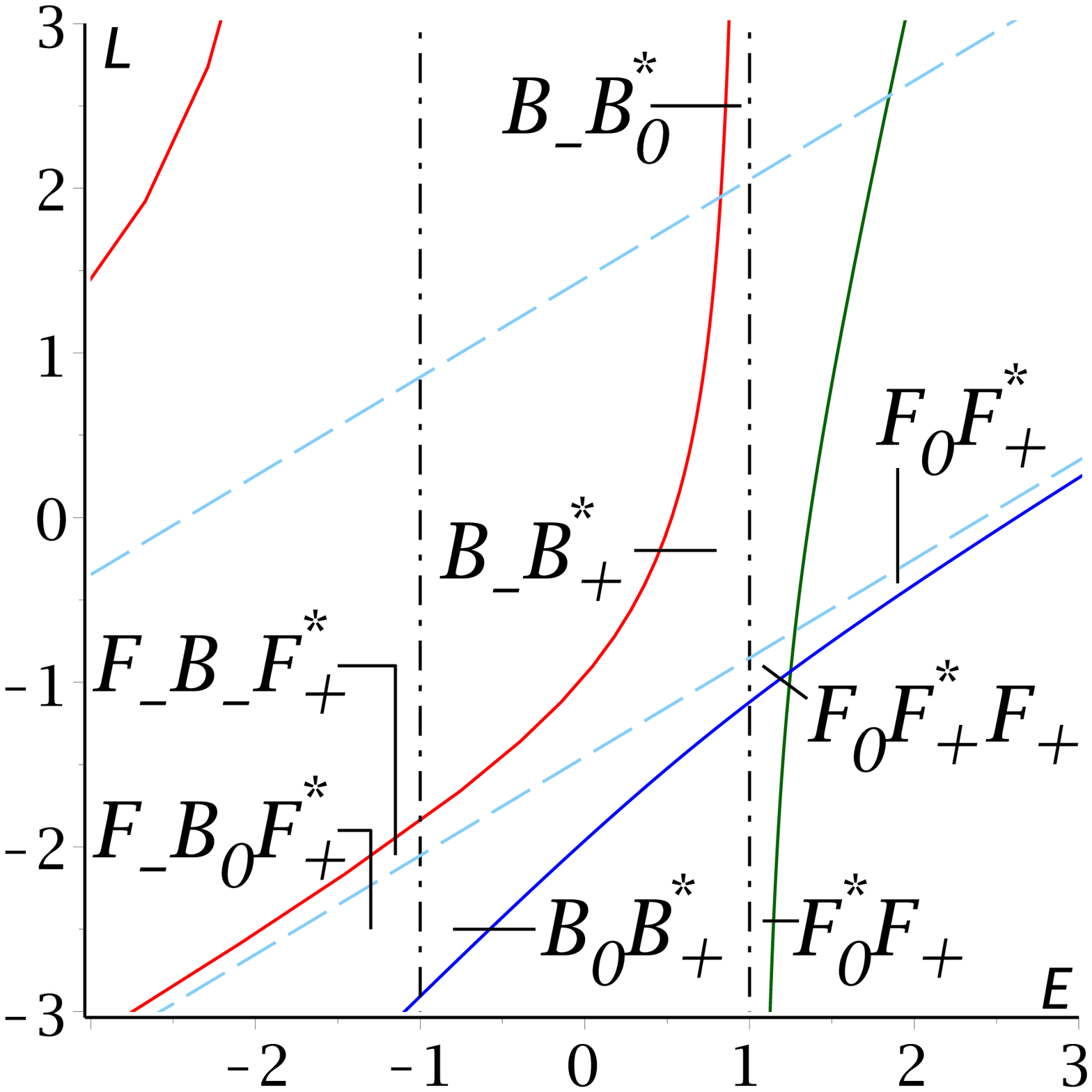}\\
\includegraphics[width=0.99\textwidth]{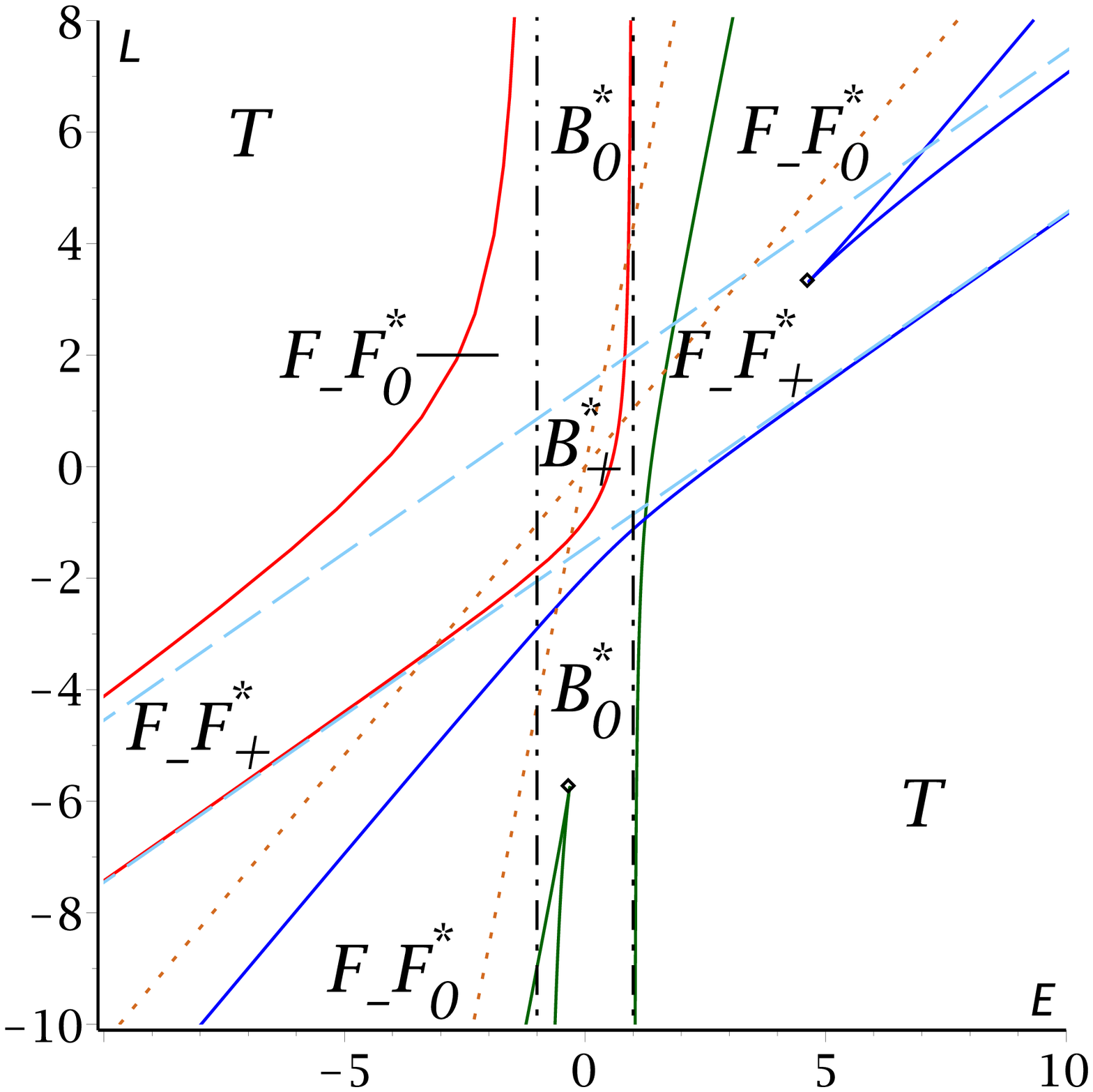}
\end{center}
\end{minipage}
}
\subfloat[(b) $e\bQ=10$]{
\begin{minipage}{0.45\textwidth}
\includegraphics[width=0.47\textwidth]{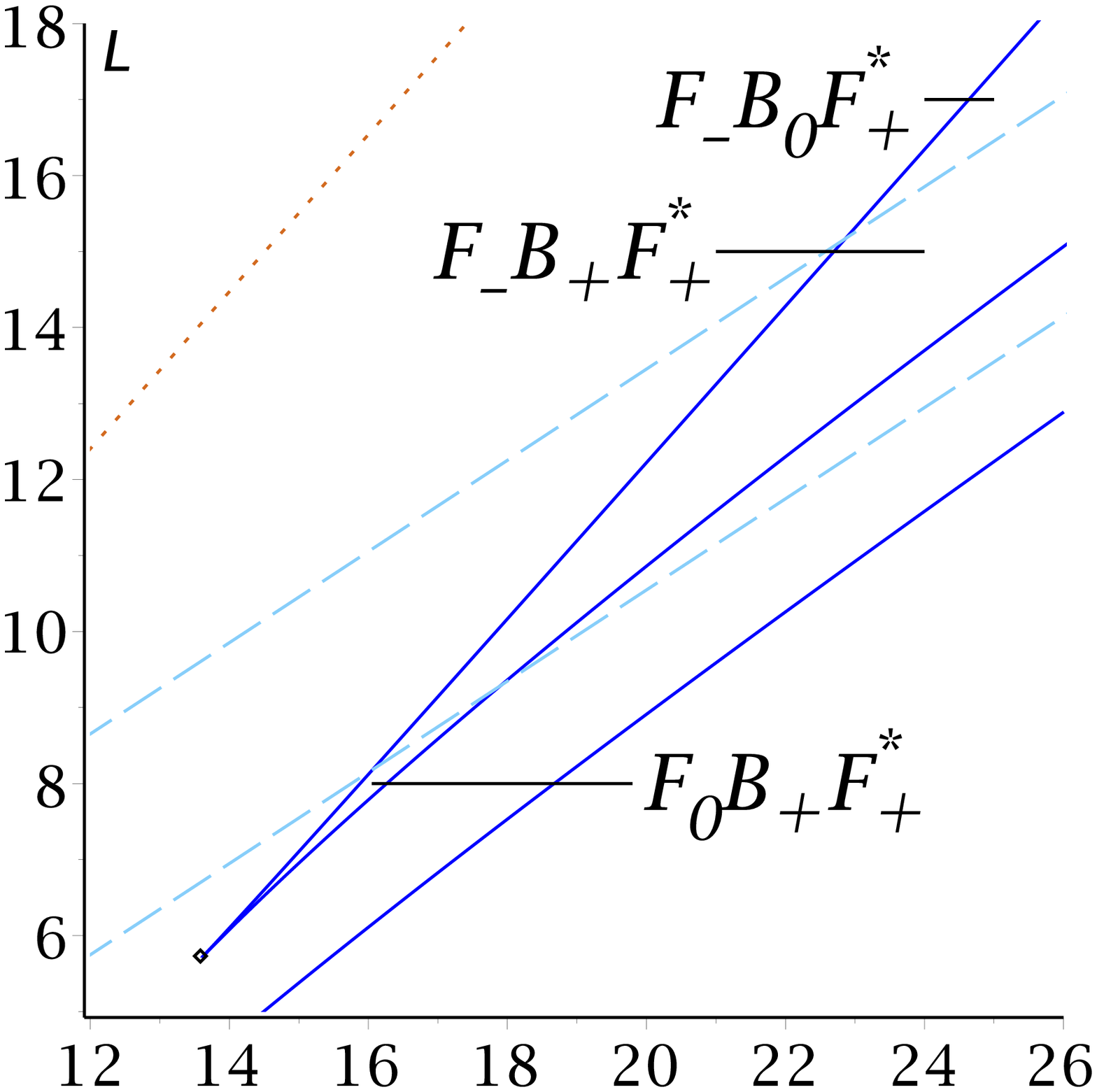}
\includegraphics[width=0.47\textwidth]{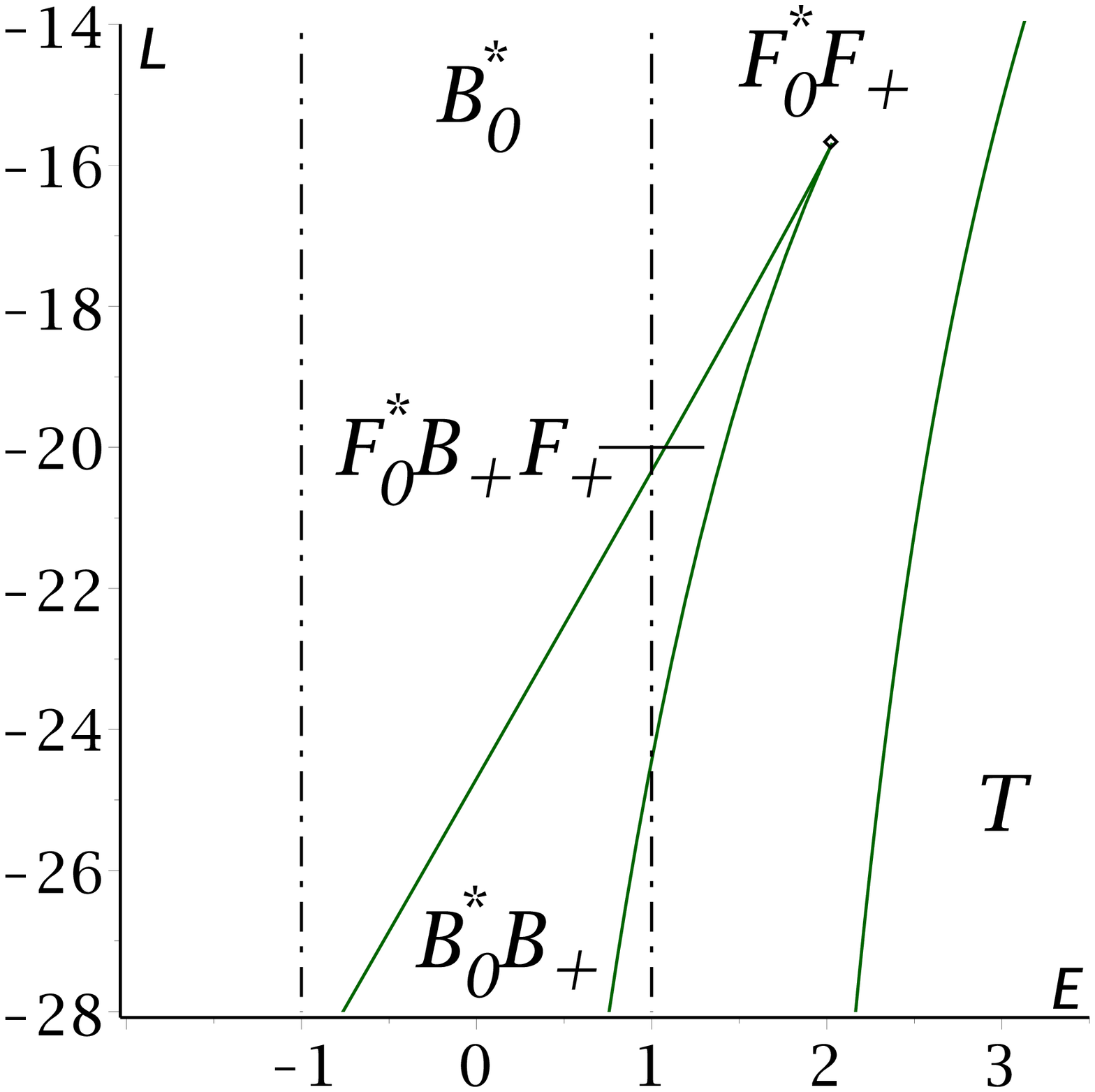}\\
\includegraphics[width=0.99\textwidth]{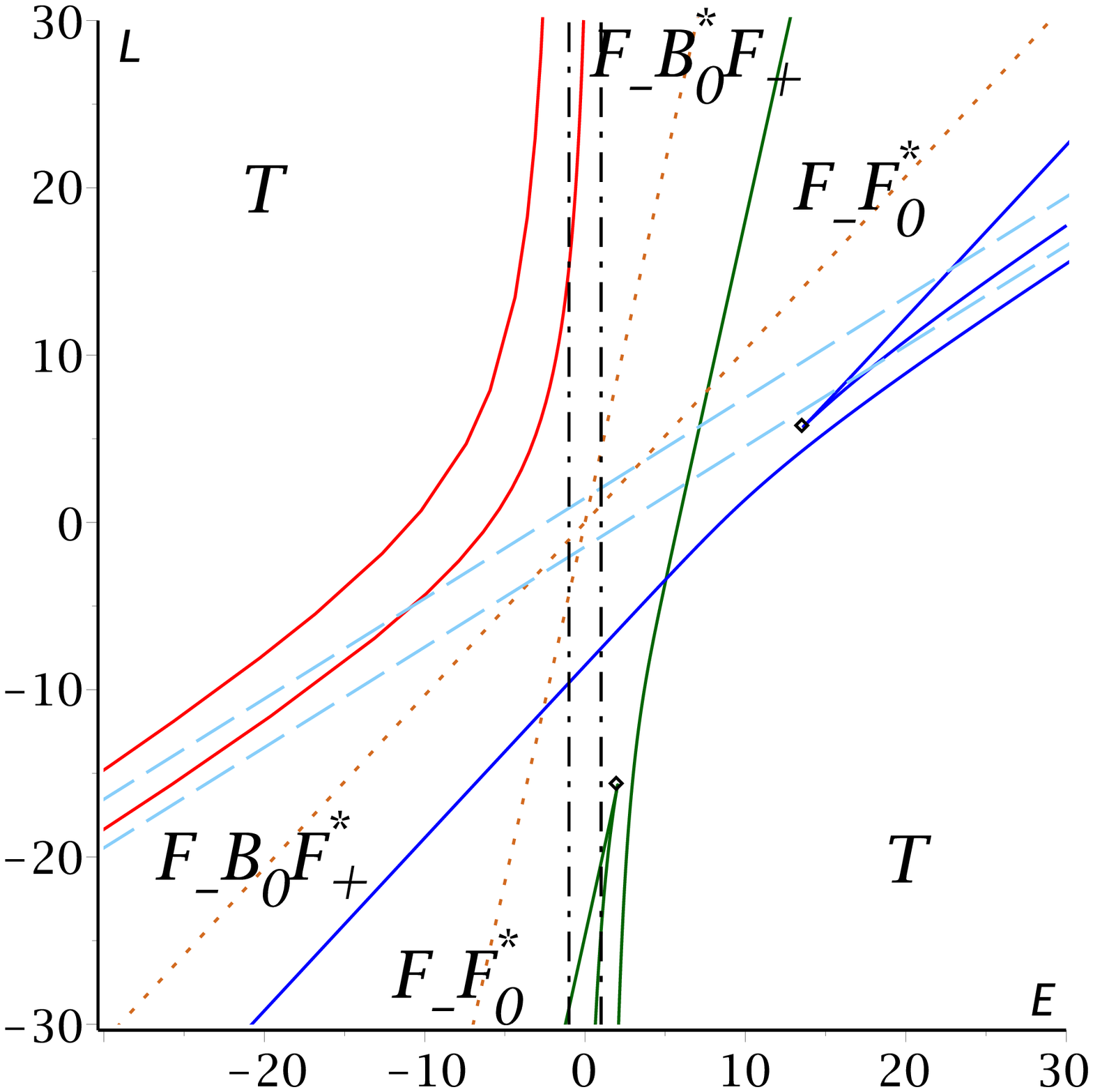}
\end{minipage}
}
\caption{Orbit configurations for the radial motion with $\ba=0.6$, $\bK=1$, $\mP^2=0.4$, and $1<\sqrt{\frac{\bK}{\bDelta(0)}}<e\bQ$. For a general description of colours and linestyles see the text. The spherical orbits on the blue (green) solid lines starting at a dot and approaching the light blue dashed (black dash dotted) lines are stable. In addition, the spherical orbits on the lower red solid line are stable but all others are unstable. Small plots on the top are enlarged details of the lower plot.}
\label{fig:r_LvsE_iv}
\end{figure}

\begin{figure}
\subfloat[$\bK=0$]{\includegraphics[width=0.3\textwidth]{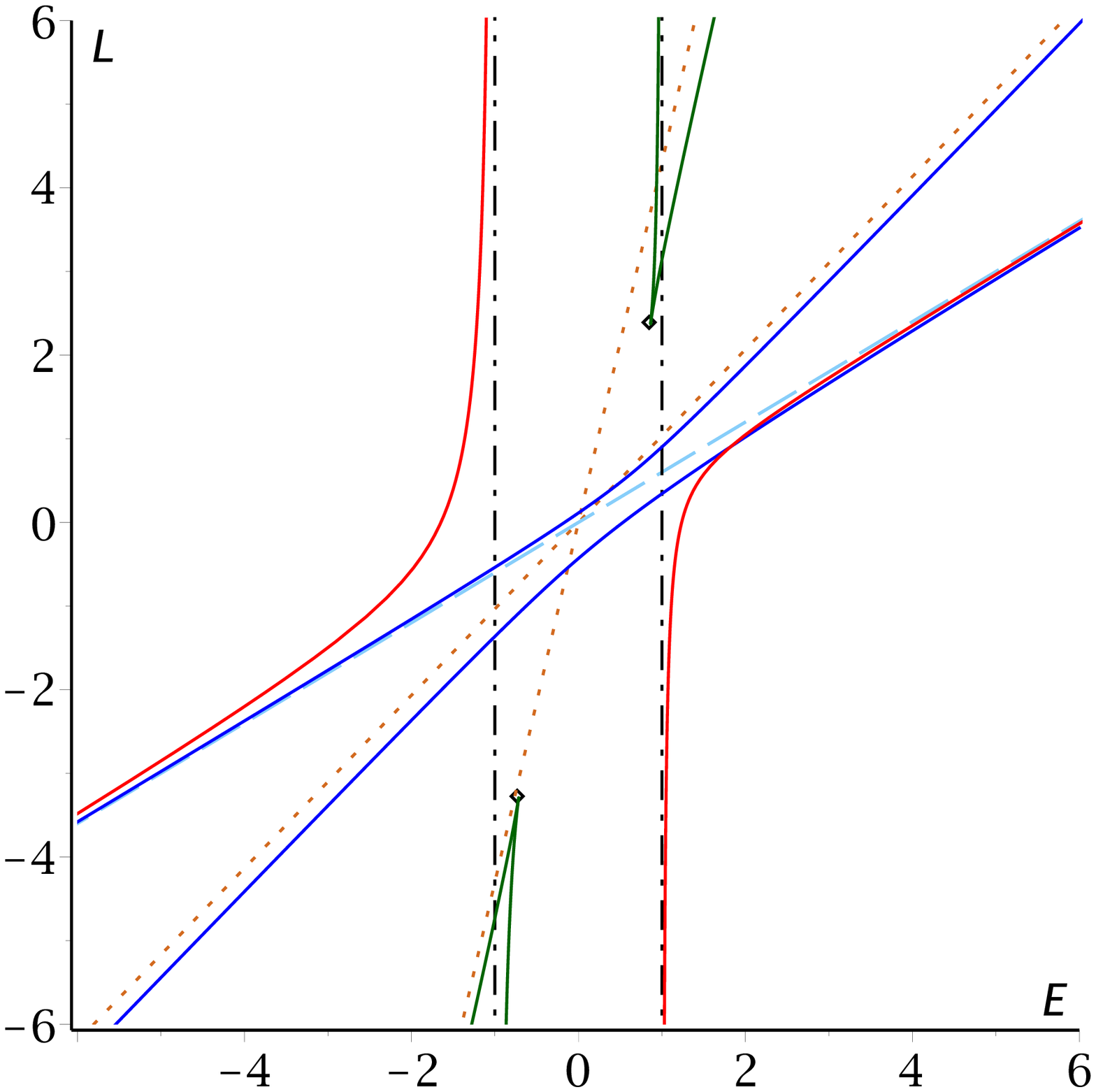}}
\subfloat[$\bK=0.02$]{\includegraphics[width=0.3\textwidth]{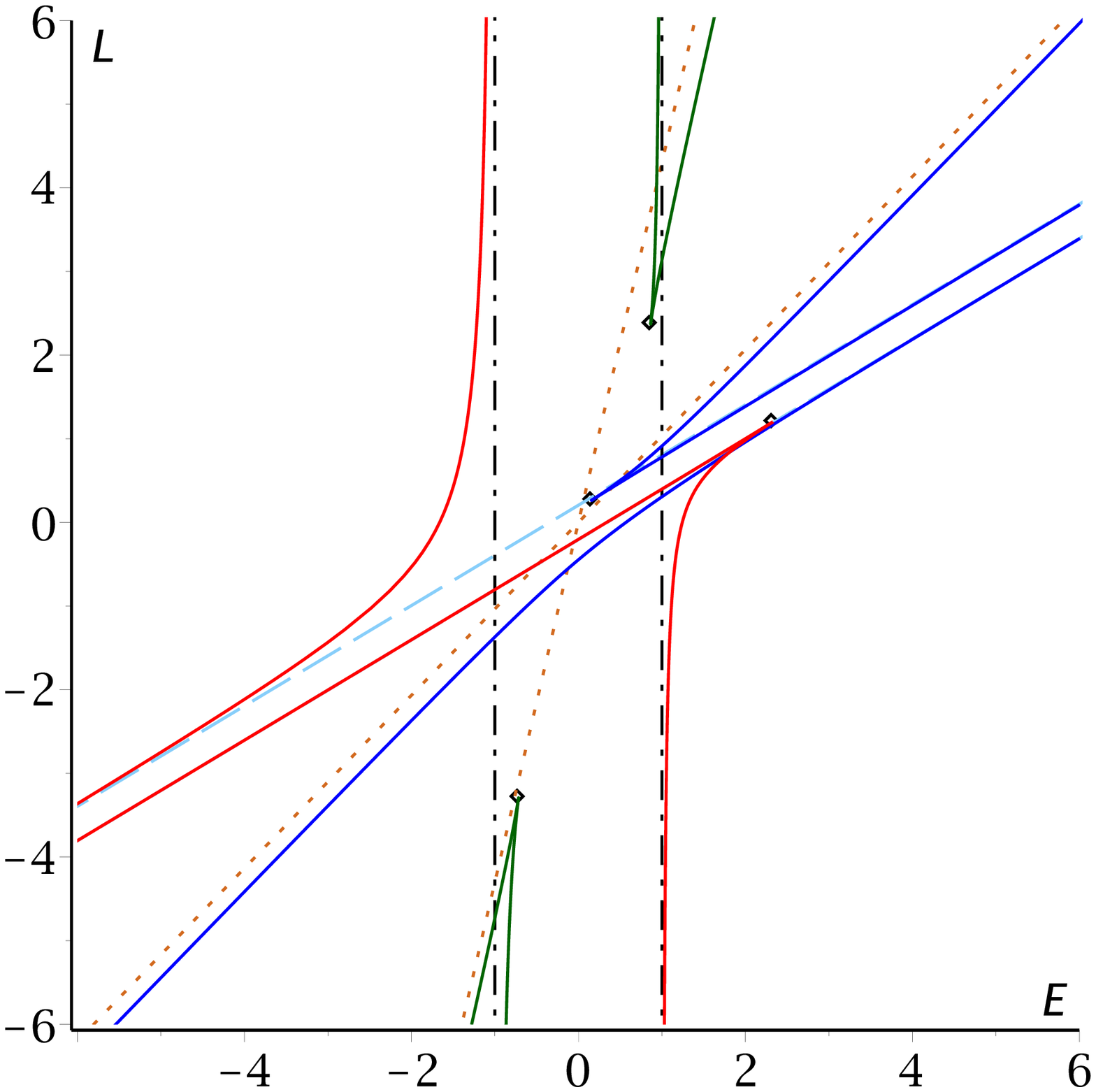}}
\subfloat[$\bK=0.0684$]{\includegraphics[width=0.3\textwidth]{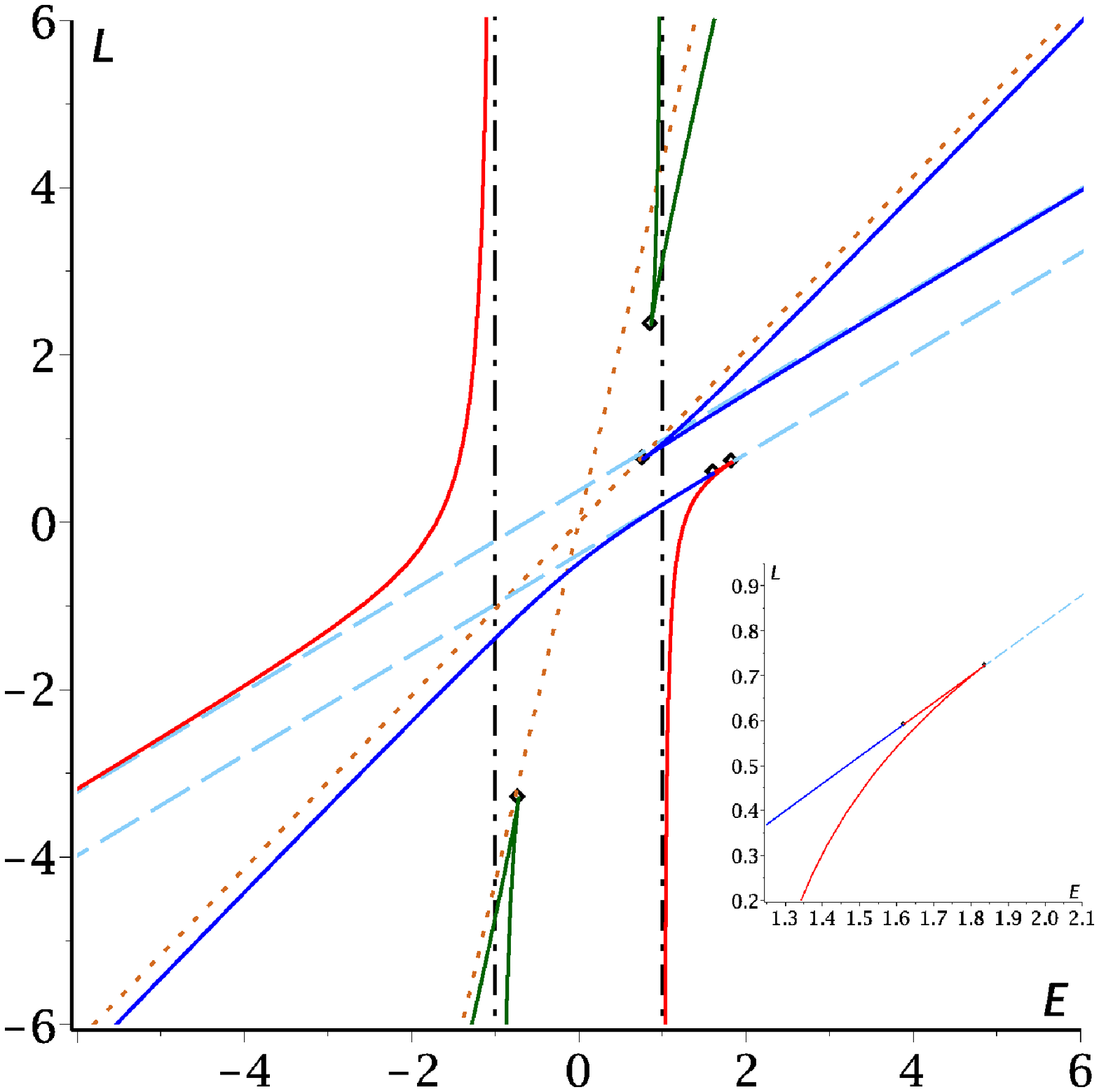}}\\
\subfloat[$\bK=0.08$]{\includegraphics[width=0.3\textwidth]{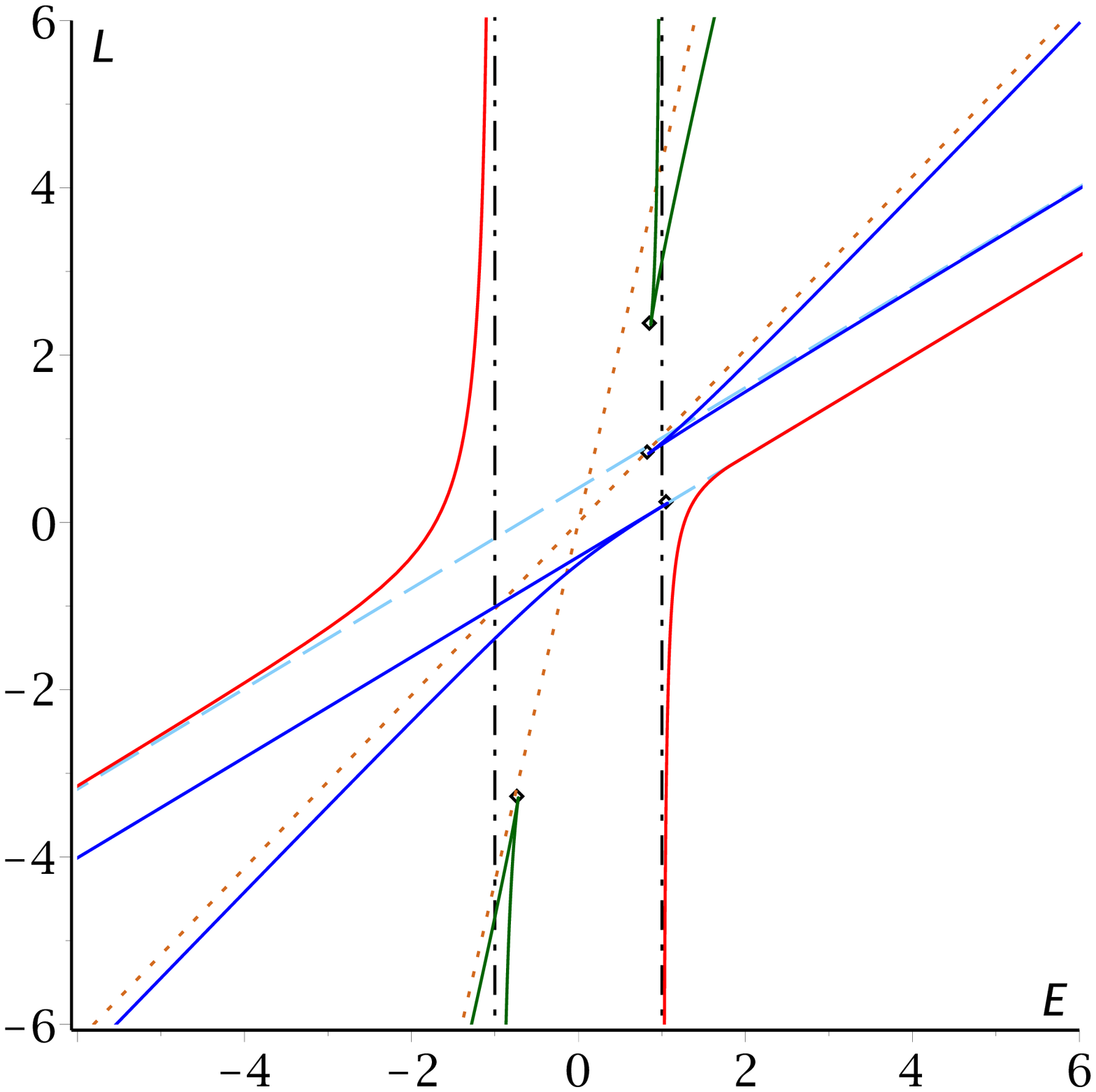}}
\subfloat[$\bK=0.6$]{\includegraphics[width=0.3\textwidth]{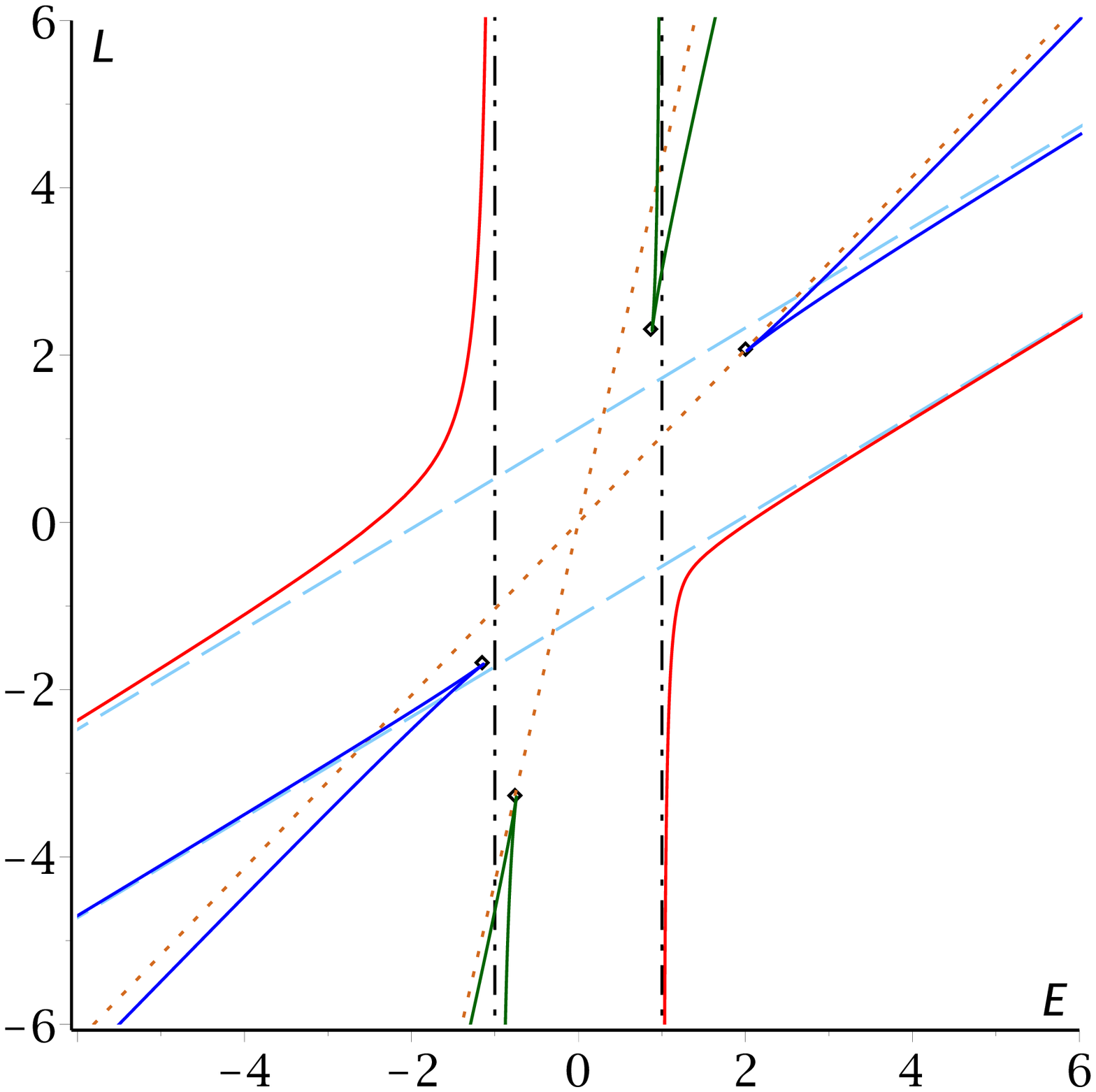}}
\subfloat[$\bK=1.0$]{\includegraphics[width=0.3\textwidth]{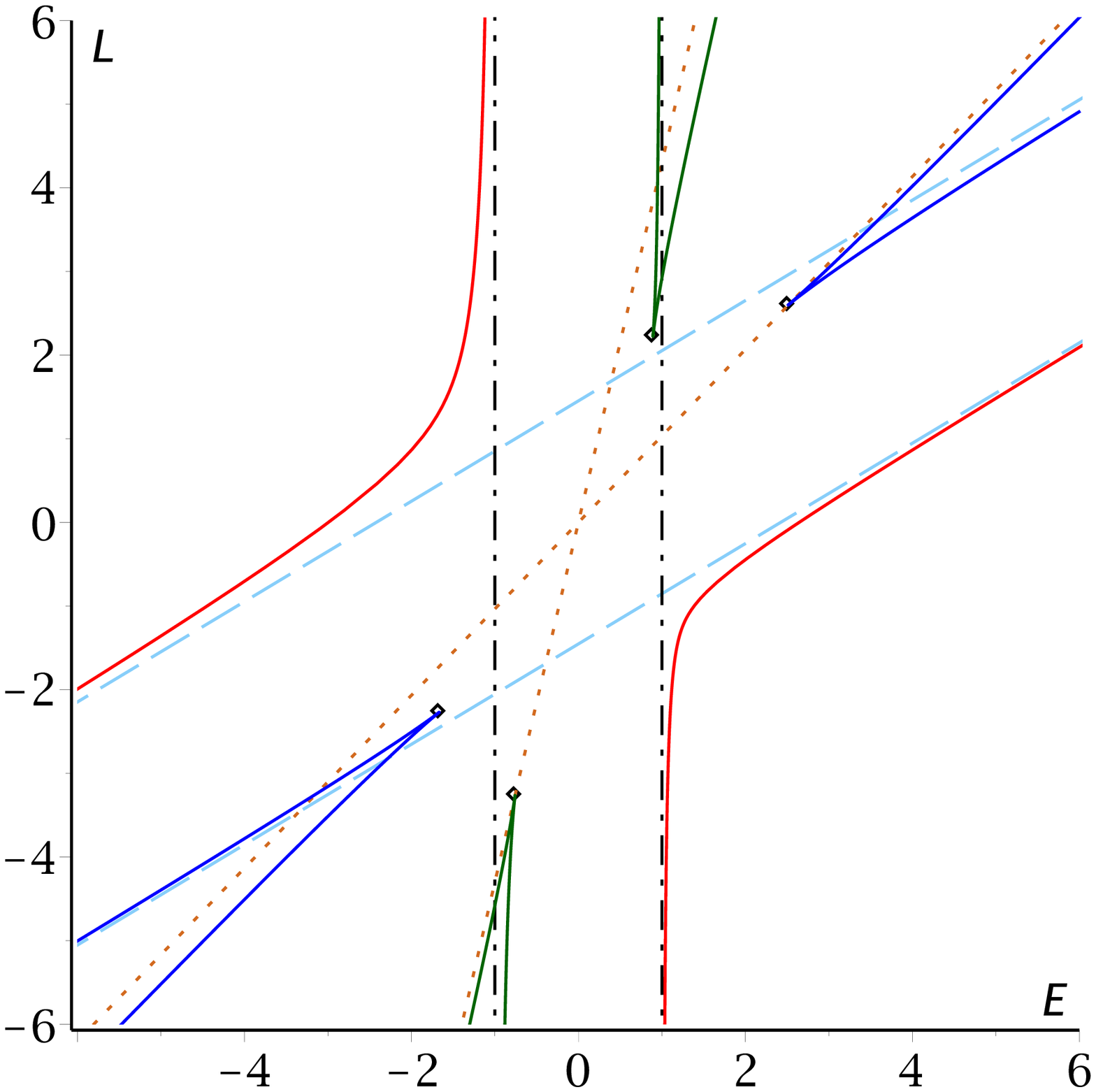}}
\caption{Orbit configurations for the radial motion with $\ba=0.6$, $\mP^2=0.4$, $e\bQ=0.3$ and varying $\bK$. For a general description of colours and linestyles see the text. Note that at $\bK = 0.0684$ the plots change from case \textit{(ii)} to \textit{(i)}.}
\label{fig:r_LvsE_series}
\end{figure}

Now let us turn back to the general case. The expressions \eqref{R_double_E} and \eqref{R_double_L} depend linearly on $\mQ=eQ$ and the parameter $\ba$ can be removed by a rescaling of the other parameters ($\frac{\bL}{\ba}$, $\frac{\mQ}{\ba}$, $\frac{\mP}{\ba^2}$, $\frac{\bK}{\ba^2}$) and the radial coordinate ($\frac{\br}{\ba}$). (Note that also distances measured in units of $M$ have to be rescaled). The dependence on $\bK$ and $\mP^2=\bQ^2+\bP^2$ is less obvious. This can be studied by solving \eqref{R_double_E} and \eqref{R_double_L} for these parameters but the expressions are quite cumbersome and we do not give them here. However, in Figure \ref{fig:r_LvsE_series} orbit configurations for varying $\bK$ are shown and all possible orbit types already appear in Figures \ref{fig:r_LvsE_i} to \ref{fig:r_LvsE_iv}.

Let us now analyze where triple zeros occur as they mark transitions from stable to unstable orbits. Solving the three conditions $\frac{d^2R}{d\br^2}(\br_0)=\frac{dR}{d\br}(\br_0)=R(\br_0)=0$ for $E$, $\bL$, and $\mQ=e\bQ$ yields
\begin{align}
E_{1,2} & = \pm \frac{1}{2} \frac{\bK\bDelta^2(\br_0)+(\br_0^2+\bK)(3\br_0^2-2\br_0+\bK)\bDelta(\br_0)-(\br_0^2+\bK)^2(\br_0-1)^2}{(\br_0^2+\bK)\bDelta(\br_0)\sqrt{(\br_0^2+\bK)\bDelta(\br_0)}}\,, \label{r_trip_E}\\
L_{1,2} & = \mp \frac{1}{2} \frac{\sqrt{(\br_0^2+\bK)\bDelta(\br_0)}}{(\br_0^2+\bK)^2\bDelta^2(\br_0)\ba} \, \big[ \bK(3\br_0^2-\ba^2+2\bK)\bDelta^2(\br_0) + (\br_0^2+\bK)(\br_0^4-\bK(\br_0^2-2\br_0+\ba^2)-\ba^2(3\br_0^2-2\br_0))\bDelta(\br_0) \nonumber \\
& \qquad   + (\br_0^2+\bK)^2(\ba^2-\br_0^2)(\br_0-1)^2\big]\,, \label{r_trip_L}\\
\mQ_{1,2} & = \mp \frac{\br_0^3\bDelta^2(\br_0) -(\br_0^2+\bK)(2\br_0^3-\br_0^2+\bK)\bDelta(\br_0) +(\br_0^2+\bK)^2\br_0(\br_0-1)^2}{(\br_0^2+\bK)\bDelta(\br_0)\sqrt{(\br_0^2+\bK)\bDelta(\br_0)}}\,.\label{TripleQ}
\end{align}
Here again, the horizons $\br_{\pm}$ are singularities and in between triple zeros are not possible. Fourfold zeros may only occur for $\br<0$, given by $E_{1,2}=\pm\frac{2\br-1}{2\sqrt{\br(\br-1)}}$, $\bL_{1,2}=\pm\frac{\br^2-2\br\mP^2-\ba^2}{2\ba\sqrt{\br(\br-1)}}$, $\mQ_{1,2}=\pm\frac{\br}{\sqrt{\br(\br-1)}}$, and $\bK=\frac{\br(\bDelta(0)-\br)}{\br-1}$, or at the ring singularity if $E=\pm\frac{1}{\sqrt{\bDelta(0)}}$, $\bK=0$, $\mQ=\frac{1}{E}$, and $\bL=\ba E$.

The regions of different orbit configurations are visualized in Figures \ref{fig:r_LvsE_i}-\ref{fig:r_LvsE_iv}. As the parameter space is six dimensional, we have to fix at least three parameters for plotting. We always choose to fix $\ba$ and $\mQ=e\bQ$ for the reasons outlined above. Here $\mQ=\pm1$ and $\mQ=\pm\sqrt{\bK/\bDelta(0)}$ will separate quite different plot structures: at $\mQ=\pm1$ the behaviour of $\bL_{1,2}$ at infinity changes and at $\mQ=\pm\sqrt{\bK/\bDelta(0)}$ the behaviour of $E$ at $\br=0$. We will restrict here to positive values of $\mQ$ but allow all values of $\bL$ and $E$. The values $e\bQ<0$ are recovered by $(E,\bL)\to(-E,-\bL)$ as noted in the section on general properties. Therefore, we distinguish between four different regions: \textit{(i)} $\mQ < \min\{1,\sqrt{\frac{\bK}{\bDelta(0)}}\}$, \textit{(ii)} $\sqrt{\frac{\bK}{\bDelta(0)}} <\mQ<1$, \textit{(iii)} $1<\mQ<\sqrt{\frac{\bK}{\bDelta(0)}}$, and \textit{(iv)} $\max\{1,\sqrt{\frac{\bK}{\bDelta(0)}}\}<\mQ$. Note that for $e\bQ=0$ the polynomial $R$ is unchanged by the transformation $(E,\bL)\to(-E,-\bL)$ and, therefore, regions of orbit configurations differ only slightly from this symmetry for small $\mQ$ in region \textit{(i)}. However, if $\mQ$ is larger than \eqref{TripleQ} with $\br_0=\frac{1}{2}(\bDelta(0)-\bK - \sqrt{(\bDelta(0)-\bK)^2+4\bK})$ (local minimum) the structure changes as an additional pair of triple zeros appears, see Figure \ref{fig:r_LvsE_i_large}.

For better comparison, we always plot $\bL$ over $E$ and indicate the dependence on $\bK$ and $\mP=\bP^2+\bQ^2$ by slowly varying them in a plot series. In each plot, we use the following conventions:
\begin{itemize}
\item Solid lines indicate double zeros of $R$ which correspond to stable or unstable spherical orbits of constant $\br$. We use red lines for orbits with constant $\br<0$, blue for constant $0<\br<\br_-$, and green for constant $\br>\br_+$.
\item Dashed lines denote orbits with turning points at $\br=0$ and are given by $\bL_{\pm}=\ba E \pm \frac{\sqrt{\bK\bDelta(0)}}{\ba}$. They mark transitions between orbits with different indices. Red or blue solid lines corresponding to orbits with constant $\br$ near $\br=0$ asymptotically approach $\bL_{\pm}$.
\item The dash dotted line marks $E^2=1$. Between $E=\pm1$ no orbit can reach infinity. In addition, $E=\pm1$ is asymptotically approached by red or green solid lines with $\br\to \pm \infty$
\item Dotted lines are asymptotes to solid lines for $\br \to \br_{\pm}$. They do NOT separate different orbit configurations.
\item Single dots mark triple zeros which separate stable from unstable orbits.
\item The labels $T$, $F$, and $B$ indicated the orbit configurations summarized in table \ref{tab:radial}. Solid, dashed, and dash dotted lines indicate transitions from one orbit configuration to another.
\end{itemize}

\begin{figure}
\subfloat[$\mP^2=0.1$]{\includegraphics[width=0.3\textwidth]{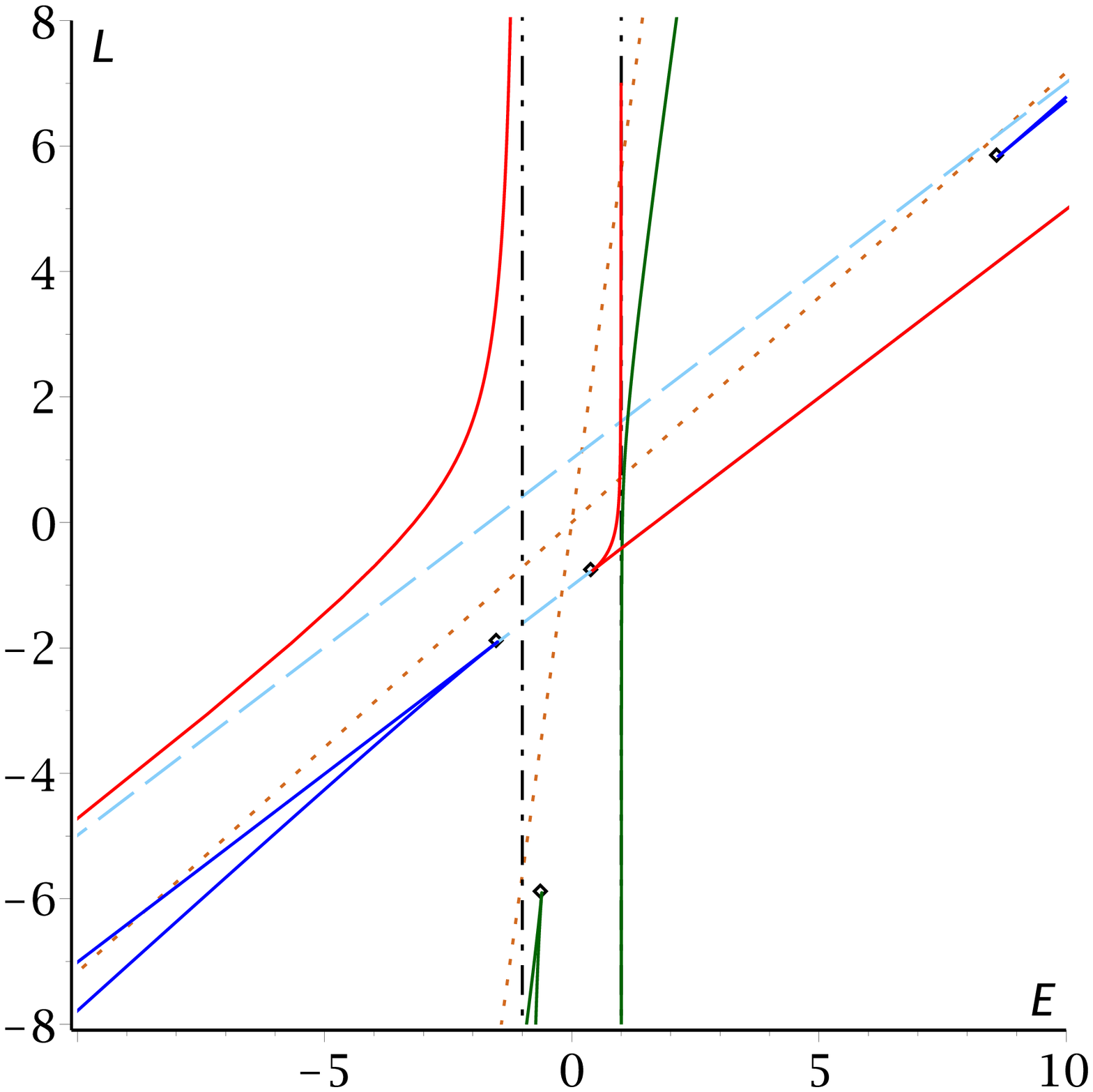}}
\subfloat[$\mP^2=0.195$]{\includegraphics[width=0.3\textwidth]{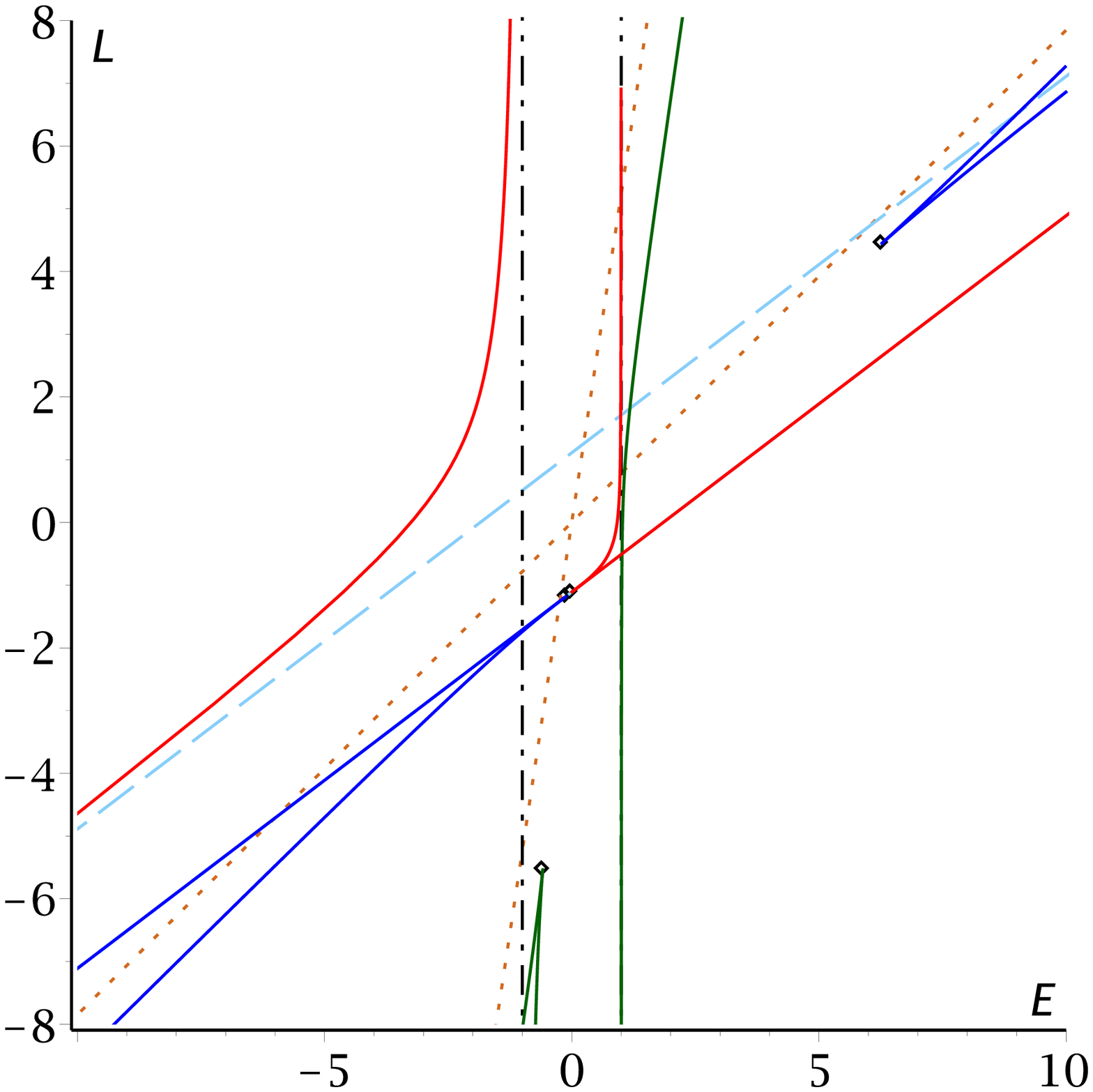}}
\subfloat[$\mP^2=0.2$]{\includegraphics[width=0.3\textwidth]{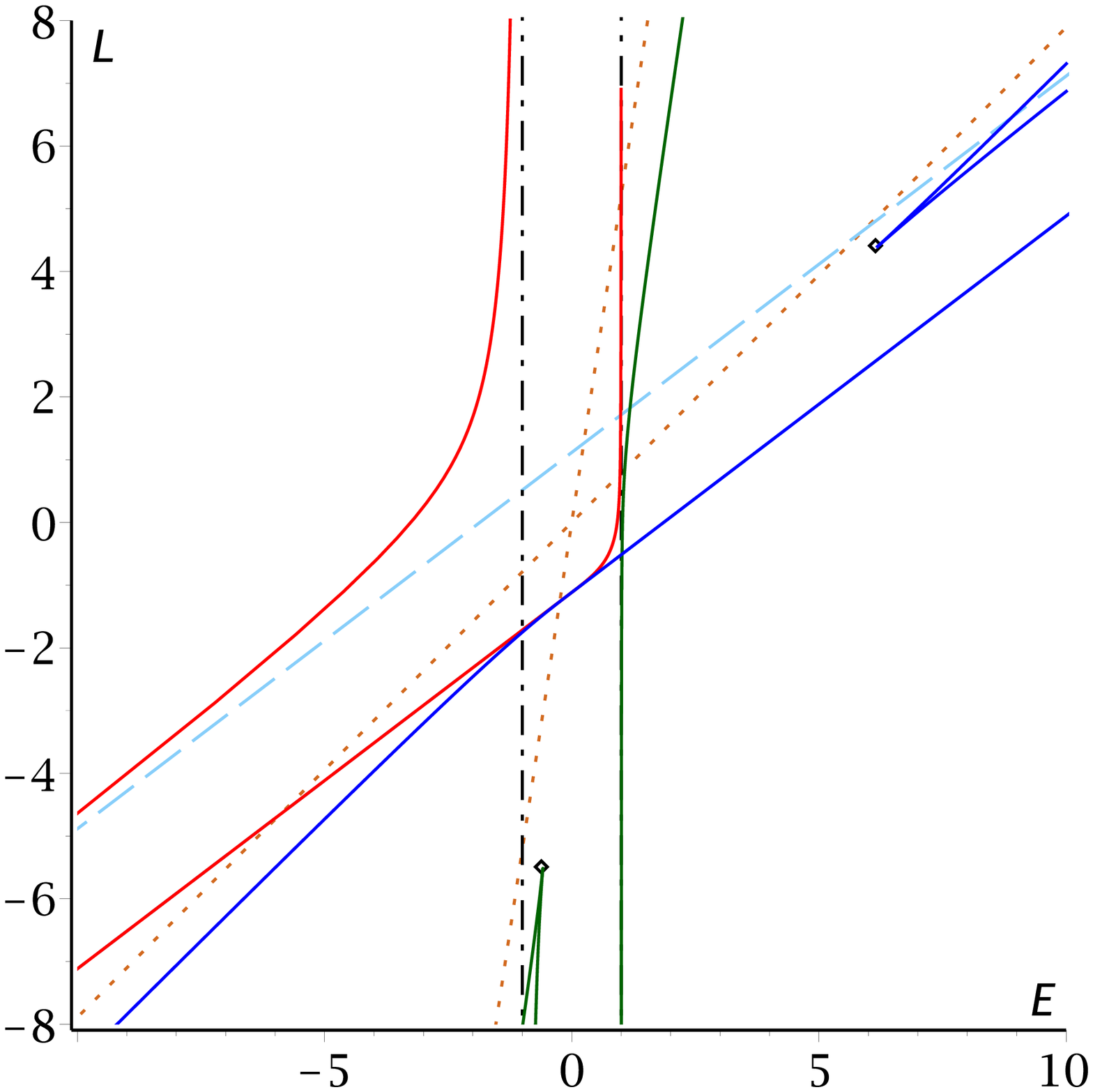}}
\caption{Orbit configurations for the radial motion with $\ba=0.6$, $\bK=0.8$, $e\bQ=1.2$ and varying $\mP^2$. For a general description of colours and linestyles see the text. Note that at $\mP^2 \approx 0.1956$ the plots change from case \textit{(iii)} to \textit{(iv)}.}
\label{fig:r_LvsE_series_2}
\end{figure}

\begin{figure}
\subfloat[$\mP^2=0.5$]{\includegraphics[width=0.3\textwidth]{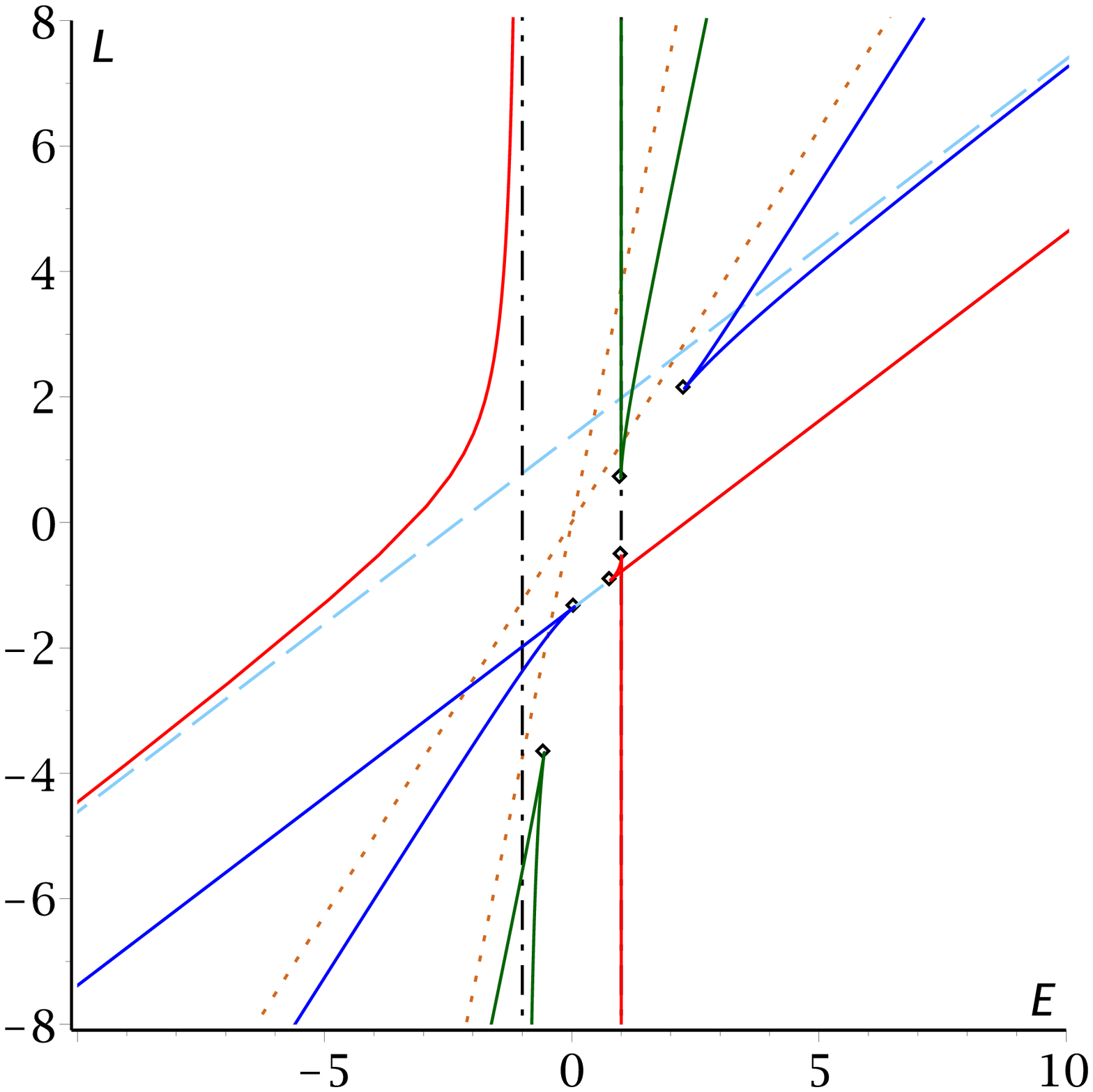}}
\subfloat[$\mP^2=0.627$]{\includegraphics[width=0.3\textwidth]{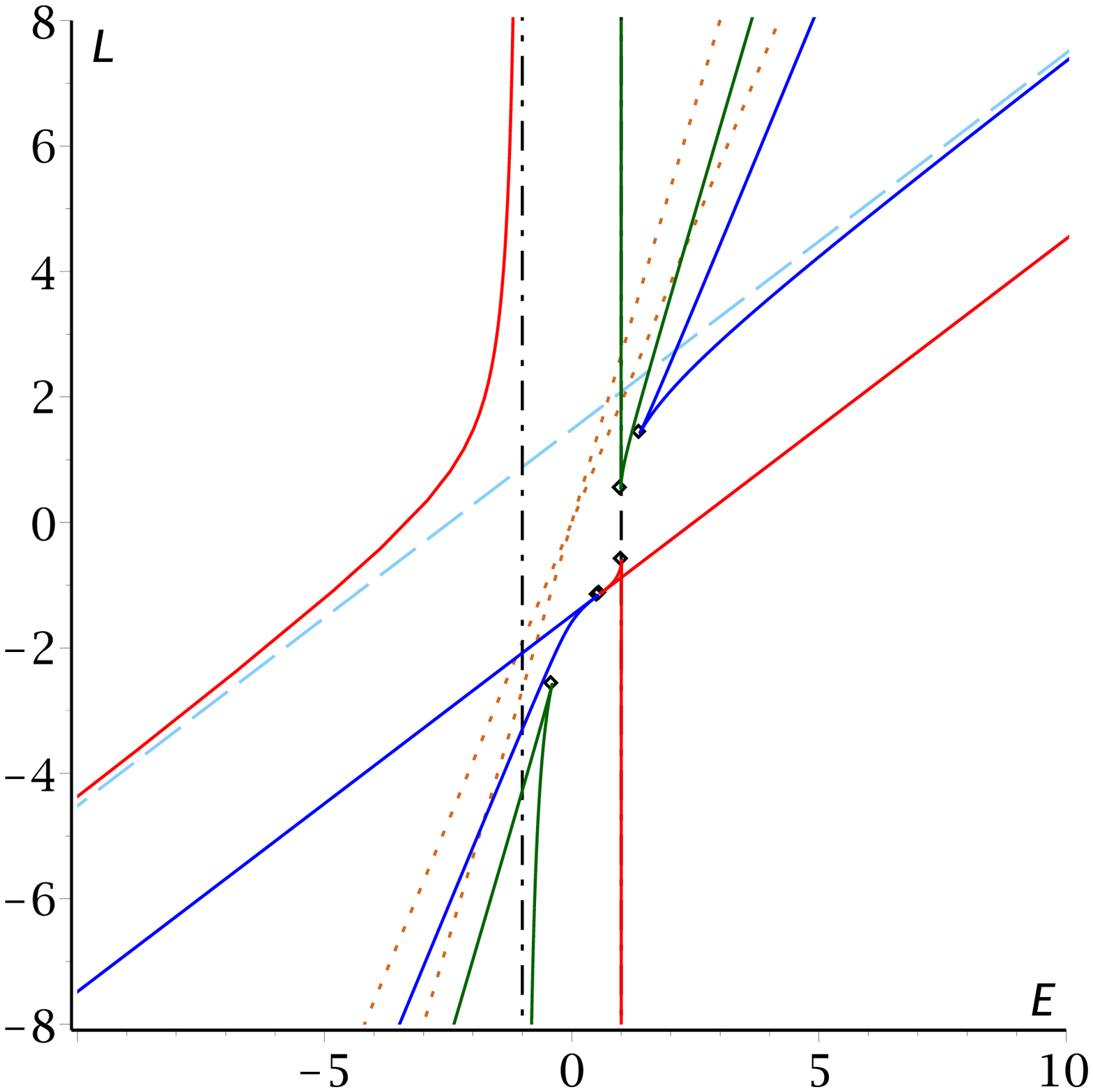}}
\subfloat[$\mP^2=0.63$]{\includegraphics[width=0.3\textwidth]{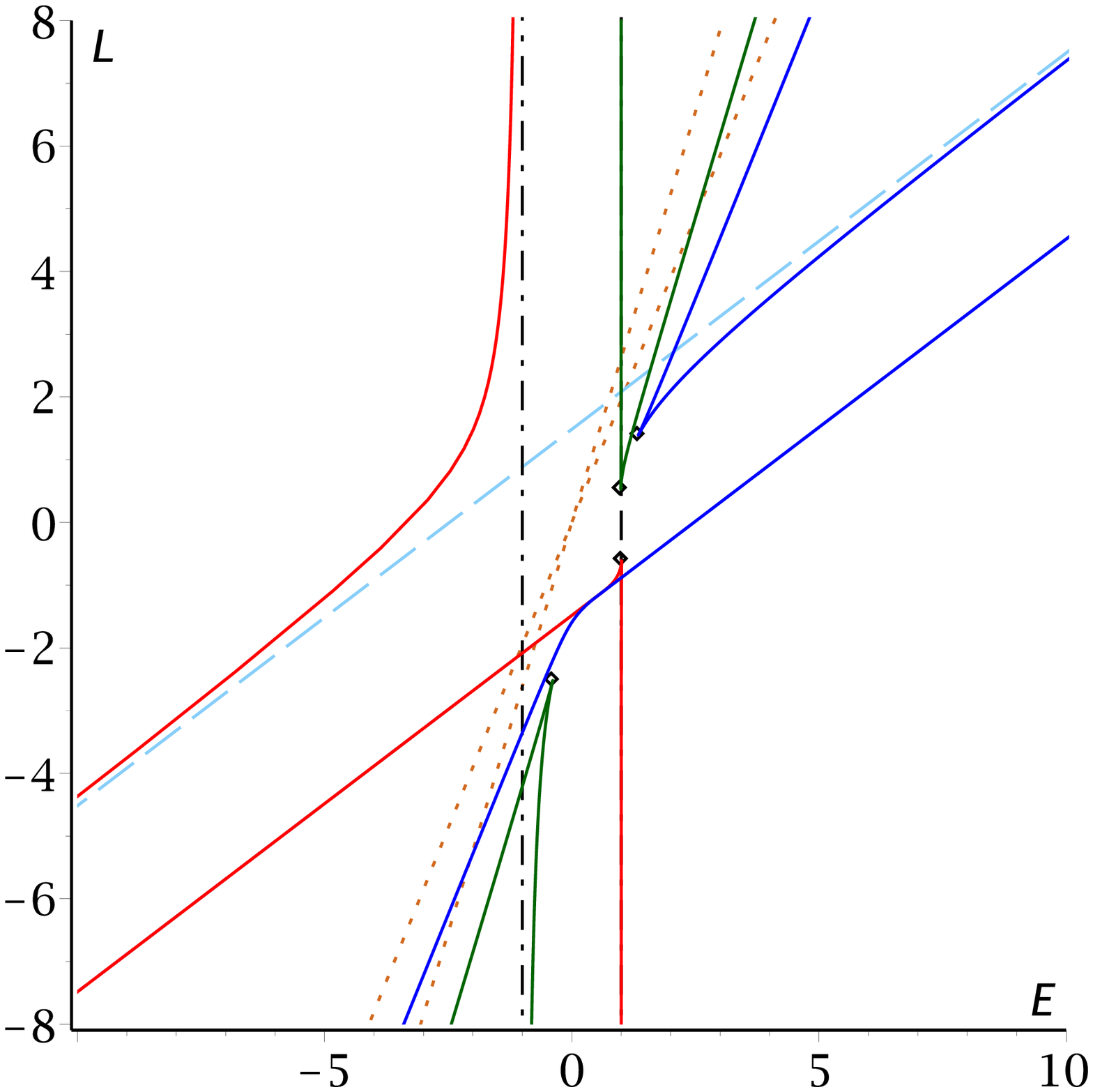}}
\caption{Orbit configurations for the radial motion with $\ba=0.6$, $\bK=0.8$, $e\bQ=0.9$ and varying $\mP^2$. For a general description of colours and linestyles see the text. Note that at $\mP^2 \approx 0.6277$ the plots change from case \textit{(i)} to \textit{(ii)}.}
\label{fig:r_LvsE_series_3}
\end{figure}

\section{Analytical solutions}
We will now solve the equations of motion (\ref{eq:eomtheta}) - (\ref{eq:eomt}) with the initial conditions 
\begin{equation}
\theta(\gamma_0) = \theta_0\,, \quad \br(\gamma_0) = \br_0\,, \quad \phi(\gamma_0) = \phi_0\,, \quad t(\gamma_0) = t_0\,,
\end{equation}
In addition the initial direction, i.e.~the sign of $\frac{dx}{d\gamma}(\gamma_0)$ for $x=\theta,\br$, has to be specified. We denote this by $\sigma_x=\sgn\left(\frac{dx}{d\gamma}(\gamma_0)\right)$.

\subsection{$\theta$-motion}
The equation of motion (\ref{eq:eomtheta}) needs to be solved. We first concentrate on the case in which the poles are not reached, i.e.~$\theta(\gamma) \in (0,\pi)$. Then it is convenient to substitute $\nu = \cos(\theta)$, compare \eqref{eq:eomnu}, and to solve the equivalent equation of motion
\begin{align}
\left( \frac{d\nu}{d\gamma} \right)^2 & = \sum_{i=0}^4 b_i\nu^i = \T_\nu(\nu)\,
\end{align}
with $\nu(\gamma_0) = \nu_0 := \cos( \theta_0) $ and $\sgn\left(\frac{d\nu}{d\gamma}(\gamma_0)\right)=\sigma_\nu:= -\sigma_{\theta}$. If $\T_\nu(\nu)$ has a zero of multiplicity two or more the solution can be solved by elementary functions. In general $\T_\nu(\nu)$ is a polynomial of order four with simple zeroes only and can be solved with the following procedure that uses the Weierstrass elliptic function $\wp$ (see \cite{Markushevich77,eva}. We transform the equation to the Weierstrass form by the substitution $\nu = (\frac{4}{a_3}\xi-\frac{a_2}{3a_3})^{-1}+\nu_{\T}$, where $\nu_{\T}$ is an arbitrary zero of $\T_\nu$ and $a_i = \frac{1}{(4-i)!}\frac{d^{(4-i)}\T_\nu}{d\nu^{(4-i)}}(\nu_{\T})$. This leads to
\begin{align}
\left(\frac{d\xi}{d\gamma}\right)^2 = 4\xi^3 - {{g_\theta}_2}\xi - {{g_\theta}_3}   \label{eq:thetaweier}
\end{align}
with
\begin{align}
{g_\theta}_2 = \frac{1}{12}a_2^2-\frac{1}{4}a_1a_3, \quad {g_\theta}_3=-\frac{1}{216}a_2^3 + \frac{1}{48}a_1a_2a_3-\frac{1}{16}a_0a_3^2\,. \label{eq:g2g3}
\end{align}
The initial conditions are $\xi(\gamma_0) = \xi_0 :=  \frac{1}{4} (\frac{a_3}{\nu_0-\nu_{\T}}+\frac{{a_2}}{3})$ and $\sgn\left(\frac{d\xi}{d\gamma}(\gamma_0)\right)=\sigma_{\xi} := -\sgn(a_3)\sigma_{\nu}$. The solution of eq.~\eqref{eq:thetaweier} can now be expressed in terms of the Weierstrass elliptic $\wp$ function,
\begin{align}
\xi(\gamma) = \wp(\gamma-\gamma_{\theta,\rm in};{g_\theta}_2;{g_\theta}_3)\,,
\end{align}
where $\gamma_{\theta,\rm in}$ is a constant such that $\wp(\gamma_0 - \gamma_{\theta,\rm in}) = \xi_0$ and $\sgn(\wp'(\gamma_0 -\gamma_{\theta, \rm in}))=\sigma_{\xi}$. This is e.g.~fulfilled by $\gamma_{\theta,\rm in} = \gamma_0 - \sigma_{\xi} \int_{\infty}^{\xi_0} \frac{d\tau}{\sqrt{4\tau^3-{g_\theta}_2\tau-{g_\theta}_3 } }$ (with the principal branch of the square root). The solution for $\theta$ is then given by
\begin{align}
\theta(\gamma)=\arccos \left(\frac{a_3}{4\wp(\gamma- \gamma_{\theta,\rm in};{g_\theta}_2;{g_\theta}_3)-\frac{a_2}{3}}+\nu_{\T} \right)  \label{eq:soltheta}.
\end{align}
Let us now consider orbits which reach the poles. First, let $\theta(\gamma)$ be in the open interval $(0,\pi)$ for $\gamma \in (\gamma_1, \gamma_2)$ but on the endpoints $\gamma_{1,2}$ the orbit may reach the poles, $\theta(\gamma_{1,2}) \in \{0,\pi\}$. Then the solution on  $(\gamma_1, \gamma_2)$, given by eq.~\eqref{eq:soltheta}, is also valid on the complete interval $[\gamma_1, \gamma_2]$ because the right hand side of eq.~\eqref{eq:soltheta} is continuous on the whole closed interval with limits $\theta(\gamma_{1,2})$ as $\gamma$ approaches $\gamma_{1,2}$. In general let $\gamma_i$, $i\geq1$, be the parameters with $\theta(\gamma_i) \in \{0,\pi\}$ and $\gamma_{i}<\gamma_{i+1}$. Define $\theta_i := \theta|_{[\gamma_{i-1},\gamma_{i}]}$ and solve the differential equation in each interval with the condition $\theta_i(\gamma_{i-1})=\theta_{i-1}(\gamma_{i-1})$, $\sgn\left(\frac{d\theta_i}{d\gamma}(\gamma_{i-1})\right)=-\sgn\left(\frac{d\theta_{i-1}}{d\gamma}(\gamma_{i-1})\right)$. The switch in sign of $\frac{d\theta_i}{d\gamma}$ in $\gamma_i$ canonically identifies $\theta$ on $[0,\pi]$.

\subsection{$\br$-motion}
The procedure to solve the equation of motion for the $\br$, see (\ref{eq:eomr}) 
\begin{align}
\left( \frac{d\br}{d\g} \right)^2 & = \mR^2(\br)-(\epsilon \br^2  + \bK) \bDelta(\br) = R(\br)
\end{align}
is analogous to that in the previous section. Again, the right hand side is a polynomial of fourth order. If it has a zero of multiplicity two or more the differential equation can be solved in terms of elementary functions. The general case can be solved with the substitution $\br = c_3 \left(4\xi-\frac{c_2}{3}\right)^{-1}+\br_R$, where $\br_R$ is a zero of $R$ and $c_i = \frac{1}{(4-i)!}\frac{d^{(4-i)}R}{d\br^{(4-i)}}(\br_R)$. This leads to
\begin{align}
\br(\g)=\frac{c_3}{4\wp(\gamma-\gamma_{\br,\rm in};{g_{\br}}_2; {g_{\br}}_3)-\frac{c_2}{3}}+\br_R \label{eq:solr}
\end{align}
with ${g_{\br}}_2$, ${g_{\br}}_3$ given as in \eqref{eq:g2g3} with $a_i = c_i$ and ${g_{\br}}_i={g_\theta}_i$. The parameter $\gamma_{\br,\rm in}$ only depends on the initial conditions, $\wp(\gamma_0-\gamma_{\br,\rm in};{g_{\br}}_2;{g_{\br}}_3)=\frac{1}{4}\left(\frac{c_3}{\br_0-\br_R}+\frac{c_2}{3}\right)$ and $\sgn(\wp'(\gamma_0-\gamma_{\br,\rm in};{g_{\br}}_2;{g_{\br}}_3))=-\sgn(c_3)\sigma_{\br}$, e.g.~$\gamma_{\br,\rm in} = \gamma_0 + \sgn(c_3)\sigma_{\br} \int_{\infty}^{\xi_0} \frac{d\tau}{\sqrt{4\tau^3-{g_{\br}}_2\tau-{g_{\br}}_3} }$.

\subsection{$\phi$-motion}
The equation describing the $\phi$-motion, see (\ref{eq:eomphi}), can be rewritten as
\begin{align}
\phi(\g)= \phi_0 + \int_{\g_0}^\g \frac{\ba\mR(\br)}{\bDelta(\br)} d\g - \int_{\g_0}^\g \frac{\mT(\theta)}{\sin^2\theta} d\g\,,
\end{align}
where the right hand side is separated in a part which only depends on $\br$ and one that only depends on $\theta$. We will now treat both integrals separately.
\subsubsection*{The $\theta$-dependent integral}
Let us start with the integral
\begin{align}
I_{\phi_\theta}(\g) = \int_{\g_0}^\g \frac{\mT(\theta)}{\sin^2\theta} d\g.
\end{align}
If we insert the expression for $\theta$ given by eq.~(\ref{eq:soltheta}), which we write symbolically as $\theta=\theta(\wp(\gamma-\gamma_{\theta, \rm in}))$, we get
\begin{align}
I_{\phi_\theta}(\g) = \int_{\g_0}^\g \frac{\mT(\theta(\wp(\gamma-\gamma_{\theta, \rm in}))}{\sin^2\theta(\wp(\gamma-\gamma_{\theta, \rm in}))}d\g = \int_{\g_0}^\g  {R_\phi}_\theta(\wp(\gamma-\gamma_{\theta, \rm in})) d\g
\end{align}
with ${R_\phi}(\theta(\wp(\gamma-\gamma_{\theta, \rm in}))$ a rational function of $\wp(\gamma-\gamma_{\theta, \rm in})$. A partial fraction decomposition then yields
\begin{align}
I_{\phi_\theta}(\g)&= \int_{\gamma_0}^\gamma \alpha_\theta+\frac{{\alpha_\theta}_1}{\wp(\gamma-\gamma_{\theta, \rm in})-{\beta_\theta}_1}+\frac{{\alpha_\theta}_2}{\wp(\gamma-\gamma_{\theta, \rm in})-{\beta_\theta}_2} d\gamma   \label{eq:phitheta}
\end{align}
with 
\begin{align}
\alpha_\theta= \ba E+\frac{e\bP\nu_{\T}-\bL}{1-\nu_{\T}^2}, \qquad {\alpha_\theta}_{1,2}=\frac{1}{8}\frac{(e\bP\pm\bL){a_4}}{(\nu_{\T}\pm 1)^2},\qquad {\beta_\theta}_{1,2}=-\frac{1}{12}\frac{3a_4\pm a_3-\nu_{\T} a_3}{\nu_{\T}\pm1}. 
\end{align}
The integral over each summand in eq.~(\ref{eq:phitheta}) can be expressed in terms of the Weierstrass $\zeta = \zeta (\cdot ; {g_\theta}_2;{g_\theta}_3) $ and $\sigma=\sigma(\cdot ; {g_\theta}_2;{g_\theta}_3)$ function (see \cite{Markushevich77,eva})
\begin{align}
I_{\phi_\theta}(\g)=& \alpha_\theta(\gamma-\gamma_0) + \frac{{\alpha_\theta}_1}{\wp'(v_1)} \Big( \ln\frac{ \sigma(\g-v_1)}{ \sigma(\g_0-v_1)}-\ln\frac{ \sigma(\g+v_1)}{ \sigma(\g_0+v_1)}+2(\g-\g_0)\zeta(v_1) \Big) \nonumber \\
& +\ \frac{{\alpha_\theta}_2}{\wp'(v_2)} \Big( \ln\frac{ \sigma(\g-v_2)}{ \sigma(\g_0-v_2)}-\ln\frac{ \sigma(\g+v_2)}{ \sigma(\g_0+v_2)}+2(\g-\g_0)\zeta(v_2) \Big) \label{eq:solphitheta}
\end{align}
where $v_1, v_2$ need to chosen such that  $\wp(v_1+\gamma_{\theta, \rm in})={\beta_\theta}_1$, and $\wp(v_2+\gamma_{\theta, \rm in})={\beta_\theta}_2$.

\subsubsection*{The $\br$-dependent integral}
The procedure to solve the $\br$-dependent integral
\begin{align*}
I_{\phi_{\br}}(\g)=\int_{\g_0}^\g \frac{\ba\mR(\br)}{\bDelta(\br)} d\g
\end{align*}
is analogous to the previous section. We substitute $\br=\br(\wp(\gamma-\gamma_{\br,\rm in}))$ of eq.~\eqref{eq:solr} and obtain a rational function as integrand, ${R_\phi}_{\br}$ of $\wp(\gamma-\gamma_{\br, \rm in})$, which we decompose into partial fractions
\begin{align}
I_{\phi_{\br}}(\g) & = \int_{\g_0}^\g {R_\phi}_{\br}(\wp(\gamma-\gamma_{\br,\rm in})) d\g\\
& = \int_{\g_0}^\g \alpha_{\br}  + \frac{{\alpha_{\br}}_1}{\wp(\g)-{\beta_{\br}}_1}+ \frac{{\alpha_{\br}}_2}{\wp(\g)-{\beta_{\br}}_2} d\g\,.
\end{align}
Here the constants are given by
\begin{align}
\alpha_{\br}= \ba\frac{\mR(\br_R)}{\bDelta(\br_R)}, \qquad {\beta_{\br}}_{1,2}= \frac{  c_3 }{12} +\frac{\pm\sqrt{-c_4^2(-1+\ba^2+\bQ^2+\bP^2) }-{c_4}{\br}+{c_4}}  {4\bDelta(\br)} \label{eq:coeffsphir}
\end{align}
and ${\alpha_{\br}}_{1,2}$ the coefficients of the partial fraction decomposition. The solution has the the form of eq.~\eqref{eq:solphitheta} with ${\alpha_\theta}_{1,2}={\alpha_{\br}}_{1,2}$, ${\beta_\theta}_{1,2}={\beta_{\br}}_{1,2}$ and $v_1, v_2$ such that  $\wp(v_1+\gamma_{\br, \rm in})={\beta_{\br}}_1$ and $\wp(v_2+\gamma_{\br, \rm in})={\beta_{\br}}_2$. Also the $\wp$, $\zeta$, and $\sigma$ functions refer to the parameters ${g_{\br}}_2$, ${g_{\br}}_3$.

\subsection{t-motion}
The solution to the differential equation describing the $t$-motion, see (\ref{eq:eomt}), can be found in a similar way as for the $\phi$ motion. An equivalent formulation of the differential equation is 
\begin{align}
t(\g)= t_0 + \int_{\g_0}^\g  \frac{(\br^2+\ba^2)\mR(\br)}{\bDelta(\br)}  d\g - \int_{\g_0}^\g  \ba \mT(\theta)  d\g\,,
\end{align}
where the right hand side is a again separated in an $\br$ and a $\theta$ dependent part. We treat these integrals separately.

\subsubsection*{The $\theta$-dependent integral}
We solve the integral
\begin{align*}
{I_t}_\theta(\g):= \int_{\g_0}^\g  \ba \mT(\theta) d\g
\end{align*}
with the same ansatz as in the foregoing section: First, we substitute $\theta = \theta(\wp(\gamma-\gamma_{\theta, \rm in})) $ of eq.~(\ref{eq:soltheta}) and then decompose the integrand ${R_t}_\phi$ which is a rational function of $\wp(\gamma-\gamma_{\theta, \rm in})$ in partial fractions,
\begin{align}
{I_t}_\theta(\g)&=\int_{\g_0}^\g {R_t}_\theta(\wp(\gamma-\gamma_{\theta,\rm in})) d\g \nonumber\\
& = \int_{\g_0}^\g \alpha_\theta  + \frac{{\alpha_\theta}_1}{\wp(\g)-{\beta_\theta}}+ \frac{{\alpha_\theta}_2}{(\wp(\g)-{\beta_\theta})^2} d\g\,,
\end{align}
where
\begin{equation}
\begin{aligned}
\alpha_\theta & =-\ba\bL-\ba^2 E\nu_{\T}^2+\ba^2E+\ba e \bP \nu_{\T}\,, \qquad \beta_\theta=\frac{1}{12} a_3\,,  \\
{\alpha_\theta}_1 & =-\frac{1}{4}\ba a_4 (2\ba E\nu_{\T}-e\bP)\,, \qquad {\alpha_\theta}_2=-\frac{1}{16}\ba^2 E a_4^2\,.
\end{aligned}
\end{equation}
The solution written in terms of $\zeta = \zeta (\cdot ; {g_\theta}_2;{g_\theta}_3) $ and $\sigma=\sigma(\cdot ; {g_\theta}_2;{g_\theta}_3)$ is given by
\begin{align}
{I_t}_\theta(\g) & =  \alpha_\theta(\g-\g_0) + \frac{{\alpha_\theta}_1}{\wp'(v)} \Big( \ln\frac{ \sigma(\g-v)}{ \sigma(\g_0-v)}-\ln\frac{ \sigma(\g+v)}{ \sigma(\g_0+v)}+2(\g-\g_0)\zeta(v) \Big) \nonumber \\
& \quad +\frac{{\alpha_\theta}_2}{\wp'^2(v)} \Big(-\zeta(\g-v)+\zeta(\g_0-v)-\zeta(\g+v)+\zeta(\g+v)-\wp(v)(\g-\g_0) \nonumber\\
& \quad - \frac{\wp''(v)}{2\wp'(v)}\Big( \ln\frac{ \sigma(\g-v)}{ \sigma(\g_0-v)}-\ln\frac{ \sigma(\g+v)}{\sigma(\g_0+v)}+2(\g-\g_0)\zeta(v) \Big)\Big)\label{eq:solttheta}
\end{align}
where $v$ such that  $\wp(v+\gamma_{\theta, \rm in})={\beta_\theta}$.

\subsubsection*{The $\br$-dependent integral}
Now the Integral 
\begin{align}
{I_t}_{\br} := \int_{\g_0}^\g  \frac{(\br^2+\ba^2)\mR(\br)}{\bDelta(\br)}  d\g	
\end{align}
will be solved. The substitution $\br=\br(\wp(\gamma-\gamma_{\br,\rm in}))$ of eq.~\eqref{eq:solr} again leads to a rational function ${R_t}_{\br}$ of $\wp(\gamma-\gamma_{\br,\rm in})$, which becomes after a partial fraction decomposition
\begin{align}
{I_t}_{\br} & = \int_{\g_0}^\g {R_\phi}_{\br}(\wp(\gamma-\gamma_{\br,\rm in})) d\g \nonumber\\
& = \int_{\g_0}^\g \alpha_{\br}  + \frac{{\alpha_{\br}}_1}{\wp(\g)-{\beta_{\br}}_1}+ \frac{{\alpha_{\br}}_2}{\wp(\g)-{\beta_{\br}}_2} d\g\,.
\end{align}
Here $\alpha_{\br}$, ${\beta_{\br}}_{1,2}$ are defined as in eq.~\eqref{eq:coeffsphir}, and ${\alpha_{\br}}_{1,2}$ are the coefficients of the partial fraction decomposition. The solution has the the form of eq.~\eqref{eq:solphitheta} with ${\alpha_\theta}_{1,2}={\alpha_{\br}}_{1,2}$, ${\beta_\theta}_{1,2}={\beta_{\br}}_{1,2}$ and $v_1, v_2$ such that $\wp(v_1+\gamma_{\br, \rm in})={\beta_{\br}}_1$ and $\wp(v_2+\gamma_{\br, \rm in})={\beta_{\br}}_2$. The $\wp$, $\zeta$, and $\sigma$ functions here refer to ${g_{\br}}_2$ and ${g_{\br}}_3$.

\subsection{Examples}
\begin{figure}
\includegraphics[width=0.4\textwidth]{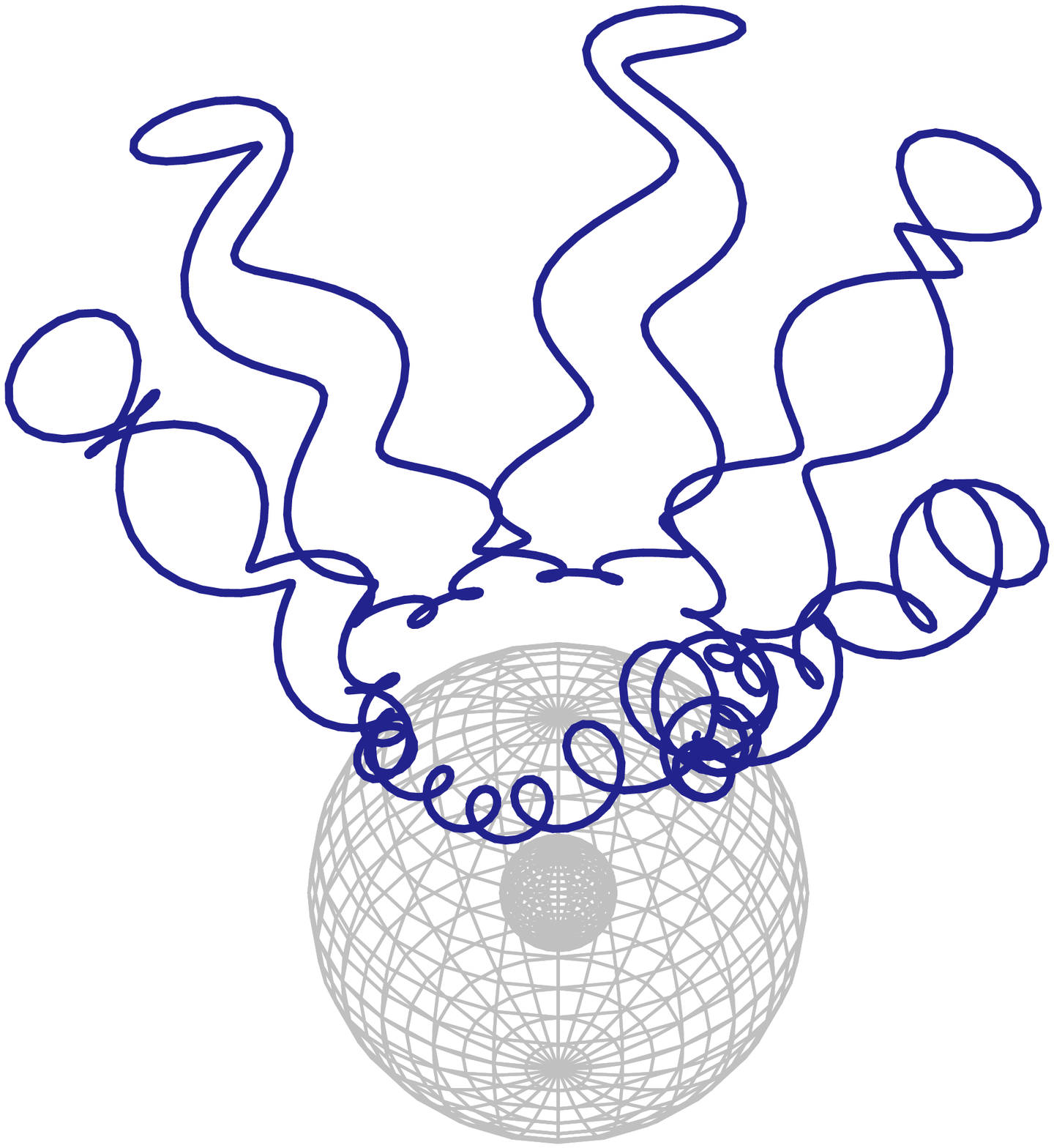}\quad
\includegraphics[width=0.45\textwidth]{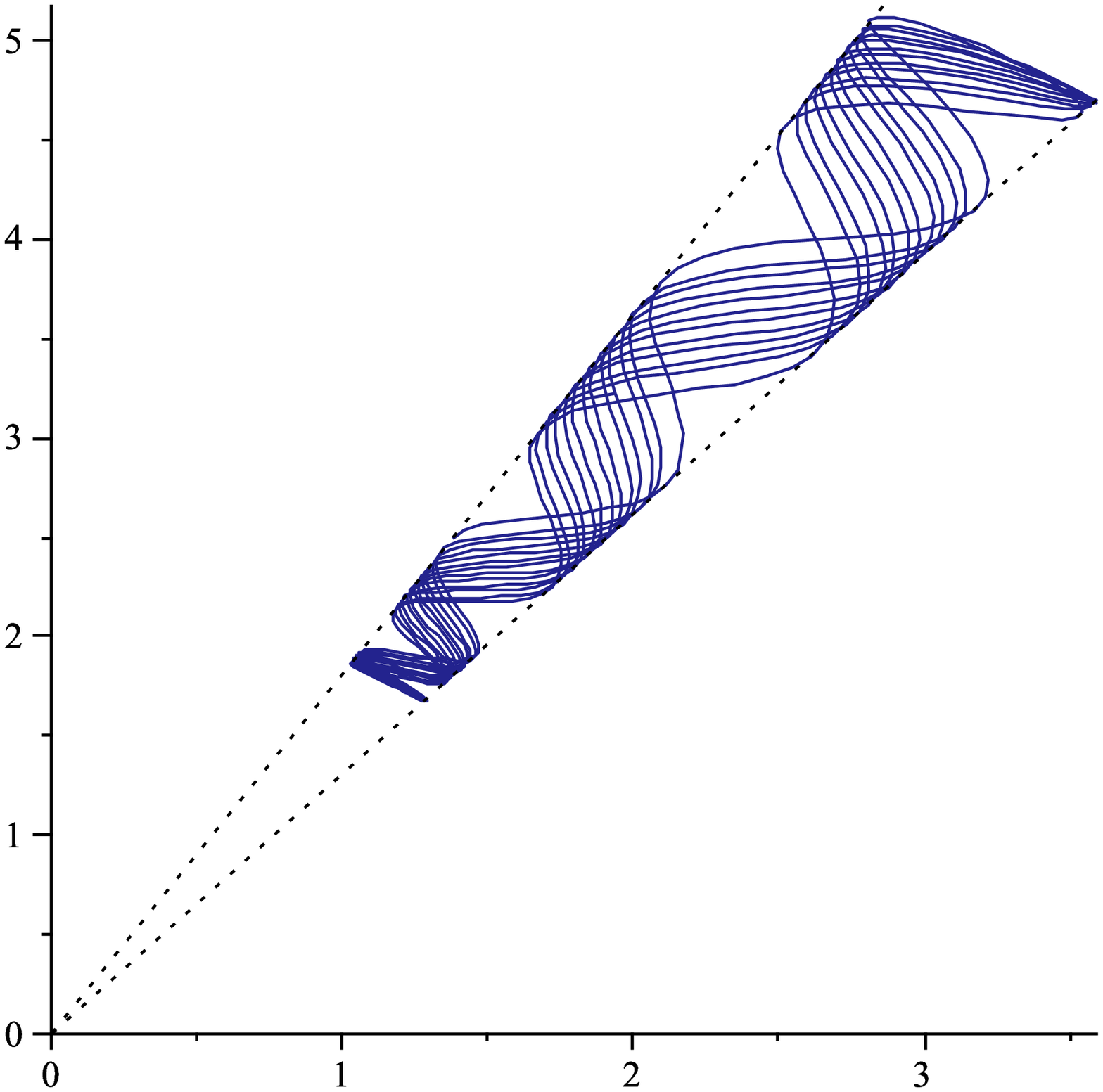}
\caption{Charged particle orbit in Kerr-Newman space-time. Here a $B_{+}$ orbit in the northern hemisphere ($N$ orbit) with $\ba=0.6$, $\bP=0.47$, $\bQ=0.16$, $e=158.11$, $E=1.98$, $\bL=-62.62$, $\bK=33.33$ is shown. Left: three-dimensional plot. The gray spheres correspond to the horizons. Right: projection on $(x,z)$ plane. The dotted lines correspond to extremal $\theta$ values.}
\label{Fig:B+N}
\end{figure}

\begin{figure}
\includegraphics[width=0.32\textwidth]{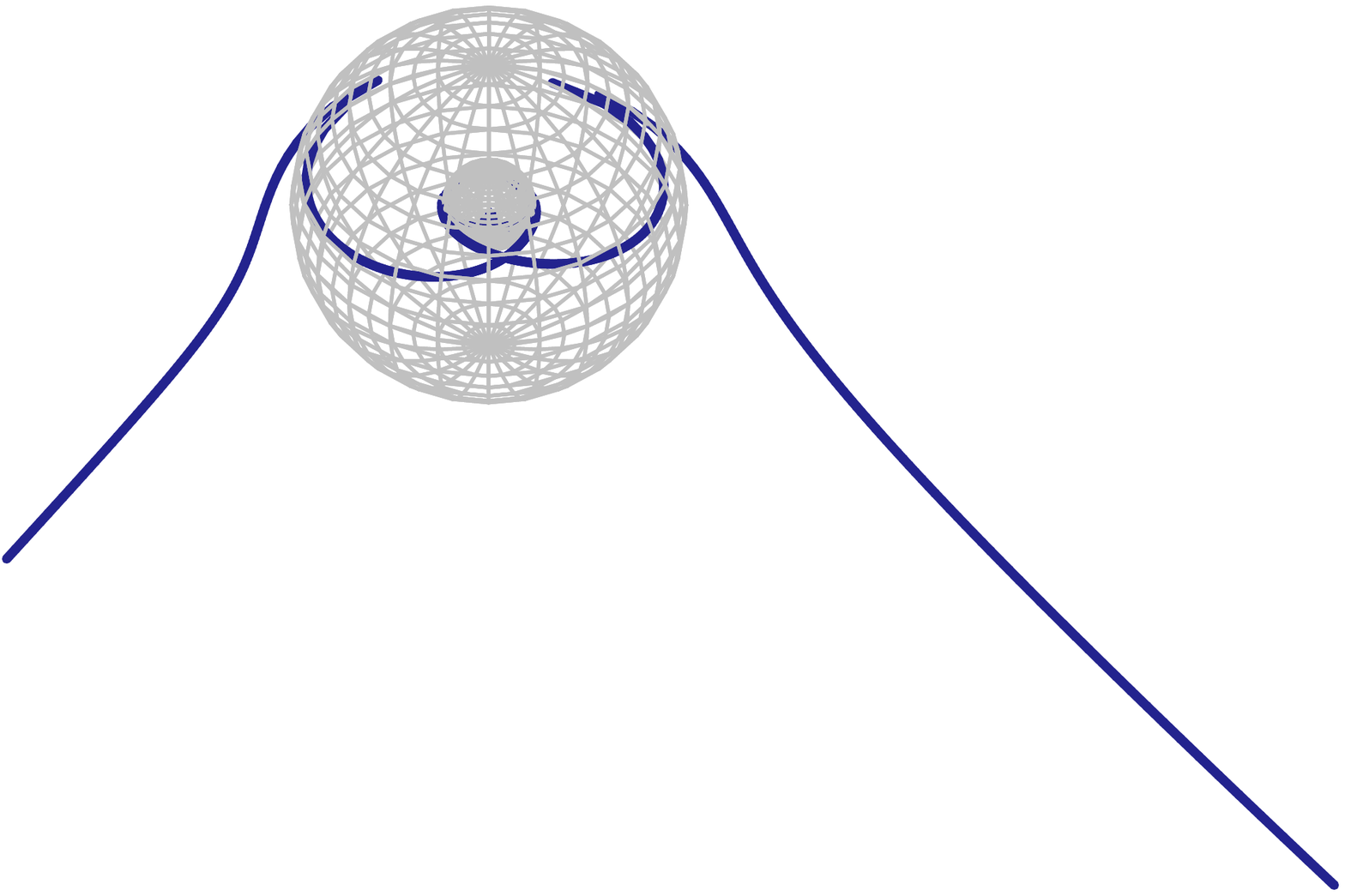}
\includegraphics[width=0.32\textwidth]{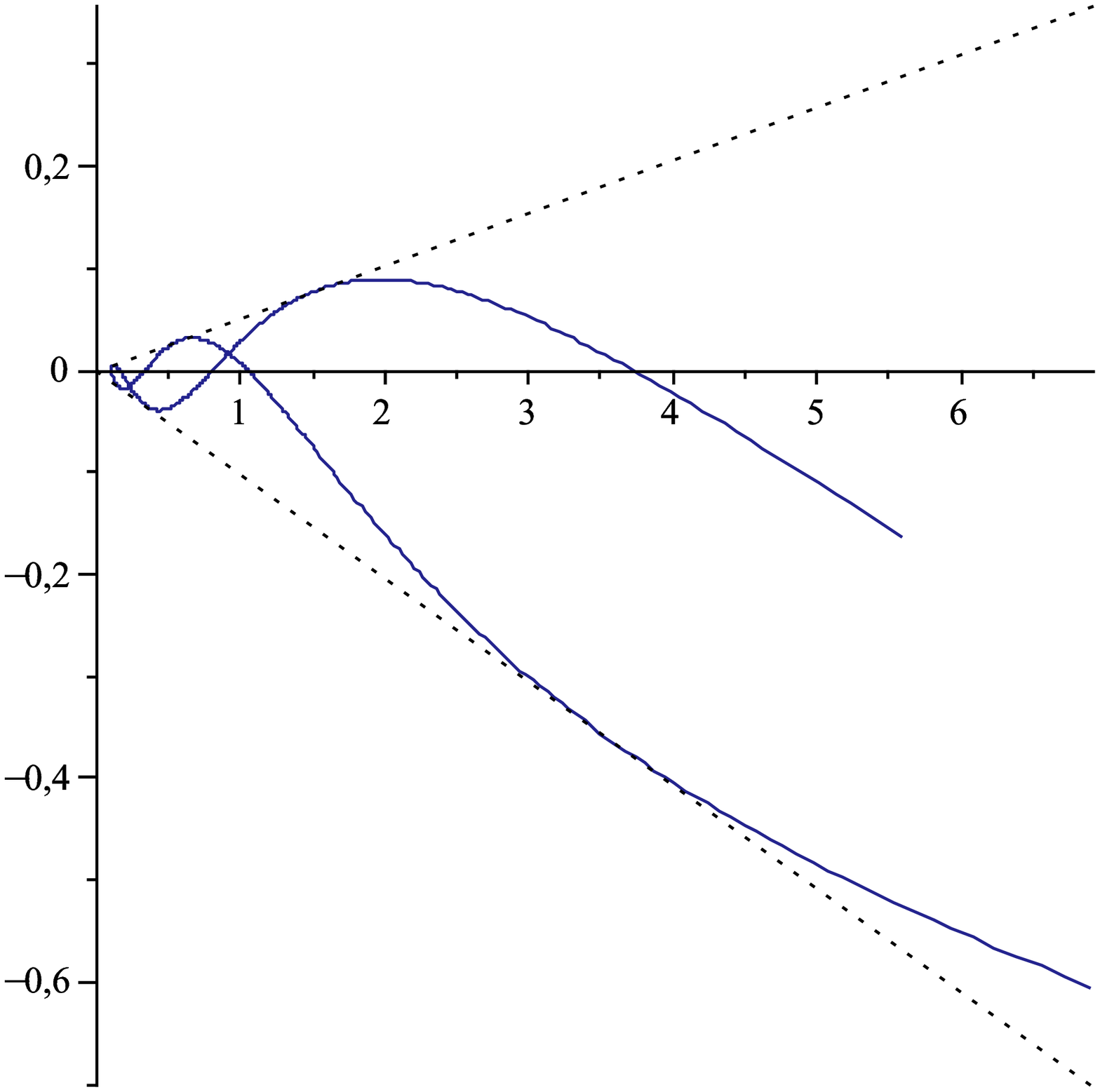}
\includegraphics[width=0.32\textwidth]{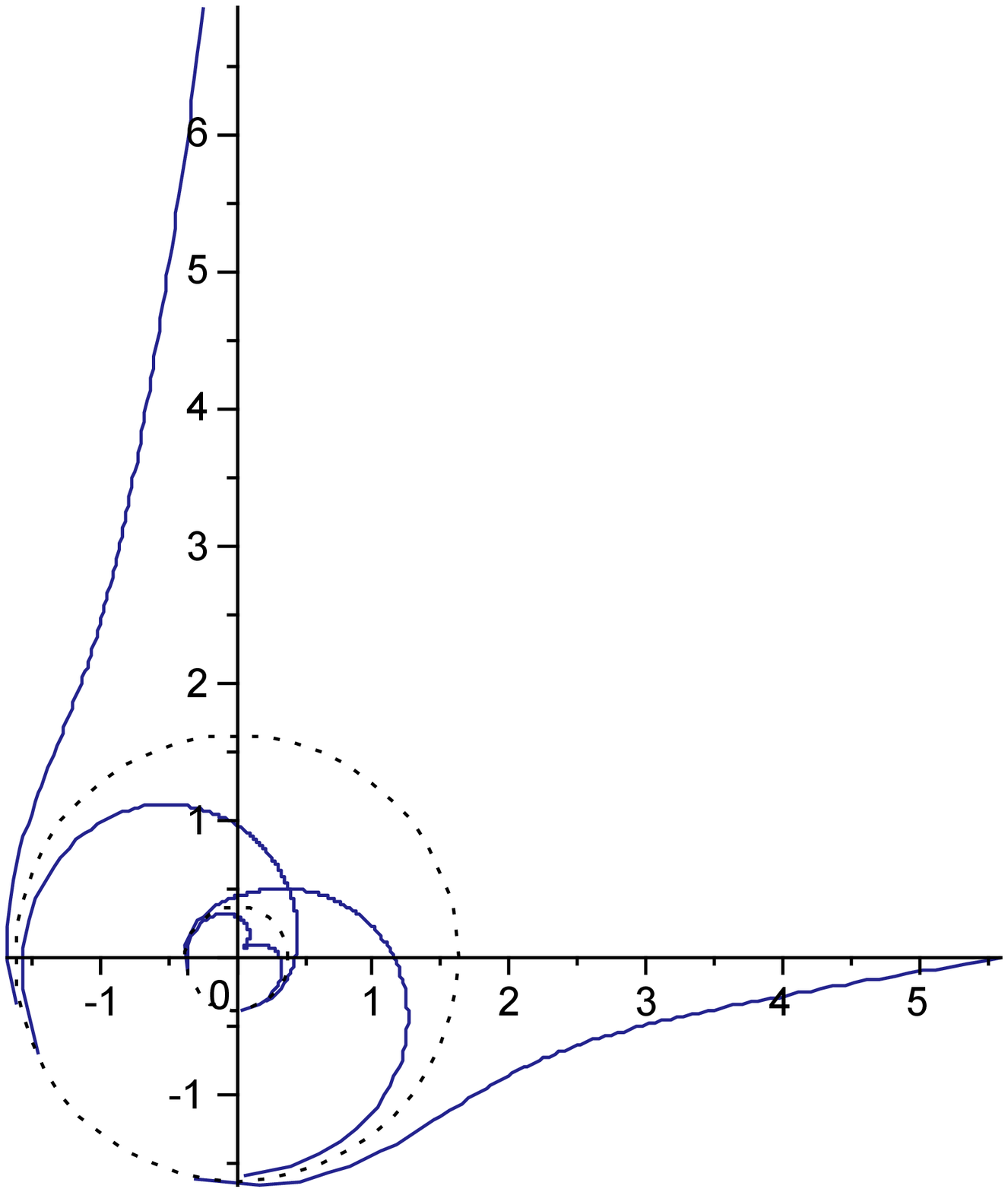}
\caption{Charged particle orbit in Kerr-Newman space-time. Here a $F_{+}^*$ orbit in the northern hemisphere ($N$ orbit) with $\ba=0.6$, $\bP=0.47$, $\bQ=0.16$, $e=158.11$, $E=-11.83$, $\bL=-5.18$, $\bK=33.33$ is shown. Left: three-dimensional plot. The gray spheres correspond to the horizons. Middle: projection on $(x,z)$ plane. The dotted lines correspond to extremal $\theta$ values. Right: projection on $(x,y)$ plane. Dotted lines correspond to the horizons. Note that $\phi(\g)$ diverges at the horizons at some finite $\g_0$. In this plot we stopped at some $\g_0-\Delta_{\g}$ ($\Delta_{\g}$ small) and continued with $\g_0+\Delta_{\g}$ on the other side of the horizon.}
\label{Fig:F+*N}
\end{figure}

The analytical solutions to the equations of motions are given by \eqref{eq:soltheta}, \eqref{eq:solr}, \eqref{eq:solphitheta}, and \eqref{eq:solttheta}, respectively with the appropriate constants \eqref{eq:coeffsphir}. Here, we use these results to exemplify the orbit structure in Kerr-Newman space-time, see Figs.~\ref{Fig:B+N} and \ref{Fig:F+*N}.

\subsection{Periastron shift and Lense-Thirring effect}
In General Relativity bound orbital motion can be much more complicated then the closed ellipses of Newtonian Gravity. However, in the weak field they are quite similar and the deviation can be characterized by a precession of the orbital ellipse, called the periastron shift, and the orbital plane, called the Lense-Thirring effect. They are caused by a mismatch between the periodicity of $r(\varphi)$ ($\theta(\varphi)$) with $2\pi$. These notion may be generalized to orbits in the strong field as demonstrated by Schmidt \cite{Schmidt02} and Fujita and Hikida \cite{FujitaHikida09} for Kerr space-time. An analogous treatment is also possible in Kerr-Newman space-times.
  
For bound orbits, the radial and colatitudinal periods $\varLambda_r$ and $\varLambda_\theta$ with respect to Mino time are defined by the smallest non-zero real value with $r(\lambda+\varLambda_r)=r(\lambda)$ and $\theta(\lambda+\varLambda_\theta)=\theta(\lambda)$ giving (see \eqref{eq:eomr} and \eqref{eq:eomtheta})
\begin{align}
\varLambda_r & = 2 \int_{\br_{\textrm p}}^{\br_{\textrm a}} \frac{d\br}{\sqrt{R(\br)}}\,, \qquad \varLambda_\theta = 2 \int_{\theta_{\textrm min}}^{\theta_{\textrm max}} \frac{d\theta}{\sqrt{\Theta(\theta)}}\,,
\end{align}
where $\br_{\textrm p}$ is the periapsis and $\br_{\textrm a}$ the apoapsis. To determine the periodicity with respect to $\varphi$, we need to know the secular accumulation of $\varphi$ with respect to the Mino time. This can be achieved by setting
\begin{align}
\varphi(\lambda)=\langle \Phi(r,\theta) \rangle_\lambda \lambda + \Phi^r_{\textrm osc}(r) + \Phi^\theta_{\textrm osc}(\theta)
\end{align}
where $\Phi(r,\theta)=\Phi_r(r)+\Phi_\theta(\theta)$ is the right hand side of \eqref{eq:eomphi}. Here 
\begin{equation}
\Upsilon_\varphi := \langle \Phi(r,\theta) \rangle_\lambda:=\lim_{(\lambda_2-\lambda_1) \to \infty} \frac{1}{2(\lambda_2-\lambda_1)} \oint_{\lambda_1}^{\lambda_2} \Phi(r,\theta) \, d\lambda
\end{equation} 
is an infinite Mino time average and $\Phi^x_{\textrm osc}(x)$ represent oscillatory deviations from this average.
Using the averaged $\varphi(\lambda)=\Upsilon_\varphi\lambda$, the periodicity of $r(\varphi)$ ($\theta(\varphi)$) is then given by $\Upsilon_\varphi\varLambda_r$ ($\Upsilon_\varphi\varLambda_\theta$). Accordingly, the periastron shift and the Lense-Thirring effect per revolution can be computed by
\begin{align}
\Delta_{\rm P} & = \Upsilon_\varphi \varLambda_r - 2\pi\,, \qquad \Delta_{\rm LT} = \Upsilon_\varphi \varLambda_\theta - 2\pi\,.
\end{align}
For neutral test particles and small $\ba$ the periastron shift and the Lense-Thirring effect was calculated to first order in \cite{Hackmann3}.

\section{Summary and Conclusions}
In this paper we discussed the motion of charged particles in the gravitational field of Kerr-Newman space-times describing stationary rotating black holes with electric and magnetic charge. We demonstrated that it is sufficient to consider test-particles with electric charge only as an additional magnetic charge would only lead to reparametrization. After that we classified the orbits in radial and colatitudinal direction. For both a large variety of orbit configurations was identified as summarized in tables \ref{tab:theta} and \ref{tab:radial}. In particular, we also identified orbits crossing the horizons or $r=0$. These configurations were then assigned to regions in the parameter space pictured in several figures. The boundaries of these regions are amongst others given by parameter combinations which represent orbits of constant $r$ or $\theta$, which were discussed in detail. For all orbit configurations analytical solutions to the equations of motion were presented in terms of elliptic functions dependent on the Mino time.

For the sake of completeness we considered here a black hole endowed with magnetic charge although such was not found until now. This has a big impact on the motion in $\theta$ direction in contrast to electric charge, which does not influence the colatitudinal motion at all. Not only deviates the motion from the symmetry to the equatorial plane for a nonvanishing magnetic charge but also additional types of orbits appear. For example, stable off-equatorial circular orbits outside the horizon do exist in this case, which are not possible else \cite{Kovaretal2008}. They are given by the intersection of a red solid line with $E^2<1$ as in Fig.~\ref{fig:theta_LvsE_K>a^2} and a green solid line with $E^2<1$ as in Fig.~\ref{fig:r_LvsE_i}. (E.g. $\ba=0.5$, $\mQ=0.3$, $\mP=0.6$, $e\bP\approx2.1$, $\bK=1$, $E\approx0.9$, and  $\bL\approx 2.4$ results in an orbit with constant $\br=3$ and $\cos\theta\approx0.8$.) On the contrary, the magnetic charge does not influence the radial motion as is appears only in the combination $\bP^2+\bQ^2$. Still, to our knowledge the discussion of the radial motion in this paper is the most complete so far for Kerr-Newman space-times.

The analytical solution presented here are largely based on the 19th century mathematics of elliptic functions already used by Hagihara \cite{Hagihara} to solve the geodesic equation in Schwarzschild space-time. However, a key ingredient here is the introduction of the Mino time which decouples the radial and colatitudinal equations of motion. We presented the results here in terms of Weierstrass elliptic functions which may be rewritten in terms of Jacobian elliptic functions. The advantage of our presentation is that one formula is valid for all orbit configurations.

\section*{Acknowledgment}
We would like to thank Claus L\"ammerzahl for suggesting this research topic and for helpful discussions. E.H. acknowledges financial support from the German Research Foundation DFG and support from the DFG Research Training Group 1620 ‘‘Models of Gravity’’.

\bibliography{literature}\bibliographystyle{unsrt}
\end{document}